\documentclass[fleqn,usenatbib]{mnras} 
\usepackage{newtxtext,newtxmath}  
\usepackage{amsmath}
\usepackage{graphicx}
\usepackage{float}
\usepackage{placeins}

\usepackage{color}
\newcommand{\adb}[1]{{#1}}

\def\rff{\par\vspace{2pt plus 1pt minus 1pt}\noindent \hangindent 9pt}

\newcommand{\equ}[1]{eq.~(\ref{eq:#1})}
\newcommand{\equs}[1]{eqs.~(\ref{eq:#1})}
\newcommand{\equm}[1]{(\ref{eq:#1})}
\newcommand{\Equ}[1]{Eq.~(\ref{eq:#1})}

\newcommand{\equnp}[1]{eq.~\ref{eq:#1}}
\newcommand{\se}[1]{\S\ref{sec:#1}}
\newcommand{\fig}[1]{Fig.~\ref{fig:#1}}

\newcommand{\Fig}[1]{Figure~\ref{fig:#1}}

\newcommand{\tab}[1]{Table~\ref{tab:#1}}
\newcommand{\be}{\begin{equation}}
\newcommand{\ee}{\end{equation}}
\newcommand{\ba}{\begin{align}}
\newcommand{\ea}{\end{align}}
\newcommand{\bad}{\begin{equation} \begin{aligned}}
\newcommand{\ead}{\end{aligned} \end{equation}}
\newcommand{\bea}{\begin{eqnarray}}
\newcommand{\eea}{\end{eqnarray}}

\def\la{\langle}

\newcommand{\bul}{$\bullet\ $}

\newcommand{\no}{\noindent}

\newcommand{\msun}{M_\odot}
\newcommand{\Msun}{M_\odot}

\newcommand{\Zsun}{Z_\odot}

\newcommand{\ifm}[1]{\relax\ifmmode#1\else$\mathsurround=0pt #1$\fi}
\newcommand{\kms}{\ifmmode\,{\rm km}\,{\rm s}^{-1}\else km$\,$s$^{-1}$\fi}

\newcommand{\Mpc}{\,{\rm Mpc}}
\newcommand{\kpc}{\,{\rm kpc}}
\newcommand{\pc}{\,{\rm pc}}
\newcommand{\cm}{\,{\rm cm}}
\newcommand{\Gyr}{\,{\rm Gyr}}

\newcommand{\Myr}{\,{\rm Myr}}

\newcommand{\yr}{\,{\rm yr}}

\newcommand{\ergs}{\,{\rm erg}\,{\rm s}^{-1}}

\newcommand{\cmc}{\,{\rm cm}^{-3}}

\newcommand{\ltsima}{$\; \buildrel < \over \sim \;$}
\newcommand{\lsim}{\lower.5ex\hbox{\ltsima}}
\newcommand{\gtsima}{$\; \buildrel > \over \sim \;$}
\newcommand{\gsim}{\lower.5ex\hbox{\gtsima}}

\newcommand{\dd}{{\rm d}}

\def\omm{\Omega_{\rm m}}
\def\oml{\Omega_{\Lambda}}

\def\Mv{M_{\rm v}}

\def\Mvtwelve{M_{{\rm v},12}}
\def\Mveight{M_{{\rm v},10.8}}

\def\Mvdot{\dot{M}_{\rm v}}
\def\Rv{R_{\rm v}}
\def\Vv{V_{\rm v}}
\def\Tv{T_{\rm v}}

\def\Ms{M_{*}}

\def\Re{R_{\rm e}}

\def\Sig1{\Sigma_1}

\def\fb{f_{\rm b}}

\def\brho{\bar\rho}

\def\rhos{\rho_{\rm s}}

\def\Rs{R_{\rm s}}
\def\ns{n_{\rm s}}
\def\Ns{N_{\rm s}}
\def\Vs{V_{\rm s}}

\def\Ts{T_{\rm s}}

\def\eps2{\epsilon_{-2}}

\def\eps{\epsilon}

\def\ssim{\!\sim\!}
\def\seq{\!=\!}
\def\ssimeq{\!\simeq\!}
\def\sequiv{\!\equiv\!}
\def\sgt{\!>\!}
\def\slt{\!<\!}
\def\sgsim{\!\gsim\!}
\def\slsim{\!\lsim\!}
\def\sgeq{\!\geq\!}
\def\sleq{\!\leq\!}

\def\sdash{\!-\!}
\def\stimes{\!\times\!}
\def\sprop{\!\propto\!}
\def\spropto{\!\propto\!}

\def\cs{c_{\rm s}}

\def\nc{n_{\rm c}}

\def\z110{(1+z)_{10}}
\def\lam25{\lambda_{.025}}
\def\Sq067{\Sigma_{Q=0.67}}
\def\Sigq067{\Sigma_{Q=0.67}}

\def\Ms{M_{*}}

\def\Mdotac{\dot{M}_{\rm ac}}

\def\zquench{z_{\rm quench}}
\def\Muv{M_{\rm uv}}

\def\Mtu{\mathcal{M}_{\rm t}}
\def\Rtu{{R}_{\rm turb}}
\def\Rturb{{R}_{\rm turb}}
\def\Rheat{R_{\rm heat}}
\def\Lbh{L_{\rm bh}}

\def\Ts{T_{\rm s}}
\def\Vs{V_{\rm s}}
\def\tshear{t_{\rm shear}}

\def\Rscrit{R_{\rm s,crit}}
\def\tshear{t_{\rm shear}}
\def\tcoolmix{t_{\rm cool,mix}}
\def\Rs{R_{\rm s}}
\def\ns{n_{\rm s}}
\def\rhos{\rho_{\rm s}}
\def\Vs{V_{\rm s}}
\def\Ts{T_{\rm s}}
\def\cs{c_{\rm s}}

\def\machb{\mathcal{M}_{\rm b}}

\defcitealias{dekel23}{D23}
\defcitealias{li23}{L23}

\title[Bimodality and Post-FFB Evolution]
{From FFB Starbursts at Cosmic Dawn to Quenching at Cosmic Morning:\\ 
Hi-z Galaxy Bimodality}

\author[Dekel et al.]
{\parbox[t]{\textwidth}
{Avishai Dekel$^{1}$\thanks{E-mail: dekel@huji.ac.il},
Nir Mandelker$^1$,
Zhaozhou Li$^1$,
Zhiyuan Yao$^1$,
Bocheng Zhu$^1$,
Sharon Lapiner$^1$,
Dhruba Dutta Chowdhury$^1$,
Omri Ginzburg$^1$
}
\\ \\ 
$^1$Racah Institute of Physics, The Hebrew University, Jerusalem 91904 Israel
}

\begin{document}

\large  

\pagerange{\pageref{firstpage}--\pageref{lastpage}} \pubyear{2002}

\maketitle

\label{firstpage}

\begin{abstract}
We \adb{propose} a mass-dependent bimodality in the early evolution of galaxies.
The massive track connects the super-bright galaxies at cosmic dawn 
($z \sgt 8$) to the super-massive quiescent galaxies and 
black holes (BHs) at cosmic morning ($z \ssim 4 \sdash 7$). 
The dark-matter halos $>\!10^{10.5}\msun$ at $z\seq 10$ are 
\adb{expected} to undergo
feedback-free starbursts (FFB) with high star-formation efficiency in 
dense star clusters within compact galaxies. 
The less massive halos avoid FFB and form stars gradually under stellar 
feedback, \adb{possibly}
leading to the peak star-forming galaxies at cosmic noon 
($z \ssim 1\sdash 3$).
The FFB and non-FFB halos originate from $>\!4\sigma$ and $2\sdash 3\sigma$
density peaks, respectively.
The post-FFB galaxies quench their star formation soon after the 
FFB phase and remain quiescent due to 
(a) gas depletion by the FFB starbursts and outflows,
(b) compaction events driven by angular-momentum loss in colliding 
streams within the high-sigma-peak FFB halos,
(c) turbulent circum-galactic medium (CGM) 
that suppresses feeding by cold streams,
and (d) BH feedback, being a key for complete quenching.
BH feedback is enhanced by FFB-driven BH seeding and growth.
\adb{It seems capable} 
of disrupting the streams by generating CGM turbulence or 
photo-heating, \adb{but this remains an open challenge}.
The cosmic-morning quiescent galaxies are expected to be massive,
compact, showing signatures of compaction, outflows and AGN, with
a comoving number density $\sim\!10^{-5}\Mpc^{-3}$, comparable to 
the super-bright galaxies at cosmic dawn and the AGN at cosmic morning.
Their UV luminosity function 
\adb{is predicted to}
peak about $\Muv \ssim -22$ and contribute $\sim\! 10\%$ of the galaxies there.
\end{abstract}


\section{Introduction}
\label{eq:intro}

\subsection{Bright Galaxies at Cosmic Dawn}

Observations using JWST reveal a very interesting excess of bright galaxies 
at ``cosmic dawn", in the redshift range $z \seq 8 \sdash 15$ (and possibly
beyond).
\adb{This is} 
compared to the predictions of standard models within the $\Lambda$CDM
cosmological framework, that were largely extrapolated from lower redshifts
\citep[e.g.,][]{naidu22,haslbauer22,finkelstein22b,finkelstein23,
boylan23,donnan23a,donnan23b,labbe23,bouwens23,mason23,lovell23,adams23,
perez23,arrabal-haro23,wilkins23,yung24,weibel24_smf}.

\smallskip 
Such an excess may partly arise if one assumes a particularly high 
luminosity-to-mass ratio, which could originate from 
(a) low dust attenuation \citep{fiore23,ferrara23,ziparo23},
(b) a top-heavy stellar initial mass function (IMF) 
\citep{zackrisson11,inayoshi22,steinhardt23,harikane23},
(c) UV from active galactic nuclei (AGN) 
\citep{tacchella23b, grudic23, yung24},
or 
(d) a bias due to a fluctuative star-formation history (SFH) 
at the declining bright end of the galaxy luminosity function 
\citep{sun23a,sun23b}.
The excess may alternatively arise by modifying the basic $\Lambda$CDM
scenario, e.g., by a fine-tuned early dark energy phase \citep{klypin21}.

\smallskip 
However, 
such a luminosity excess or changes to the basic framework do not need to be 
advocated, because an excess in the stellar masses themselves is naturally 
expected in the high-density conditions at cosmic dawn based on a 
straightforward physical model within the $\Lambda$CDM framework.
\adb{This model}
predicts high star-formation efficiency due to 
{\it feedback-free starbursts} (FFB) in thousands of star clusters within 
early compact galaxies \citep{dekel23,li24}.
The trivial key idea is that when the gas density is above a threshold of 
$n \ssim 3\stimes 10^3 \cm^{-3}$, the free-fall star formation
is largely complete in $\sim\!1\Myr$. 
\adb{This is} before the onset of suppressive 
feedback from supernovae and stellar winds from  
the massive stars with lifetimes of a few Megayears. 
At moderately low metallicities, such a density is also 
associated with a cooling time $<\!1\Myr$ that guarantees a short $\sim\!1\Myr$
burst followed by a feedback-free window of opportunity.
Once the surface density is also above a threshold, 
$\Sigma \ssim 3\stimes 10^3 \Msun\pc^{-2}$, the radiative feedback is confined
by the self-gravity of the star-forming clouds and does not suppress the 
star-formation rate (SFR) 
in the first free-fall times \citep{menon23, grudic23}.

\smallskip 
The FFB phase is \adb{predicted} to favorably occur at $z\sgt 8$ in halos of
masses $\sim\! 10^{11}\msun$ and to last for $\sim\!100\Myr$ with an average
SFR $\sim\! 50 \sdash 100\msun\yr^{-1}$.
The SFH is predicted to be bursty, consisting of several generations of 
$\sim\!10\Myr$, each starting with a burst followed by a quenched period.
The metallicity and dust attenuation are predicted to be low.
Together, these features also contribute to the elevated brightness 
which has been primarily caused by the suppressed feedback.

A hint that is potentially consistent with the FFB predictions is provided by
JWST observations using gravitational lensing, which reveal high-redshift 
galaxies, possibly disks, where the star formation is predominantly in 
massive, dense star clusters
\citep{vanzella23,claeyssens23,mowla24,mowla24_nat,mowla24_gems,
adamo24,messa24,fujimoto25}.
The detailed relation between these observed systems and the FFB scenario is 
yet to be investigated \citep{mayer25}, and is beyond the scope of the current
paper.

\subsection{Massive Quiescent Galaxies at Cosmic Morning}
\label{sec:quiescent}

The focus of the current paper is the growing evidence from JWST observations 
for an excess of massive quiescent galaxies several hundred million years 
after the FFB bright phase at cosmic dawn,
in the period we term ``cosmic morning", $z \seq 4 \sdash 7$
\citep{carnall23, carnall23_z35, carnall24, degraaff24, wang24, 
setton24, antwidanso25, turner25, weibel25}.
The preliminary stellar mass estimates of these galaxies are in the ball park 
of $\Ms \ssim 10^{10.5}\msun$, sometimes even $10^{11}\msun$ 
\adb{and higher}.
With a global, integrated star-formation efficiency (SFE)
of $\sim\!0.1$ at cosmic morning
(the ratio of stellar mass to baryonic mass that has been 
accreted onto the halo), 
the corresponding halo masses are expected to be of $\Mv \gsim 10^{12}\msun$.
The observed galaxies are typically compact, with a stellar half-mass radius
of $\Re \ssim 200\pc$, and with a high stellar surface density 
$\Sigma \sgt 10^4\Msun\pc^{-2}$ \citep[e.g. Fig.~3 of][]{weibel25}.
The number density of such galaxies seems to be one-to-two orders of magnitude
above the expectations from 
\adb{current cosmological hydro-gravitational simulations} 
\citep[e.g. Fig.~6 of][]{weibel25}.
\adb{This is}
in the ball park of the excess of bright galaxies at cosmic dawn, namely at
a level of $n(>\Mv)\sim\! 10^{-5} \Mpc^{-3}$.

\begin{figure} 
\centering
\includegraphics[width=0.49\textwidth,trim={0.2cm 0.2cm -0.1cm 0.0cm},clip]
{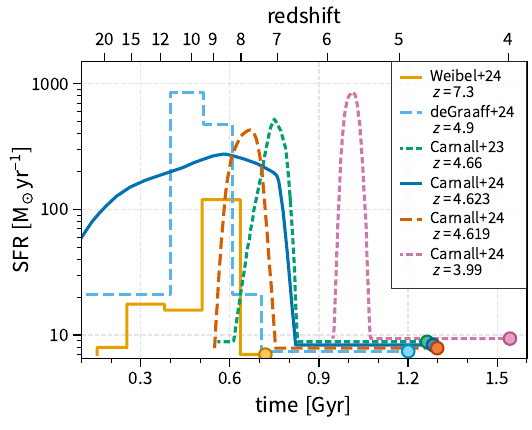} 
\vspace{-5pt}
\caption{
FFB-like starbursts in the histories of massive quiescent galaxies.
Shown are star formation histories as estimated from the spectra of 
six characteristic massive quiescent galaxies observed by JWST at cosmic
morning, $z \seq 4\sdash 7$ \adb{(marked by circles)}.
The galaxies are ZF-UDS-6496 at $z \seq 3.99$ \citep{carnall24},
PRIMER-EXCELS-117560 at $z \seq 4.619$ \citep{carnall24},
PRIMER-EXCELS-109760 at $ z \seq 4.623$ \citep{carnall24},
GS-9209 at $z \seq 4.66$ \citep{carnall23},
RUBIES-EGS-QG-1 at $z \seq 4.9$ \citep{degraaff24},
and RUBIES-UDS-QG-z7 at $z \seq 7.3$ \citet{weibel25}.
\adb{Small vertical offsets are introduced to enable distinction between the 
curves below $10 \msun\yr^{-1}$, where they are below the detection limit.}
Each recovered SFH seems to consist of a starburst of a global duration 
$100\sdash 200\Myr$, 
at a level of a few hundred $\Msun \yr^{-1}$, 
followed by a robust quiescent period of several hundred Megayears
till the observed time.
The crudely estimated stellar masses are 
between $10^{10.3}\msun$ and $10^{11}\msun$ for RUBIES-UDS-QG-z7
and RUBIES-UDS-QG-1 respectively.
The surface densities are well above $10^{4}\msun\pc^{-2}$ \citep{weibel25}.
The estimated cosmological number density of these galaxies
is between $10^{-6}$ and $10^{-4} \Mpc^{-3}$, well above the expectations from 
\adb{current cosmological simulations} \citep{weibel25}.
}
\vspace{-10pt}
\label{fig:sfh}
\end{figure}

\smallskip 
The spectra of these galaxies, under certain assumptions (e.g., concerning
metallicity and dust), are used to deduce the star-formation histories.
In particular, the NIRSpec/PRISM spectra reveal deep Balmer absorption lines 
and a strong spectral break at a rest-frame wavelength of $4000\AA$,
indicating a lack of star formation in the recent history
under simultaneous modeling with NIRCam and MIRI photometry.
\Fig{sfh} shows examples of SFHs that were derived from such galaxies by
different authors.
The characteristic finding is that most of the stars were formed at 
cosmic dawn, $z \ssim 8 \sdash 11$, likely in a massive peak of duration  
$100 \sdash 200 \Myr$, followed by relatively abrupt quenching of star 
formation, which is maintained till the detection time at cosmic morning.
\adb{
An extreme example of a quiescent galaxy more massive than 
$\Ms \seq 10^{11}\msun$ at $z \seq 3.2$ indicates integrated SFE higher than
$0.5$ at $z \seq 7 \sdash 12$, with peak SFR extending to $z \ssim 20$
\citep{turner25}. 
}

\smallskip 
While the exact properties of the early peaks of star formation that are 
deduced by cosmic archaeology are less solid than the subsequent quiescent 
period,
one cannot resist the temptation to identify these early SFR peaks with the 
excessively bright galaxies observed at cosmic dawn with similar abundances. 
The high stellar masses of these quiescent galaxies are indeed
consistent with the high stellar masses produced at cosmic dawn due to high 
global star-formation efficiency in dense clusters within compact galaxies
along the lines of the FFB scenario \citep{dekel23,li24}.

\subsection{Bimodality and Quenching}

Later in time, ``cosmic noon" at $ z \ssim 1 \sdash 3$ is known to be the
peak period of massive star-forming galaxies (SFGs), 
where the cosmological star-formation density reaches a maximum level 
\citep{madau14}.
At this epoch, the typical forming halos are of mass 
$\Mv \ssim 10^{11\sdash 12}\msun$.
\adb{
This defines the ``golden mass" of galaxy formation \citep{dekel19_gold}, 
which is expressed as a peak in the commonly quoted stellar-to-halo mass ratio 
\citep{behroozi13,moster18,behroozi19,behroozi20,moster20,stefanon21,shuntov22}.
}
The emerging questions are how the cosmic-noon SFGs relate to the 
cosmic-morning quiescent galaxies, and how the latter relate to the 
super-luminous galaxies at cosmic dawn.

\smallskip 
Here we attempt to understand the evolutionary tracks of massive galaxies from 
cosmic dawn through cosmic morning to cosmic noon.   
\adb{
We propose a bi-modality as a function of mass into massive galaxies that 
undergo an FFB phase at cosmic dawn and galaxies of somewhat lower masses 
that do not undergo an FFB phase. 
The FFB galaxies suffer severe quenching during the post-FFB phase at cosmic 
morning while the non-FFB galaxies evolve to SFGs at cosmic noon. 
}

\smallskip 
\adb{The challenging issue addressed in this paper
is to explore the possible post-FFB quenching mechanisms.}
What gives rise to the end of the FFB phase?
What causes the abrupt quenching?
What is responsible for the maintenance of the long-term, almost-complete 
quenching?
In turn, what is responsible for the peak of SFR at cosmic noon?

\smallskip 
To help answer these questions, we consider several processes that favorably 
occur at high redshifts and may be relevant. 
Most of them tend to take place near or above a threshold halo mass, 
comparable to the golden mass.
They include 
(a) effective supernova feedback at 
    $\Mv \slt 10^{11.8}\msun V_{100}^3 (1+z)^{-3/2}$ \citep{ds86},
(b) virial shock heating of the circum-galactic medium (CGM)
     at $\Mv \sgt 10^{11.8}\msun$ \citep{db06},
(c) feeding of galaxies by cold streams at $z \sgt 1$ \citep{db06,dekel09}, 
    enhanced by entrainment of hot CGM gas through turbulent radiative
mixing layers \citep{aung24}.
(d) compaction events at $\Mv \ssim 10^{11.3}\msun$ \citep{zolotov15,lapiner23},
and
(e) formation of post-compaction long-lived disks and rings at 
$\Mv \sgt 10^{11.2}\msun$ \citep{dekel20_flip,dekel20_ring}.

\subsection{Black Holes and AGN Feedback}

A key relevant process of distinct importance for quenching
is AGN feedback associated with rapid 
\adb{black-hole (BH)} growth.
\adb{AGN feedback} is
also seen to be mostly relevant at cosmic noon above a threshold mass 
$\Mv \sgt 10^{11.8}\msun$ \citep{forster19}.
This onset of BH growth may be triggered by a compaction event,
and reflect the transition from the low-mass halo regime where supernova 
feedback suppresses the BH growth to the massive-halo regime where the 
CGM is hot \citep{dekel19_gold,lapiner21}. 

\smallskip 
JWST observations at cosmic morning reveal the existence of central
massive black holes of masses \adb{$10^6 \sdash 10^8\msun$}.
\adb{They have excessively high BH-to-stellar mass ratios of $\sim\! 0.01$
\citep{harikane23,ubler23,maiolino24},
compared to the standard values of
$10^{-4} \sdash 10^{-3}$ at lower redshifts \citep{reines15}.
}
A feasibility study \citep{dekel25} indicates that 
the FFB scenario may provide a natural setting for such a rapid black hole  
growth. This involves the formation of seed BHs by sped-up core collapse in 
the young, rotating FFB star-clusters. It is followed by migration driven by
dynamical friction  
to the galactic-disk centers where super-massive BHs first grow by mergers, 
overcoming gravitational-wave recoils. A post-FFB accretion phase onto
the merger-driven massive central BH can boost the accretion-driven AGN 
feedback in these galaxies, potentially resulting in a major quenching 
mechanism. 

\subsection{Outline}

The paper is organized as follows.
In \se{tracks} we propose a bimodality into two different evolutionary tracks, 
for post-FFB galaxies and non-FFB galaxies.
In \se{q1_compaction} to \se{q4_ffb}
we consider the possible quenching mechanisms that may be
responsible for the post-FFB quiescent galaxies and the interplay between them.
In particular,
in \se{q1_compaction} we address compaction-driven quenching,
in \se{q2_bh} we focus on the role of AGN feedback in the quenching, 
in \se{q3_streams} we evaluate possible ways to suppress gas supply by cold
streams for complete quenching, also largely driven by AGN feedback,
and in \se{q4_ffb} we consider the particularly efficient quenching 
in post-FFB galaxies.
In \se{obs} we discuss observable predictions for the quiescent galaxies.
In \se{conc} we summarize our results and conclude.

\begin{figure*} 
\centering
\includegraphics[width=0.60\textwidth] 
{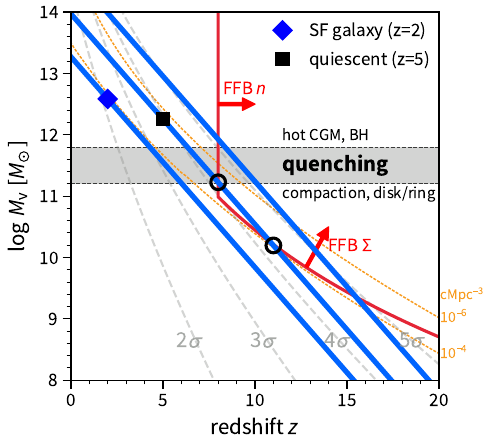} 
\vspace{-5pt}
\caption{
Bimodal evolutionary tracks of galaxies: FFB and non-FFB.
Shown in a diagram of halo mass versus redshift 
are average growth tracks for halos that at $z_0 \seq 10$
have masses $\Mv \seq 10^{9.84}, 10^{10.54}, 10^{11.24}\msun$
\adb{(three blue thick diagonal lines)}. 
In comparison, the threshold for FFB in a disk (red curve, \adb{marked ``FFB"}) 
consists of thresholds for gas
density $n$ and surface density $\Sigma$ following Fig.~6 of \citet{dekel23}.
The $N \sigma$ peaks in the density fluctuation field (dashed grey curves)
indicate that the middle halo is roughly a $4\sigma$ peak.
Galaxy comoving number densities of $10^{-4}$ and $10^{-6} \Mpc^{-3}$ 
are marked (thin dashed orange curves), 
confining the middle halo at $z \sleq 11$.
We learn that the halos of mass $\geq\!10^{10.5}\msun$ at $z \seq 10$
($\geq\!4\sigma$ peaks) undergo an FFB phase, while halos of lower masses
(lower-$\sigma$ peaks) never become FFB.
The ``fiducial", most abundant FFB galaxies, those near the middle blue curve,
enter the FFB phase near $z \ssim 11$ (rightmost open black circle)
and exit at $z \ssim 8$ (leftmost open black circle), 
after $\sim\! 170\Myr$ and when $\Mv \sgsim 10^{11}\msun$.
Soon after, the fiducial FFB halos cross the ``golden mass"
\citep{dekel19_gold} $\Mv \ssim 10^{11.5}\msun$, 
\adb{in the shaded area} between the horizontal dashed
lines that marks the onset of quenching by several different mechanisms.
These include a hot CGM \citep{db06},
wet compaction events \citep{zolotov15,lapiner23},
the rapid growth of super-massive BHs and AGN \citep{lapiner21},
as well as long-lived extended disks and rings
\citep{dekel20_flip,dekel20_ring}.
The quenching of the FFB galaxies is efficient because of the intense
compaction activity in high-sigma halos \citep{dubois12}
and the compaction-driven AGN feedback \citep{lapiner21} that is boosted
by FFB BH seeding and growth \citep{dekel25}.
By cosmic morning, at $z \ssim 5$, these quenching mechanisms lead
to quiescent galaxies of stellar mass $\Ms \ssim 10^{10.5}\msun$ in halos of 
$\Mv \sgsim 10^{12}\msun$ (black square), as observed.  
These halos eventually become quiescent groups and clusters of galaxies.
The lower-sigma halos with masses below the FFB threshold accrete gas and form
stars more gradually as they grow. 
At cosmic noon, $z \ssim 2\sdash 3$, when these
halos are above $10^{12}\msun$, cold gas streams penetrate
efficiently through the hot CGM \citep{dekel09}, boosted by entrainment of
hot CGM gas onto the cold streams \citep{aung24}, 
giving rise to star forming galaxies at their peak (blue diamond).
At $z \sleq 1$, cold streams do not penetrate the hot halos 
\citep[][Fig.~7]{db06}, such that these galaxies end up with lower SFRs.
}
\vspace{-10pt}
\label{fig:post}
\end{figure*}

\begin{figure} 
\centering
\includegraphics[width=0.49\textwidth, trim={0.2cm 0.2cm -0.1cm 0.0cm},clip]
{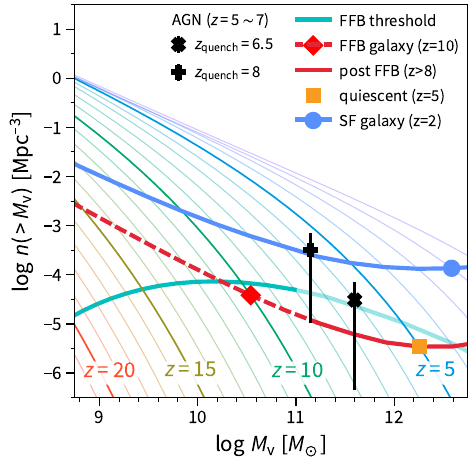} 
\vspace{-5pt}
\caption{
Two evolutionary tracks in the plane of halo mass function, comoving number
density versus halo mass, complementing \fig{post}.
The halo mass functions at different redshifts are shown in the background,
based on \citet{watson13}.
The red curve \adb{(the bottom curve at high masses)}
refers to the evolving fiducial FFB halos 
that reach $\sim\! 10^{10.5}\msun$ at $z\seq 10$ (roughly $4\sigma$ peaks).
The cyan curve\adb{(the bottom curve at low masses)}
 is the threshold for FFB, $\Mveight (1+z)_{10}^{6.2} \seq 0.8$.
The curves stretching on the left of $z \seq 8$ 
(dashed, $\Mv \slt 10^{11.2}\msun$) refer to the FFB phase,
and the curves to the right of $z \seq 8$ 
(solid, $\Mv \sgt 10^{11.2}\msun$) correspond to the post-FFB quiescent phase.
The typical halo number density in the post-FFB zone is  
$n \sgsim 10^{-5}\Mpc^{-3}$.  
The uppermost, \adb{blue curve} near $n \ssim 10^{-3}\Mpc^{-3}$ 
corresponds to the
non-FFB evolution track reaching the peak of star-forming galaxies at 
$z \ssim 2$.
Colored symbols mark the fiducial starbursting FFB halos at cosmic dawn
($\Mv \ssim 10^{10.5}\msun$ at $z \ssim 10$ where $n \ssim 10^{-4.3}\Mpc^{-3}$),
the fiducial quiescent halos at cosmic morning
($\Mv \ssim 10^{12.2}\msun$ at $z \ssim 5$ where $n \ssim 10^{-5.3}\Mpc^{-3}$),
and the fiducial non-FFB star-forming galaxies at cosmic noon
($\Mv \ssim 10^{12.6}\msun$ at $z \ssim 2$ where $n \ssim 10^{-3.5}\Mpc^{-3}$).
Shown for comparison (black symbols \adb{with error bars}) 
are number densities of AGN observed by JWST
\citep{harikane23_agn} at $z \seq 5 \sdash 7$ in quiescent galaxies brighter
than the corresponding predicted UV magnitudes of post-FFB galaxies
with quenching onset at $z_{\rm quench} \seq 8$ or $6.5$ (see \fig{Muv}).
The abundance of BHs is in the ball park of that predicted for the quiescent
galaxies when the onset of quenching is at $z \ssim 6.5$, and is somewhat 
higher for higher $z_{\rm quench}$.
(See also the number densities in \fig{n_post} below.)
}
\vspace{-10pt}
\label{fig:hmf}
\end{figure}

\section{Bimodal evolutionary tracks}
\label{sec:tracks}

\adb{We show in this section that}
the conditions for a halo to undergo FFB hints for a division of the halo
population into two different evolutionary tracks, the FFB track and the
non-FFB track. These tracks are primarily distinguished by halo mass,
or equivalently by the rareness of the halo as a peak in the density 
fluctuation field. 
This can be deduced from \fig{post}, which addresses the evolution of 
galaxies in the plane of halo mass versus redshift.

\smallskip
\Fig{post} first displays three examples of average growth tracks for 
halo masses that have a range of values at a given redshift, 
$\Mv \seq M_0 \ssimeq 10^{9.8}, 10^{10.5}$ and $10^{11.2}\msun$ at 
$z \seq z_0 \seq 10$. 
The average mass growth is computed based on the simple analytic expression 
derived by \citet{dekel13} in a straightforward way based on the invariant
timescale in the Press-Schechter (PS) and Extended Press Schechter (EPS) theory
\citep{press74,bond91}, 
\be
\Mv(z) = M_0\, e^{-\alpha(z-z_0)} \, , \quad \alpha \simeq 0.79 \, .
\label{eq:Mvz}
\ee
This expression was derived in the Einstein-deSitter regime at high redshifts
$z \sgt 1$, and it has been shown to be a good approximation in comparison with
\adb{cosmological simulations} \citep[see also][]{wechsler02,zhao09}. 

\smallskip 
\Equ{Mvz} is derived by integrating eq.~7 of \citet{dekel13}, 
which boils down to
\be
\frac{\Mvdot}{\Mv} = s_{10}\, \Mvtwelve^{\beta}\, (1+z)_{10}^{5/2} \, ,
\label{eq:beta}
\ee
with the simplifying approximation $\beta \seq 0$ (as in eq.~8 there).
The power $\beta$ depends on the slope of the power spectrum of density
fluctuations near the mass scale of interest.
A value of $\beta \ssimeq 0.14$ was advocated in \citet{dekel13}, and shown 
to provide a best fit to \adb{cosmological simulations} for massive galactic 
halos at cosmic noon. 
However, we find that $\beta \seq 0$ (actually $\beta \ssimeq 0.03$)
provides a better approximation for the halos of $\sim\!10^{11}\msun$ at 
$z \sgt 5$. This is verified in comparison to \adb{cosmological simulations}
by \citet{zhao09}, which are reliable out to $z \ssim 10$. 
We thus adopt $\beta \seq 0$, and thus gain simplicity in \equ{Mvz} and the 
subsequent analysis.
In \equ{beta} the halo mass is $\Mv \equiv 10^{12}\msun\, \Mvtwelve$
and $1+z \equiv 10\,(1+z)_10$.
The normalization factor in \equ{beta} is calibrated from simulations
\citep{dekel13} to be 
$s_{10} \ssimeq 9.5 \Gyr^{-1}$, and this translates in the Einstein-deSitter
regime to the value of $\alpha$ quoted in \equ{Mvz}.

\smallskip 
\Fig{post} shows in the background of the same diagram the curves 
corresponding to $N\sigma$
peaks in the Gaussian random field of linear density fluctuations, based on the
Press-Schechter formalism \citep{press74}, for $N \seq 2 \sdash 5$.
We learn that the middle growth curve, which corresponds to 
$\Mv \seq 10^{10.5}\msun$ at $z\seq 10$, roughly matches the  
$4\sigma$-peaks curve in a broad redshift range, $z \sgeq 5$. 
The upper growth curve of higher-mass halos, 
with $\Mv \seq 10^{11.2}\msun$ at $z\seq 10$,
is close to the $5\sigma$-peaks curve. 
Thus, these high-mass halos at and above the middle growth curve
correspond to rare high-sigma peaks. 

\smallskip 
Also shown in the background of \fig{post} are contours of constant  
comoving number density of halos, $n \seq 10^{-6}$ and $10^{-4} \Mpc^{-3}$.
These contours confine the middle growth curve at $z \sleq 11$.
We caution that the exact interplay between the contours of constant $n$ and 
constant $N$ depends on the assumed value of $\beta$ in the specific accretion 
rate of \equ{beta}.

\smallskip 
The number densities of the halos above the middle growth curve match the low  
abundance of the super-luminous galaxies as observed using JWST at cosmic dawn.
\adb{These galaxies have}
halo masses of $\sim\! 10^{11}\msun$ based on the $\Lambda$CDM halo mass
function shown in \fig{hmf}, given the observed  
number density of $n\ssim 10^{-5}\Mpc^{-3}$ \citep{finkelstein23}.
We will see in \fig{hmf} below that
these number densities similarly match the abundance of the massive quiescent 
galaxies at cosmic morning, with $\Mv \ssim 10^{12}\msun$ \citep{weibel25}.
On the other hand, the lower growth curve of lower masses shown in \fig{post}, 
those with $\Mv \seq 10^{9.8}\msun$ at $z\ssimeq 10$, is below the
$4\sigma$-peaks curve, coinciding with the $3\sigma$ curve at $z\ssimeq 4$
and with the $2\sigma$ curve at $z \ssim 2$. These more typical 
(but still high) peaks match the higher abundance of the typical 
star-forming galaxies at cosmic noon.

\smallskip 
Next, and most important,
\fig{post} displays the threshold for FFB according to the disk version
of the FFB scenario, following Fig.~6 of \citet{dekel23}.
The vertical part of the threshold at $z \ssim 8$ refers to the critical gas 
number density, $n \ssim 3\times 10^3 \cmc$, 
in the star-forming clouds within the disks. 
Higher densities allow for free-fall collapse times of 
$\sim\!1\Myr$ and thus efficient feedback-free star formation prior to the 
onset of feedback by supernovae and stellar winds. 
This threshold stems from eq.~42 of \citet{dekel23}, where the density in the
clumps is estimated to be
\be
\nc = 5.6\times 10^3\cmc\, c\, \lambda_{.025}^{-3}\, (1+z)_{10}^{3} \, .
\ee
Here $\lambda \sequiv 0.025\,\lambda_{.025}$ is the contraction factor from the 
halo to the disk, which is partly associated with the halo spin parameter
\citep{fall80,jiang19}.
\adb{The parameter}
$c \sgt 1$ is the density contrast between the star-forming clumps and the
background density in the disk.
Such densities, with moderately low metallicities of $Z \ssim 0.02 \Zsun$,
also correspond to cooling on free-fall timescales which permits a short 
starburst followed by a well-defined time window free of feedback
\citep[][eq.~12]{dekel23}.
The other part of the threshold mass that declines with redshift refers to the
critical surface density, $\Sigma \ssim 3\stimes 10^3 \Msun\pc^{-2}$.
\adb{Above this threshold,}
the stellar radiative pressure is balanced by the self-gravity   
of the clouds, minimizing the suppression of star formation by radiative
feedback \citep{menon23,grudic23,menon24}.
This threshold, based on eqs.~62 and 67 of \citet{dekel23}, is roughly
\be
\Mveight\, (1+z)_{10}^{6.2} > 0.8 \, ,
\label{eq:FFB_threshold}
\ee
where $\Mv \equiv 10^{10.8}\msun\,\Mveight$.

\smallskip 
Taking the threshold curve for FFB at face value 
(which is clearly an over-simplification considered for a qualitative estimate
only), we learn from \fig{post} that the halos of mass 
$\geq\!10^{10.5}\msun$ at $z \seq 10$, representing $\geq\!4\sigma$ peaks,
undergo an FFB phase. 
\adb{On the other hand,} 
halos of lower masses, which emerge from lower-$\sigma$ peaks, never become FFB.
Thus, the most abundant FFB galaxies at $z \ssim 10$ are those of
$\Mv \sgsim\!10^{10.5}\msun$, which we therefore refer to as the 
``fiducial" FFB galaxies.
We see from \fig{post} that
they enter the FFB phase near $z \ssimeq 11$, while the more massive,
less abundant halos enter the FFB phase at higher redshifts.   
This defines a bi-modality between galaxies that undergo FFB and galaxies that
do not, separated near halo mass $\Mv \ssim 10^{10.5}\msun$ at $z \ssim 10$, 
namely near $4\sigma$ peaks.

\smallskip 
The FFB phase ends when the density drops below the critical value at
$z \ssim 8$. For the fiducial FFB galaxies the overall duration of the FFB 
phase, between the two black open circles in \fig{post},
is $\sim\! 170\Myr$, similar to the estimate in \citet{dekel23} 
\citep[see also Fig.~9 of][]{li24}. At the end of the FFB phase, the fiducial 
halos are somewhat more massive than $10^{11}\msun$. 

\smallskip 
Soon after the end of the FFB phase, the fiducial FFB halos reach the 
``golden mass" \citep{dekel19_gold}, in the vicinity of 
$\Mv \ssim 10^{11.5}\msun$.
\adb{This} marks the onset of quenching of star formation by several different
quenching mechanisms, to be discussed below.
These include the onset of a hot CGM \citep{db06},
the rapid growth of super-massive black holes and AGN \citep{lapiner21},
\adb{gas-rich (``wet") compaction events} \citep{zolotov15,lapiner23},
and the formation of long-lived extended disks and rings
\citep{dekel20_flip,dekel20_ring}.
The range of halo masses expected at the onset of quenching 
is crudely marked by a shaded area between horizontal lines in \fig{post}.
We learn that the fiducial post-FFB halos start quenching soon after 
$z \ssim 8$.
By $z \ssim 5$, they grow above $10^{12}\msun$, where all the quenching
mechanisms mentioned above have been potentially at play. If the post-FFB
quenching is indeed efficient, both in terms of abrupt onset and long-term
maintenance, these galaxies would resemble the observed massive quiescent
galaxies at cosmic morning.   

\smallskip 
With an integrated global star-formation efficiency SFE of 0.2 to 1 in the FFB
phase \citep{li24}, where $\Mv \ssim 2\stimes 10^{11}\msun$, 
and with low SFR since then in the post-FFB phase, 
the stellar mass is expected to be roughly 
constant at $\Ms \sgsim 10^{10} \sdash 10^{10.5}\msun$.   
A similar estimate of $\Ms \ssim 10^{10.5}\msun$ is obtained from the halo 
mass at the quiescent phase assuming a stellar-to-halo-mass ratio of $0.1$ 
\citep[e.g.,][]{behroozi19}. 
These estimates of stellar mass are comparable to the typical observed 
estimates.
As these halos grow further at lower redshifts, they are expected to become 
the hosts of quiescent groups and clusters of galaxies.

\smallskip 
Lower mass halos, which never enter the FFB phase, can accrete gas and form 
stars gradually as they grow. They enter the quenching regime at $z \sleq 7$, 
but their quenching is likely to be less efficient than in the post-FFB 
galaxies (see \se{q44_ffb}). 
At cosmic noon, near $z \ssim 2\sdash 3$, when these non-FFB
halos are above $10^{12}\msun$, cold gas streams penetrate efficiently 
through the hot CGM \citep{dekel09,mandelker20a}.
\adb{They are}
boosted by entrainment of hot CGM gas onto the cold streams through a turbulent
mixing layer \citep{aung24}, giving rise to star forming galaxies at the peak 
of cosmic star formation density.
At $z \ssim 1$ and later, these galaxies find themselves in the regime where
cold streams do not penetrate the hot halos
\citep[][Fig.~7]{db06,mandelker20a,mandelker20b,daddi22a,daddi22b}, such that
they end up with lower SFRs today.

\smallskip 
\Fig{hmf} complements \fig{post} in introducing the two evolutionary tracks of
FFB and non-FFB galaxies, 
focusing on the halo masses and their cosmological number densities.
The halo mass functions at different redshifts (separated by $\Delta z \seq 1$)
are plotted in the background based on simulations by \citet{watson13},
\adb{which attempt to be reliable at high redshifts.}
The assumed cosmology is $\Lambda$CDM with
$\omm\seq 0.3$, $\oml\seq 0.7$, $h\seq 0.7$ and $\sigma_8\seq 0.82$, 
and the halos are assumed to be spherical with a mean density contrast of 200.
The red curve refers to the evolving fiducial halos
that reach $10^{10.5}\msun$ at $z\seq 10$ (roughly $4\sigma$ peaks).
The cyan curve marks the threshold for FFB due to the surface density
criterion, \equ{FFB_threshold}, based on \citet{dekel23}.
The curves stretching on the left of $z \seq 8$  
($\Mv \slt 10^{11.2}\msun$) refer to the starbursting FFB phase,
while the curves extending to the right of $z \seq 8$ 
($\Mv \sgt 10^{11.2}\msun$) correspond to the subsequent quiescent phase
of the post-FFB halos.
The typical halo number density for this FFB and post-FFB zone is 
in the ball park of $n \ssim 10^{-5}\Mpc^{-3}$. 
\adb{This highlights} the similar
abundances of bright FFB galaxies at cosmic dawn and quiescent post-FFB 
galaxies at cosmic morning (see also \fig{n_post} below), 
with the associated halo masses growing from $10^{11}\msun$ to above 
$10^{12}\msun$.

\smallskip
In \fig{hmf},
the non-FFB evolutionary track is represented by the 
upper curve near $n \ssim 10^{-3.5}\Mpc^{-3}$, leading to 
the peak of star-forming galaxies at cosmic noon, with halos above
$10^{12}\msun$ at $z \ssim 2$.
Symbols mark the fiducial starbursting FFB halos at cosmic dawn
($\Mv \ssim 10^{10.5}\msun$ at $z \ssim 10$ where $n \ssim 10^{-4.5}\Mpc^{-3}$),
the fiducial quiescent halos at cosmic morning
($\Mv \ssim 10^{12.25}\msun$ at $z \ssim 5$ where $n \ssim 10^{-5.4}\Mpc^{-3}$),
and the fiducial non-FFB star-forming galaxies at cosmic noon
($\Mv \ssim 10^{12.6}\msun$ at $z \ssim 2$ where $n \ssim 10^{-3.9}\Mpc^{-3}$).

\smallskip
\adb{
If the evolutionary track of FFB galaxies as envisioned based on \fig{post}
is valid, these galaxies are supposed to quench rather abruptly over 
$\sim\!100\Myr$ soon after the FFB phase, and subsequently the quiescence
should be maintained for at least several hundred Megayears.
The challenge attempted in much of the rest of this paper is to identify
the physical mechanism(s) that can be responsible for this quenching.
We note in \fig{post} that the post-FFB halo masses are near the golden mass 
of galaxy formation \citep{dekel19_gold}, between $10^{11}$ and $10^{12}\msun$,
above which several different quenching mechanisms are capable of being
at play.
We first discuss in \se{q1_compaction} compaction-driven quenching.
We then briefly address quenching by AGN feedback in \se{q2_bh}. 
We focus on the challenge of stopping the supply by cold streams
in \se{q3_streams}.  
We finally emphasize the possible dependence of the quenching mechanisms
on having an earlier FFB phase in \se{q4_ffb}.
}

\section{Compaction-driven quenching}
\label{sec:q1_compaction}

\adb{
In this section we consider the quenching mechanisms that operate
at or above the golden mass. In \se{golden}, we argue that supernova feedback 
and the presence of a hot CGM are not likely to be the main mechanisms 
that we are after at cosmic morning.
We then address in \se{direct_compaction} the direct ways by which wet 
compaction events may cause quenching.
In \se{q2_bh}, we will consider the indirect effect of compaction events
on quenching via their role in the growth of black holes and the associated AGN
feedback.
} 

\subsection{Golden mass between supernova feedback and hot CGM}
\label{sec:golden}

At cosmic morning, the golden mass seems to be above the maximum halo mass for 
effective supernova feedback on the scale of the galaxy as a whole.
\adb{This} can be estimated by
\citep{ds86} 
\be
\Mv \simeq 3.3\times 10^{10}\msun V_{100}^3 (1+z)_{7}^{-3/2} \, ,
\ee
where $\Vv = 100\kms V_{100}$ is the halo virial velocity 
and $1+z \seq 7\,(1+z)_{7}$.
To the extent that this estimate is valid, 
we may not expect supernova feedback to play a direct major role
in the overall quenching of post-FFB galaxies where $\Mv \sgeq 10^{11}\msun$
\citep{boylan23,dekel23}, the way they do in less massive galaxies.
This is confirmed in 
\adb{cosmological simulations} at $z \ssim 6$ \citep[e.g.][]{dubois13}.

\smallskip 
In more massive halos,
the golden mass serves as a threshold for a hot CGM near the virial 
temperature, bounded by a stable viral shock \citep{bd03,db06}.
At low redshifts, $z \slt 2$, where the accretion tends to be more isotropic,
the hot CGM suppresses the gas supply to the galaxy and thus limits the SFR.
However, at cosmic noon, $z \sgeq 2$, the hot CGM is expected to be
penetrated by narrow cold streams from the cosmic web 
\citep[Fig.~7 of][]{db06,keres09}, as seen in 
\adb{zoom-in cosmological simulations} 
\citep{dekel09}. At such epochs, the hot CGM is not expected to serve 
as a direct quenching mechanism. In fact, 
the cold streams are likely to be intensified by entrainment from the hot CGM, 
which is partly fed by recycling of gas from the galaxy \citep{aung24}. 
This gives rise to the observed peak of SFR density as a 
function of redshift \citep{madau14} and the associated 
observed peak of stellar-to-halo mass ratio as a function of mass at 
cosmic noon
\citep{behroozi20,moster20,stefanon21,shuntov22}.
In general, supernova feedback at small masses and a hot CGM at large masses
confine the golden mass in between. 
\adb{However, these mechanisms are not likely to be the major quenching 
mechanisms that we are after near the golden mass at cosmic morning.}

\subsection{Direct quenching by compaction events}
\label{sec:direct_compaction}

\adb{Cosmological hydro-gravitational simulations} 
\citep{zolotov15,lapiner23} and observations
\citep{barro13,dokkum15,barro17} reveal that most galaxies undergo 
\adb{
gas-rich (``wet") compaction events into compact starbursts (termed ``blue
nuggets").
} 
These events induce major transitions in the galaxy properties, and in 
particular trigger quenching of star formation in several different ways.  
\adb{The} compaction processes are due to drastic angular-momentum (AM) losses,
about half of which are caused by mergers and the rest by colliding
counter-rotating
streams, recycling fountains from the galactic disks, and other mechanisms
\citep{lapiner23}. 
\adb{The} major compaction events, those that trigger a 
decisive long-term quenching process and a transition from central
dark-matter to baryon dominance, tend to occur near the golden
mass at all redshifts. 
\adb{This} is seen in \adb{cosmological simulations} 
\citep{tomassetti16} 
and in machine-learning-aided comparisons to observations \citep{huertas18}.
This preferred mass scale for major compaction events seems to be due, at least
partly,
to the two physical processes mentioned above of supernova feedback and
hot CGM (with AGN feedback - see below), which tend to suppress compaction 
attempts at lower and higher masses.
At cosmic dawn, the FFB halos are indeed of masses $\sim\!10^{11}\msun$,
in the vicinity of the golden mass, and they correspond to high-sigma peaks of
$>\!4\!\sigma$, which tend to have more frequent and stronger compaction events
(see \se{q44_ffb} below).

\smallskip 
The compaction events drive quenching in several different ways, directly and
indirectly.
Quite straightforwardly, as the gas that is pushed to the galaxy center 
triggers a central starburst in the form of a blue nugget,
it causes central gas depletion due to star formation and the associated 
outflows,
thus leading to inside-out quenching \citep{tacchella16_ms, tacchella16_prof}.
Secondly, compaction events drive the subsequent formation of extended gas 
disks and rings by new incoming streams \citep{tacchella16_prof, dekel20_ring}. 
The compact massive nugget, or bulge, stabilizes the disk/ring against
gravitational instability. This suppresses star formation in the outskirts
\citep[``morphological quenching",][]{martig09}
and limits the rate of gas supply to the center by instability-driven
inward radial transport \citep{dekel20_ring,dutta25}.
It turns out that long-lived extended disks are coincidentally favored above 
the same golden mass due to the fact that merger-driven disk flips become 
less frequent than the disk orbital time above this mass at all redshifts 
\citep{dekel20_flip}. 
This contributes to limiting the gas supply for star formation in the center
and thus generating inside-out quenching.

\smallskip
It remains to be verified to what extent these direct effects of compaction can 
lead to complete quenching, as observed. Otherwise, complete quenching may
require another mechanism, such as enhanced AGN feedback, which can be an 
indirect outcome of the compaction process (\se{q2_bh} below).    

\begin{figure*} 
\centering
\includegraphics[width=0.99\textwidth]
{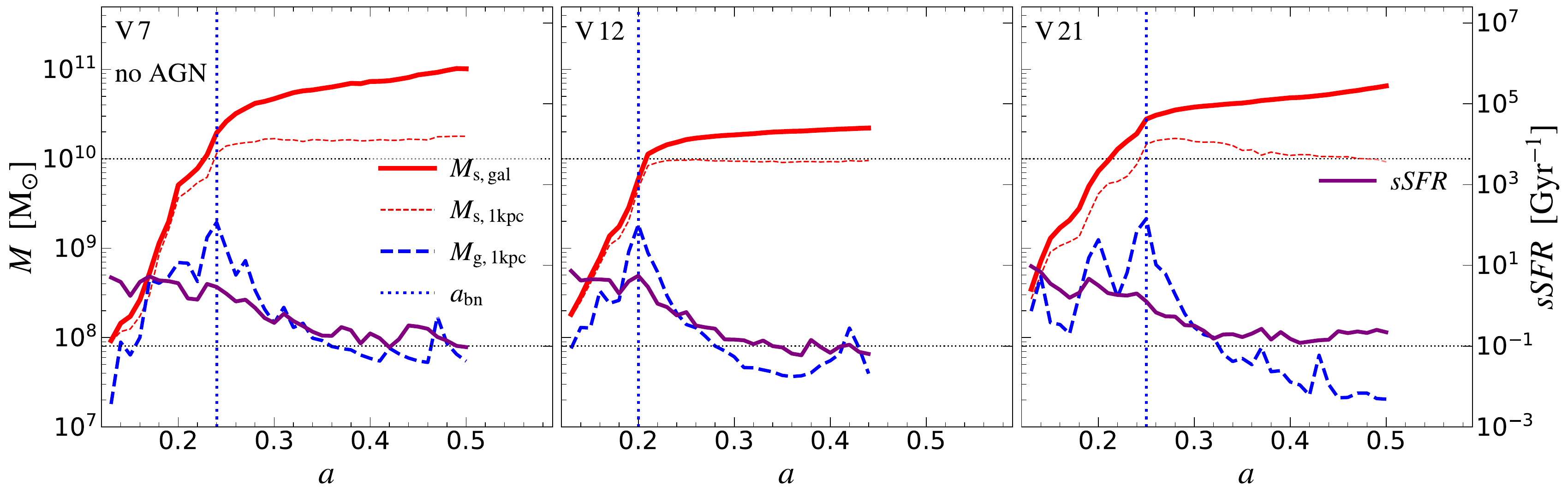} 
\includegraphics[width=0.99\textwidth]
{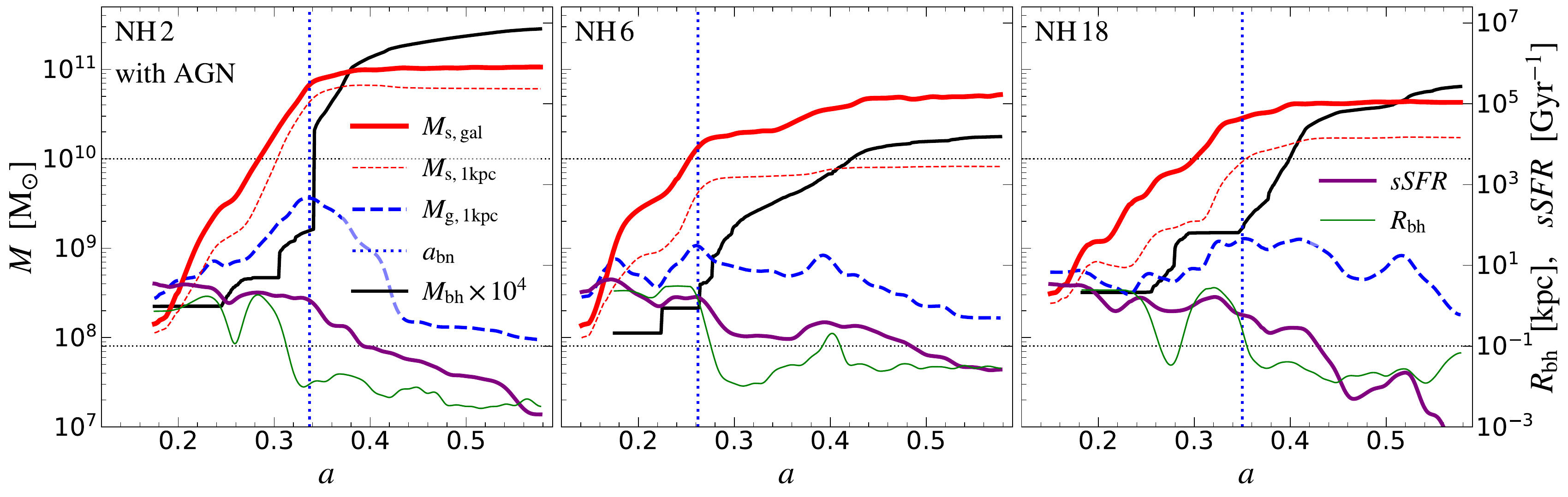} 
\vspace{-5pt}
\caption{
Compaction-driven quenching without (top) and with (bottom) AGN feedback.
Shown are the evolution of quantities of interest in the VELA simulations
without AGN feedback \citep{lapiner23} (top)
and in the NewHorizon simulations that incorporate AGN feedback
\citep{lapiner21} (bottom).
The compaction event is characterized by the peak in gas mass
within the central $1\kpc$ (blue \adb{dashed} line,
peak marked by a vertical line)
and the subsequent decay of the BH orbit to the
center \citep{lapiner21} (green, \adb{thin} line, \adb{right axis for BH
orbital radius}).
Following the compaction, in the presence of AGN the SFR drops
(magenta, \adb{gradually declining line, right axis for sSFR}) 
and the stellar mass (solid red line \adb{at the top}) 
tends to remain rather constant
over the whole galaxy (as well as in the inner $1\kpc$, dashed red).
Galaxy NH6, with two successive compaction events, shows two successive
quenching events.  On the contrary,
in the absence of AGN feedback, the quenching is only partial, limited to the
inner $1\kpc$, while the stellar mass over the whole galaxy keeps growing.
We learn that AGN feedback is crucial for the deep quenching at cosmic morning,
consistent with the findings of other simulations (see text).
We thus suspect that the boosted BH growth due to the FFB phase may be
a key for post-FFB quenching.
}
\vspace{-10pt}
\label{fig:quenching}
\end{figure*}

\section{Black-hole growth and AGN feedback}
\label{sec:q2_bh}

\adb{In this section we consider the quenching by AGN feedback, and argue that
it may be the key for the required near-complete quenching at cosmic morning.}

\subsection{Black-hole growth}
\label{sec:BH_growth}

A major source of quenching in massive galaxies is commonly assumed to be
AGN feedback, driven by the rapid growth of a supermassive central accreting 
black hole. Observationally at cosmic noon, high-luminosity AGN tend to occur 
above a threshold mass that is comparable to the golden mass \citep{forster18}.
Based on the \adb{NewHorizon cosmological simulations}, the onset of BH growth
has been shown to originate in a compaction event 
\citep{dekel19_gold,lapiner21}, as follows.
Below the golden mass, black-hole growth is suppressed by supernova feedback.
Above the golden mass, the central black hole is free to grow once the halo is 
massive enough to lock the supernova ejecta by its deep potential well
and the hot CGM.
Once the halo is in this mass range,
the BH growth is triggered by a compaction event, which brings gas to the 
galaxy center and deepens the potential well. 
\adb{This makes} the black hole orbit 
sink to the galaxy center and thus allowing rapid BH growth by accretion.
In turn, the ignited AGN feedback helps keeping the CGM hot and maintaining 
long-term quenching.

\smallskip 
In addition to the general scenario of compaction-driven BH growth,
it turns out that the FFB scenario may provide a natural setting for 
particularly rapid black-hole growth.
\adb{This is}
both through the formation of intermediate-mass BH seeds and their runaway 
merging into super-massive BHs with high BH-to-stellar mass ratios.
The feasibility of this scenario has been addressed by \citet{dekel25}, as
follows.
BH seeds can form by core collapse in a few free-fall times within the building 
blocks of the FFB galaxies, namely, thousands of young, compact, rotating
star clusters within each galaxy. 
The core collapse is sped up by the migration of massive stars to the
cluster center due to the 
young, broad stellar mass function (and more so if the IMF is top heavy). 
It is expected to be stimulated by a ``gravo-gyro" 
instability \citep{hachisu79} due to internal cluster rotation and flattening 
within the compact galactic disk that fragmented to form the clusters. 
BHs of $10^4\msun$ are typically expected in $10^6\msun$ FFB clusters within 
sub-kpc galactic disks at $z \ssim 10$. 
The orbiting clusters rapidly disrupt each other, generating a smooth stellar 
disk capable of exerting dynamical friction on the BHs.
The BHs then migrate to the galaxy center by dynamical friction
in a few $100\Myr$, hastened 
by the compact FFB stellar galactic disk configuration. 
Efficient mergers of the BH seeds produce $10^{6-8}\msun$ 
BHs with a BH-to-stellar mass ratio $\sim\!0.01$ by $z \seq 4\sdash 7$, 
resembling the JWST observations \citep{harikane23_agn,maiolino24}. 
Once the central BH has grown to above $10^6\msun$ by this merger-driven 
process, post-FFB gas accretion can efficiently boost the BH growth further 
and generate AGN capable of effective feedback.

\smallskip 
The wet compaction events can further assist in the post-FFB  
black-hole growth in three important ways.
\adb{
First, they can prevent 
suppression of dynamical-friction inspiral of the BH seeds into the
galactic center \citep[``core stalling",][]{read06,goerdt06,kaur18,banik21},
by steepening the inner galactic density profile.
}
Second, they can help the final parsec approach \citep{begelman80}
by providing central gas.
Third, and most critical, they can help overcoming the bottleneck in the 
initial growth of the central BH that is introduced by gravitational-wave 
recoils of the merged BH \citep{madau04,blecha08}.
\citet{dekel25} found that the recoils may be under control
if the BHs inspiral within a relatively cold disk, such that their spins tend
to be aligned with the orbital angular momentum. However, in more realistic 
thick disks, the merged BH could avoid ejection by recoil only if the escape 
velocity from the central galaxy is boosted to a few hundred $\kms$ by wet 
compaction events.

\subsection{AGN quenching in simulations and observations}

The crucial role of AGN feedback in reaching complete compaction-driven
quenching over the whole galaxy can be illustrated by comparing cosmological
simulations with and without AGN feedback.
\Fig{quenching} shows such a comparison, using example galaxies from the VELA
simulations with no AGN and the NewHorizon simulations where AGN are
incorporated.
These simulations and their relevant analyses are described  
in \citet{lapiner23} and \citet{lapiner21} and references therein, respectively.

\smallskip
Very briefly,
the VELA simulations \citep{ceverino14,zolotov15}
utilize the ART AMR code \citep{krav97,ceverino09} with $\sim\!25\pc$
maximum resolution to zoom in on galactic halos that reach 
$\sim\!10^{12}\msun$ at 
$z \ssim 1$.  Supernova and radiative stellar feedback are incorporated,
as described in appendix A of \citet{lapiner23} and in \citet{ceverino14},
but AGN feedback is not included.
The NewHorizon simulation \citep{dubois21}, on the other hand,
utilizes the RAMSES AMR code \citep{teyssier02} with $\sim\!40\pc$ resolution
to zoom in on a group of galaxies with eight galaxies of stellar masses above 
$10^{10}\msun$ by $z \seq 0.7$.
The subgrid physics incorporates both supernova and stellar feedback as well as
black holes with AGN feedback as described in appendix A of \citet{lapiner21},
in \citet{dubois21} and in the NewHorizon website.

\smallskip
We see in \fig{quenching} that the compaction events in the VELA simulations
with no AGN lead to inside-out quenching. This is quenching in the inner 
$1\kpc$, with a declining SFR and a constant stellar mass in time, 
but the overall stellar mass keeps rising, namely the quenching
of the whole galaxy is incomplete.
We learn that
the compaction-driven gas consumption, the emergence of a disk/ring,
and the hot CGM do trigger quenching in the inner galaxy, but fail to
completely shut down the overall star formation. The specific SFR
gradually declines from sSFR$\sim\!0.3 \Gyr^{-1}$ but it remains 
non-vanishing at sSFR$\gsim\!0.1 \Gyr^{-1}$.
On the other hand, when AGN feedback is incorporated in the NewHorizon
simulations, the whole galaxy is quenched, with $\Ms$ rather constant in time.
The specific SFR declines more steeply and reaches low values of 
sSFR$\sim\!0.01 \Gyr^{-1}$.
Here, the compaction triggers rapid BH growth and AGN feedback, which in turn
suppresses further gas supply and allows for the complete quenching. 

\smallskip
Qualitatively similar results were obtained 
when comparing other cosmological simulations with and without AGN feedback.
For example, the Horizon-AGN and no-AGN simulations \citep{dubois16}
and similarly the NIHAO simulations \citep{blank19}.
These simulations highlight the role of merger-driven AGN activity in leading 
to complete quenching.
\citet{degraaff24_compaction}, who compared TNG50 simulations with and
without AGN feedback at cosmic morning, 
found that as soon as AGN feedback switches on,
the central $1\kpc$ becomes completely stellar-dominated, with all gas 
expelled, indicating at least central quenching.
We note however that the existing simulations mainly address galaxies at 
cosmic noon, at $z \slsim 3$. Extending the conclusions concerning
 the role of AGN 
feedback to cosmic morning will require simulations that reliably explore the 
galaxy and black hole evolution at higher redshifts, resolving the relevant 
physical processes with sub-pc resolution and appropriate physical recipes. 
\adb{
A hint for the 
crucial role of AGN feedback in the first quenched galaxies at $z \ssim 7$ 
is provided by the semi-analytic simulations GAEA \citep{xie24}, but conclusive
results should await proper high-resolution simulations.
} 

\smallskip 
Observationally, there is potential evidence for outflows and 
AGN feedback in compact quiescent galaxies at cosmic morning 
\citep{onoue24,degraaff24,valentino25,weibel25} 
and early cosmic noon \citep{belli24,deugenio24}.
These are consistent with compaction-driven AGN feedback as the main quenching
mechanism.

\section{Shutoff of gas supply by cold streams}
\label{sec:q3_streams}

\adb{In this section we address the non-trivial challenge of stopping the 
gas supply by cold streams, which is required for complete quenching. 
In \se{challenge} we introduce the challenge.
In \se{properties} we describe the expected relevant stream properties.
In \se{other} we comment on other, largely unsuccessful mechanisms for 
suppressing the streams.
In \se{turbulence} we present preliminary estimates for possible stream 
disruption by CGM turbulence.
In \se{heating} we bring estimates for possible stream disruption by AGN 
photo-heating.
In \se{disk_turbulence} we comment on the effect of disk turbulence.
}

\subsection{The challenge of stream disruption \adb{and the critical radius}}
\label{sec:challenge}

The feeding of massive galaxies at high redshifts is primarily by cold
streams from the cosmic web \citep{db06, dekel09, keres09}.
The average baryonic accretion rate (mostly through streams) onto a halo
of mass $\Mv$ at $z$, based on \citet{dekel13}
\citep[also eq.~31 of][]{dekel23}, is
\be
\Mdotac \simeq 620 \msun\yr^{-1}  M_{12}^{1.14} (1+z)_7^{5/2} \, ,
\label{eq:Mdot_acc}
\ee
where $\Mv \sequiv 10^{12}\msun M_{12}$ is the halo virial mass and
$1+z \sequiv 7\, (1+z)_7$.
The observed quiescent galaxies at cosmic morning, as well as
the fiducial post-FFB halos according to \fig{post}, are indeed in the ball
park of $\Mv \ssim 10^{12}\msun$ at $z \ssim 6$.     
Thus, the accretion rate onto them is well above the observational estimates
of the upper limits on their SFR, 
which are on the order of $\sim\! 10 \msun\yr^{-1}$.
Therefore, a crucial requirement for quenching in these
galaxies is that the cold streams are largely disrupted
and the cold gas mass flux is greatly reduced before they feed 
the central galaxy. As we see below, this turns out to be a very non-trivial 
requirement.

\smallskip 
Early work on the purely hydrodynamical evolution of cold streams showed that 
sufficiently narrow streams, with small radii that are related to the virial
radius as $\Rs \slsim 0.05\,\Rv$, 
would fully disrupt in the hot CGM due to Kelvin-Helmholtz Instabilities 
\citep[KHI,][]{mandelker16,mandelker19a,padnos18}. 
However, these instabilities are stabilized by radiative cooling, self-gravity 
and magnetic fields. When radiative cooling is considered,
hot CGM gas cools and condenses onto the stream through a
turbulent radiative mixing layer that forms at the interface between the
stream and the CGM due to the initial onset of KHI \citep{mandelker20a}.
This causes the cold gas mass in the streams to grow rather than decline 
due to KHI, provided the cooling time in the turbulent mixing layer, 
$\tcoolmix$, is shorter than the timescale for KHI to fully disrupt 
the stream in the pure-hydro case (hereafter
the shearing timescale, $\tshear$).
Analogous results were obtained in the context of spherical clouds or planar
shear layers 
\citep{gronke18,ji19,gronke20b,sparre20,fielding20,tan21,kanjilal21}.

\smallskip   
These considerations, confirmed by idealized  
\adb{hydro-simulations of cylindrical or conical streams}
\citep[][and references therein]{aung24}, 
yield a critical stream radius above which entrainment of hot CGM gas onto the 
cold stream dominates over KHI-induced mixing of stream gas into the CGM. 
\adb{Thus,}, 
streams with $\Rs \sgt \Rscrit$ are expected to survive the journey 
to the central galaxy. 
The critical radius is estimated by \citep{mandelker20a,mandelker20b}
\be
\Rscrit\simeq 0.45\kpc\, \alpha_{\rm 0.1}\, \delta_{100}^{3/2}\, \machb\,
\frac{ {\Ts}_{,4} } { {\ns}_{,0.01}\, \Lambda_{\rm mix,-22.5} } \, .
\label{eq:Rstream_crit}
\ee
\adb{Here,}
$\delta_{100} \sequiv \rhos/(100\,\rho_{\rm b})$ is the density ratio 
between the cold stream and hot CGM (background).
Then
$\machb \sequiv \Vs/c_{\rm b}$ is the stream velocity normalized by the 
CGM sound speed, and 
${\Ts}_{,4} \sequiv \Ts/(1.5\stimes 10^4{\rm K})$ is
the stream temperature relative to thermal equilibrium with the UV background. 
The quantity 
${\ns}_{,0.01} \sequiv \ns  /(0.01\cmc)$                 
is the hydrogen number density in the stream.
Then
$\Lambda_{\rm mix,-22.5} \sequiv 
\Lambda(T_{\rm mix},\rho_{\rm mix})/10^{-22.5} \ergs \cm^3$ 
is the net cooling rate in the turbulent mixing layer assumed to have 
temperature and density $T_{\rm mix} \seq \delta^{1/2} \Ts$, and 
$\rho_{\rm mix} \seq \delta^{-1/2} \rhos$.
The factor
$\alpha_{0.1}$ is a dimensionless number parameterizing the growth rate of 
KHI in the non-linear regime in the absence of cooling, which
depends on $\Vs/(\cs +c_{\rm b})$, where $\cs$ and $c_{\rm b}$ are the
sound speeds in the stream and in the CGM. It is in the range 
$\alpha_{0.1}\ssim (0.5\sdash 2)$ \citep{padnos18,mandelker19a}.

\subsection{Stream properties and disruption by KHI}
\label{sec:properties}

A key input for evaluating $\Rs/\Rscrit$ and therefore
our analysis of stream disruption
is an estimate of the range of properties for the cold streams as they enter 
the virial radius, which we discuss here in some detail.
The stream properties at given halo mass and redshift
are derived from a model that assumes pressure confinement of the cold 
streams by the hot CGM \citep{mandelker20b}.
According to this model, the stream-to-CGM density ratio is 
\be
\delta_{100} \simeq 2.3\, \Mvtwelve^{2/3}\,(1+z)_7\, 
\frac{T_{\rm b,v}}{{\Ts}_{,4}} \, ,
\label{eq:delta_s}
\ee
where $T_{\rm b,v} \sequiv T_{\rm b}/\Tv$ is the temperature in the outer CGM 
normalized by the halo virial temperature.
The stream density is 
\be
{\ns}_{,0.01} \simeq 15\, \Mvtwelve^{2/3}\, (1+z)_7^4\, f_{\rm h,0.3}\,
\frac{T_{\rm b,v}}{{\Ts}_{,4}} \, . 
\label{eq:n_Hs}
\ee
Here, 
the mass of the hot CGM is assumed to be $f_{\rm h} \fb \Mv$ 
with $f_{\rm b}\ssim 0.17$ the universal baryon fraction 
and 
\adb{$f_{\rm h} \sequiv 0.3\,f_{\rm h,0.3}$ 
the mass fraction of baryons in the hot CGM.} 
The hot gas density profile in the outer CGM is assumed to follow an NFW 
profile with a concentration of $c=10$. 
Finally, the stream radius is estimated as
\be
\Rs \simeq 7\kpc \,(1+z)_7^{-1} 
\left( \frac{ f_{\rm acc,s}\, {\Ts}_{,4} }
{ \eta\,f_{\rm h,0.3}\,T_{\rm b,v} } \right)^{1/2} \, .
\label{eq:R_s}
\ee
Here, 
$\eta \sequiv \Vs/\Vv \ssim 0.9\, \machb\, T_{\rm b,v}^{1/2}$ is the stream 
velocity normalized by the halo virial velocity,
also related to the stream Mach number $\machb$. 
The factor $f_{\rm acc,s}$ represents the accretion rate onto the halo along 
a given filament normalized by the average baryonic accretion rate onto the 
halo (\equnp{Mdot_acc}) divided by three streams. 

\smallskip
Taken altogether, we have
\be
\frac{\Rs}{\Rscrit} \simeq 60\, \Mvtwelve^{-1/3}\, (1+z)_7^{3/2}
\frac{f_{\rm acc,s}^{1/2}\,f_{\rm h,0.3}^{1/2}\,
\Lambda_{\rm mix,-22.5}\, T_{\rm b,v}^{1/2}}
{\eta^{3/2}\,\alpha_{\rm 0.1}\,T_{\rm s,4}}.
\label{eq:R_s_crit}
\ee
In general, for stream disruption this ratio has to be smaller than unity.

\smallskip
The Hydrogen column density in the stream is given by $\Ns \sequiv \ns\, \Rs$.
Using \equs{n_Hs} and \equm{R_s}, it can be expressed in terms of halo mass
and redshift in units of $10^{21} \cm^{-2}$ as
\be
N_{\rm s, 21} \simeq 3.25\,\Mvtwelve^{2/3}\, (1+z)_{7}^3\, 
\left( \frac{ f_{\rm acc,s} \, f_{\rm h,0.3} \, T_{\rm b,v} } 
            { \eta\, T_{\rm s,4} } \right)^{1/2} \, .
\label{eq:Ns}
\ee
When the stream is photo-heated by external radiation, $\Ns$ is monotonic with 
the optical depth, and $\Rs/\Rscrit$ is correlated with $\Ns$ 
\adb{(see \se{heating})}.

\smallskip 
\adb{
Is there a range of stream properties that
may give rise to stream disruption by KHI?
}
We see from \equ{R_s_crit} that, for the fiducial case (where all the
parameters as normalized in the equation are set to unity), 
streams are significantly larger than the 
critical radius for survival and growth. In fact, for fiducial streams, 
entrainment of hot CGM gas onto the cold stream can roughly triple the cold 
gas inflow rate onto the galaxy compared to the inflow onto the halo 
\citep{mandelker20b,aung24}. 
However, there is significant uncertainty 
in the various normalization parameters, and these can significantly lower 
$\Rs/\Rscrit$. For example, the number of cosmic web filaments 
that connect to a halo (halo connectivity) is expected to be a function of 
the peak-height of the halo \citep{codis18}. Therefore, the number of 
filaments feeding a halo of a given mass can increase with redshift, such that 
it is reasonable to assume that six filaments or more feed a halo of 
$\sim\! 10^{12}\msun$ at $z\sim 6$ \citep{codis18}, 
which also seems consistent with cosmological simulations 
\citep[e.g.][]{mandelker21}. This would naturally decrease $f_{\rm acc,s}$ 
by a factor of $\gsim\! 2$. Combined with reasonable scatter in the total 
accretion rate implies that  $f_{\rm acc,s}\ssim 0.25$ is reasonable. 
In post-FFB galaxies, where the stellar-to-halo mass ratio is high, the 
baryonic mass fraction in the hot CGM may be low, so $f_{\rm h,0.3}\ssim 0.5$ 
is reasonable. The CGM temperature near the virial shock is set by the shock 
jump conditions and is often less than $\Tv$ \citep{bd03} which reflects the 
average temperature in a hydrostatic halo. Cosmological simulations also show 
that the CGM temperature in the outer halo is typically less than $\Tv$ 
\citep{lochhaas21}.  Therefore, $T_{\rm b,v}\sim 0.5$ is also reasonable. 
If the streams are heated either by local external sources of radiation from 
nearby galaxies or by internal turbulence, one 
could also reasonably obtain ${\Ts}_{,4}\ssim 2$. Taking all of these together, 
we still have $\Rs/\Rscrit\ssim 7\,\Mvtwelve^{-1/3}\,(1+z)_7^{3/2}$. 
If the streams enter the halo at the escape velocity rather than the virial 
velocity, then $\eta\ssim 1.4$ and $\Rs/\Rscrit\ssim 2.5$. We thus find 
that even with assumptions that are favorable for stream disruption, streams 
at $z\sgsim 6$ are still expected to be in the fast-cooling entrainment regime 
and thus hard to be disrupted by KHI. 

\smallskip 
At $z\ssim 1 \sdash 4$, the turnover in the observed star-forming main 
sequence, where the sSFR begins declining, is found to be well fit by the 
mass for which the fiducial $\Rs/\Rscrit\ssim 20$ rather than unity
\citep{daddi22a}.  This rather high threshold ratio 
could be explained
by the fact that idealized shear-layer simulations suggest that when the 
ratio is $\lsim\! 10$ the cold gas mass initially declines before eventually 
rising due to late onsetting entrainment
\citep[e.g.][]{gronke20b,mandelker20a,kanjilal21}. 
Furthermore, supersonic velocities with $\machb \sgsim 2$ 
(as expected for streams in free-fall, \citealp{mandelker20b,aung24}) 
can suppress entrainment by a factor of $\gsim\! 2$ and lead to cold gas 
evaporation even in the nominal fast-cooling regime \citep{yang23}. 
If correct, this implies that a ratio of $\Rs/\Rscrit \ssim 7$ may be small
enough to initiate quenching. 
On the other hand, idealized simulations of cold streams, flowing through a 
hot CGM in hydrostatic equilibrium within an NFW-halo potential, show that 
even if $\Rs/\Rscrit \slt 1$ at $\Rv$, this ratio increases as the stream 
flows down the potential well due to the larger ambient pressure.
\adb{Thus}, while the cold gas mass declines near the virial radius due to KHI,
there is always some radius $r \slt \Rv$ where entrainment begins and the 
cold gas mass starts  increasing \citep{aung24}. 
Furthermore, even weak self-gravity in the streams 
can reduce KHI-induced mixing due to buoyancy forces \citep{aung19},
while magnetic fields stabilize KHI by magnetic tension \citep{berlok19}, 
with each process extending stream lifetimes by factors of a few. 
These make the shutoff of feeding by streams more challenging. 

\subsection{Other disruption mechanisms}
\label{sec:other}

Other potential avenues of stream disruption besides KHI have been explored.
\adb{
For example, one might think that this could be associated with fragmentation 
of the streams into stars and star clusters.
}
Streams at $z\sgsim 5$ are expected to be gravitationally 
unstable \citep{mandelker18} in the sense that their mass-per-unit-length
is greater than the maximal value for hydrostatic equilibrium 
\citep{ostriker64}. In this case, the streams can fragment into
gravitationally bound clumps prior to reaching the central galaxy,
potentially leading to star formation and 
cluster formation
within the CGM \citep{mandelker18,aung19}. Certain cosmological simulations
indeed show star formation in streams in a manner consistent with this model
\citep{mandelker18,bennett20}, though it is unclear how reliable these results
are given the limited resolution in the CGM and the very crude star-formation 
recipes used in the simulations. Even if this were the case, it is unclear 
why this would lead to complete galaxy quenching
\adb{as required}, since the cold gas in the
streams would still contribute to star formation even if outside the central
galaxy. 

\smallskip 
Another possibility is that the gas in intergalactic cosmic web filaments
is heated up by a cylindrical accretion shock to virial temperatures of 
$\sim\! 10^6$K prior to entering the halo virial radius.
This is expected to occur in cosmic web filaments above a critical line-mass
 \citep{lu24}, 
in much the same way as the CGM in halos above a critical mass is expected to
be hot \citep{db06}. However, preliminary estimates suggest that
while filaments feeding halos of $\Mv\ssim 10^{12}\msun$ at $z \sgt 4$
indeed support a stable accretion shock at the cylindrical virial radius
of the dark 
matter filament (referred to as a hot circum-filamentary medium, or CFM),
the filaments still harbor a significant cold gas fraction in their centers.
\adb{This is} due to 
the cooling time being only slightly longer than the free-fall time
\citep{lu24,aung24}. It thus seems unlikely that the cold gas supply onto
such halos can be completely shut off by cylindrical virial shocks in cosmic
web filaments. 

\begin{table*}
\centering     
\begin{tabular}{c|c|c|c|c|c|c|c|c|c|c|c|c}
\hline
Model          & $\eta$     
               & $f_{\rm acc,s}$  & $f_{\rm h,0.3}$  & $T_{\rm s,4}$
               & $T_{\rm b,v}$  & $\delta_{100}$  & $n_{\rm s,0.01}$
               & ${\Rs}_{,1}$  & $\Rs/\Rscrit$ & $N_{\rm s,21}$
\\ 
\hline      
{\it fiducial}     & $1$    
                   & $1$          & $1$                  & $1$
                   & $1$          & $2.3$                & $15$
                   & $7$          & $60$                 & $3.2$
\\          
{\it slow cooling} & $\sqrt{2}$ 
                   & $0.1$        & $0.3$                & $2$
                   & $0.4$        & $0.45$               & $0.9$
                   & $7.6 $       & $1.16$               & $0.2$
\\    
{\it fast cooling} & $1$        
                   & $2$          & $2$                  & $0.5$
                   & $1$          & $4.6$                & $60$
                   & $5$          & $240$                & $9.3$
\\
{\it low gas}      & $1$        
                   & $0.1$        & $0.4$                & $1$
                   & $1$          & $2.3$                & $6$
                   & $3.5$        & $12$                 & $0.65$
\\           
\hline
\end{tabular} 
\caption{
Stream models. Listed are the
normalization parameters and the corresponding stream properties based on the
\citet{mandelker20b} model (\equnp{delta_s} to \equnp{Ns}), 
for a halo with $\Mvtwelve \seq 1$ and $(1+z)_7 \seq 1$. 
In the {\it fiducial} model all the parameters are at 
their fiducial values. The {\it slow cooling} and {\it fast cooling} models 
minimize and maximize $\Rs/\Rscrit$, thus maximize and minimize the chances for 
stream disruption, respectively. 
The {\it low gas} has small $\Ns$ and $\Rs$ that increase the stream survival, 
by reducing $f_{\rm acc,s}$ and $f_{\rm h}$.
In all models $\alpha_{0.1} \seq 1$ and $\Lambda_{\rm mix,-22.5} \seq 1$ 
(with no AGN).
}
\vspace{-13pt}
\label{tab:models}
\end{table*}

\smallskip 
Finally, it has been speculated that the cold gas streams at the
centers of hot cosmic-web filaments 
may ``shatter" when penetrating the virial shock around massive halos.
This is because the confining pressure in the hot CGM is more than an order of
magnitude larger than the confining pressure in the hot CFM \citep{lu24,yao25}.
This can lead to a shock penetrating the cold stream radially (in a 
cylindrical sense) which then causes the stream to fragment into small
cloudlets of size  
$l_{\rm cool}\ssim {\rm min}(c_{\rm s}t_{\rm cool})\ssim 10\pc$
due to Richtmeyer-Meshkov and thermal instabilities 
\citep{mccourt18, gronke20b, yao25}. 
These small cold clouds then disperse in the outer CGM and can be disrupted 
by hydrodynamic instabilities as they are smaller than the critical size for 
cloud survival \citep{gronke18,gronke20,gronke22}.
However, preliminary results testing this
hypothesis in idealized simulations suggests that this does not shut off the 
flux of cold gas towards the central galaxy. This is because the rapid cooling
of high density gas in the inner halo leads to strong coagulation of the 
resulting cloudlets, while the strong shear between the initial stream and the
hot CGM focuses the coagulated gas into a (possibly more clumpy) 
stream-like structure (Yao et al., in prep.). 
Based on the aforementioned studies, we estimate that
the critical stream radius above which coagulation will 
dominate over shattering is roughly a factor of $\sim\! 2$ larger 
than the critical radius above which entrainment dominates over mixing 
(\equnp{Rstream_crit}). As discussed above, typical streams are expected 
to be significantly larger than this (\equnp{R_s_crit}), 
which limits the viability of shattering as a mechanism for stream disruption.

\smallskip
Two more promising avenues for stream disruption have yet to be considered in
detail, both of which may be enhanced by strong AGN feedback. 
These are stream disruption by strong turbulence in the CGM
and photo-heating of the streams by AGN radiation.
Recall that AGN feedback is understood to operate in two different modes
depending on the black-hole accretion rate.
The thermal (or quasar) mode is dominant 
when the BH growth rate exceeds $1\%$ of the Eddington limit, while the jet
(or kinetic) mode operates at lower accretion rates \citep{heckman14}.  
The thermal mode may be particularly relevant during the early stages of
quenching, as it is initiated by compaction or merger events which can trigger
large gas inflow rates towards the center of the galaxy.
As the quenching proceeds and the BH accretion rate drops below the $1\%$
Eddington threshold, the AGN is assumed to transition to the kinetic (or jet)
mode. In this
regime AGN outflows, including jets and the associated winds, can inject
turbulence into the CGM which could suppress the cold inflows as we see below.
This may be
particularly relevant in the denser inner regions of the halo, where the
enhanced coupling between the outflows and the ambient gas may more effectively
disrupt the cold streams in the turbulent environment. 
\adb{
Simulations by \citet{dubois13} indeed demonstrated that cold streams at high 
redshift are susceptible for disruption by AGN feedback 
\citep[see also][]{costa14,costa22}.
}

\subsection{Stream disruption by CGM turbulence}
\label{sec:turbulence}

The first promising mechanism for stream disruption is due to a highly 
turbulent CGM, with turbulent Mach numbers $\Mtu$ of order unity
or slightly higher with respect to the hot gas. 
Such turbulence may disrupt streams by generating 
shocks at the stream interface, potentially enhancing shattering, and
also reducing the efficiency of entrainment.

\smallskip 
A crude estimate for the disruptive effect of CGM turbulence on a stream
can be obtained based on the analysis of \citet{gronke22} 
of cold clouds that are embedded in a hot turbulent medium.
They find that the clouds disrupt once they are smaller than a 
critical size for survival, which increases rapidly with turbulent Mach number
for supersonic turbulence.
\adb{
We note in passing that \citet{jennings23} included thermal conduction in 
conditions pretty close to what the AGN bubble ($10^8$K) and cold streams 
($10^4\sdash 10^5$K) could be \citep[see also][]{armillotta16,armillotta17}.   
}
As shown in eq.~9 of \citet{gronke22}, the survival criterion of clouds in 
a turbulent box was calibrated similarly to the classical cloud-crushing 
framework in a wind tunnel
but modified with an additional Mach number dependence. 
Specifically, the critical size for survival can be approximated as
\be
\Rtu \sim  10^{0.6\mathcal{\Mtu}} R_\mathrm{crit,w} \, ,
\label{eq:Rturb}
\ee
where $R_\mathrm{crit,w}$ denotes the critical radius for cloud survival
in a wind-tunnel setup, as defined in eq. 2 of \citet{gronke22}.

\smallskip
This wind-based criterion arises from equating the mixing-layer cooling 
timescale with the cloud-crushing timescale, and is analogous to the
corresponding radius for streams (\equnp{Rstream_crit}) which was calibrated by 
equating the mixing-layer cooling timescale with the KHI mixing timescale.
For clouds, the Mach number $\mathcal{M}$ in the expression for the wind 
$R_\mathrm{crit,w}$ is replaced by the turbulent Mach number $\Mtu$, 
reflecting the enhanced mixing and dynamical disruption driven by turbulence 
compared to a laminar wind. This is sensible for a spherical cloud where 
much mixing occurs behind the bow shock near the cloud head in wind-tunnel 
setups \citep{kaul25}, and is therefore likely to similarly arise behind 
turbulent shocks distributed isotropically. However, for streams, the primary 
mixing and disruption mechanism is KHI induced by the bulk flow rather than 
a bow shock or something similar \citep{mandelker20a,kaul25}. 
Therefore, we speculate as an educated guess based on previous results 
that the Mach number $\mathcal{M}$ in the expression 
for the wind $R_\mathrm{crit,w}$ should remain the bulk Mach number as in 
\equ{Rstream_crit}, with the turbulent critical radius given by \equ{Rturb}. 
\adb{This means} 
multiplying $R_\mathrm{crit,w}$ by $10^{0.6\Mtu}$. 
For $\Mtu \seq 1$, this will lower $\Rs/\Rscrit$ by a factor of $\sim\! 4$ 
with respect to \equ{R_s_crit}. We stress that this increased survival 
radius due to turbulence is in addition to the aforementioned reduced 
entrainment efficiency for supersonic flows based on \citet{yang23}, 
with the relation between them not yet clear. 
This may also help explain the observed association of the threshold 
$\Rs/\Rscrit\ssim 20$ with the turnover of the star-forming main sequence 
\citep{daddi22b}. 

\begin{figure} 
\centering
\includegraphics[width=0.49\textwidth, trim={0.3cm 0.5cm 0.0cm 0.0cm},clip]
{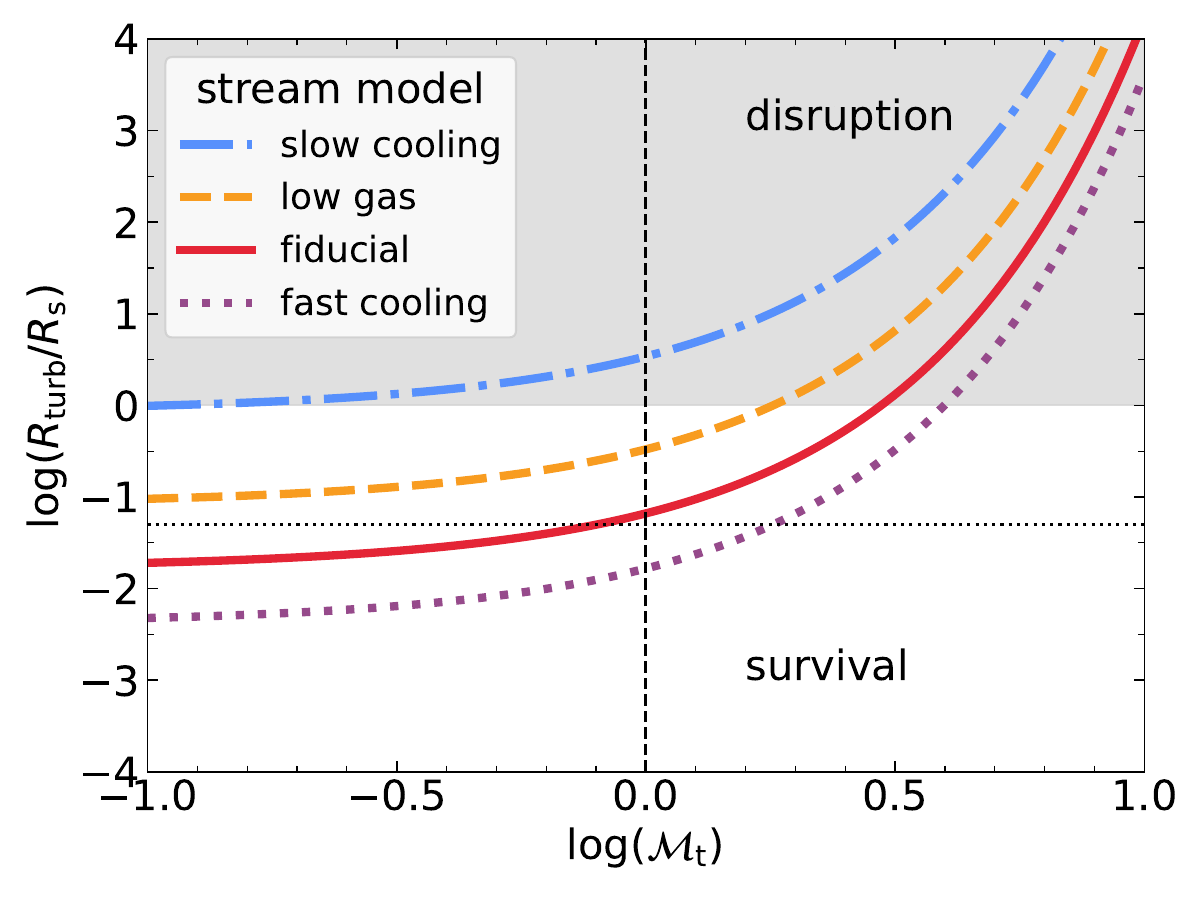}   
\vspace{-5pt}
\caption{
Stream disruption by CGM turbulence.
Shown is an estimate of the critical stream radius for disruption by CGM 
turbulence, $\Rtu$, with respect to the actual stream radius, 
as a function of the turbulent Mach number.
The transonic point is marked by a vertical dashed line.
This estimate is based on a crude extrapolation of the results for clouds
\citep{gronke22} to streams.
The cases shown refer to stream properties in $\Mv \ssim 10^{12}\Msun$
halos during the central phase of the post-FFB quiescent regime at cosmic
morning, $z \ssim 6$.
The relevant parameters of the four cases are listed in \tab{models}.
We learn that,
with respect to $\Rtu/\Rs \seq 1$ (the horizontal line separating the shaded
and un-shaded areas), the {\it slow cooling} stream is expected to be capable 
of being disrupted even by subsonic turbulence. 
The {\it low gas} stream may disrupt even by transonic or slightly supersonic
turbulence
On the other hand, the {\it fiducial} stream and the {\it fast cooling} stream 
seem to require supersonic turbulence of $\Mtu \ssim 3$ for disruption.
With respect to the more relaxed threshold $\Rtu/\Rs \seq 0.05$ 
(which is motivated by observations - the horizontal doted line), 
the fiducial streams are expected to be disrupted even by transonic turbulence.
}
\vspace{-10pt} 
\label{fig:Rturb}
\end{figure}

\smallskip 
The stream disruption is evaluated for four different models of stream
properties, presented in \tab{models}. 
They are based on the \citet{mandelker20b} model 
(\equnp{delta_s} to \equnp{Ns}, for a halo with mass
$\Mv \seq 10^{12}\msun$ at $z \seq 6$.
In these models the normalization 
parameters in \equ{R_s_crit} are varied, to minimize or maximize 
$\Rs/\Rscrit$, thus spanning a {\it fiducial} case along with streams more 
prone to disruption and streams more prone to survival. 
In the {\it fiducial} model all the normalization parameters are at
their fiducial values. The {\it slow cooling} model chooses parameters in order
to minimize the ratio of $\Rs/\Rscrit$ thus maximizing the chances for stream
disruption. The {\it fast cooling} model chooses parameters in order to 
maximize $\Rs/\Rscrit$ and thus maximize the chances for stream survival
and growth.
The {\it low gas} model chooses parameters to have a small $\Ns$
and $\Rs$ by reducing the accretion factor per stream $f_{\rm acc,s}$
and the fraction of baryons in the hot CGM $f_{\rm h}$.
In all models we keep $\alpha_{0.1} \seq 1$ and
$\Lambda_{\rm mix,-22.5} \seq 1$ (with no AGN).
While each of these can provide an extra factor of two in $\Rs/\Rscrit$
independent of AGN, they have no effect on the physical properties such as 
$\delta$, $\ns$, $\Rs$ and $\Ns$.
The correlation of $\Rs/\Rscrit$ and $\Ns$
indicates that streams with slow/fast  
cooling, which are easier/harder to disrupt by KHI, are also more/less
susceptible to heating by AGN.

\smallskip
\Fig{Rturb} shows the estimated critical stream radius for disruption by 
turbulence with respect to the stream radius, $R_{\rm turb}/\Rs$, as a 
function of the turbulent Mach number $\Mtu$,
based on \equ{R_s_crit} and \equ{Rturb}.
While all four cases of stream properties 
seem to be in the survival regime based on the fact that
$\Rs/\Rscrit$ is larger than unity according to \equ{R_s_crit}, 
\fig{Rturb} shows that they are all pushed into the disruption regime when  
$\Mtu$ is sufficiently large. 
If one considers the observationally motivated threshold of $\sim\! 0.05$ 
rather than $\sim\! 1$, this is made even more pronounced. 
We comment that due to the steep increase of $\Rturb$ with Mach number in 
the supersonic regime, the threshold Mach number for stream disruption 
is only weakly dependent on redshift and halo mass. 
While sustaining significant turbulence in the outer CGM would probably require
intense accretion or mergers, stream disruption may be more feasible 
in the inner CGM, where the supersonic fraction can be substantial,
driven by the interplay between cooling flows and feedback processes. 

A specific scenario for a turbulent CGM has been proposed by Stern et al. 
(private communication) based on spherical simulations and cosmological FIRE
simulations. Following the results for cold inflow in halos 
of $\Mv \sleq 10^{11}\msun$ and for hot halos when $\Mv \sgeq 10^{12}\msun$
\citep{bd03,keres05,db06,keres09},
they find that halos of intermediate mass, $\Mv \ssim 10^{11.5}\msun$, develop 
a cold, turbulent inner region that may aid in disrupting the stream.
The turbulence may
be driven either by the inflowing streams themselves, by contracting  
cooling flows in the CGM, or by mergers through compaction events. 
AGN feedback from the galaxy (not included in these simulations)
may increase the turbulence to even higher levels than inferred in the 
simulations and/or push it up to higher halo masses.
However, the CGM in this scenario, as studied at low redshifts, is relatively 
cold, with temperatures $T\ssim (2\stimes 10^5){\rm K}$. 
It is unclear how to extrapolate their results to high redshifts and
whether it would be consistent with the assumption of \citet{mandelker20b}, 
which is the basis for our current model,
that the CGM is hot and in pressure equilibrium with the cold stream.

Several caveats are worth noting regarding the speculative crude extrapolation 
from clouds to streams.
First, the steep increase in $R_\mathrm{turb}$ 
with supersonic Mach number warrants further exploration since
\citet{gronke22} calibrated their parameters primarily in the subsonic regime.
Second, if the clouds are much larger 
than the minimal size for cloud survival or the turbulence is subsonic or
transonic, the cloud growth can actually be enhanced by entrainment 
\citep{gronke22}. 
We thus caution that the net effect of CGM turbulence on realistic
streams is yet to be investigated more accurately using simulations 
(Yao et al., in prep.).

\subsection{Stream disruption by photo-heating from AGN}
\label{sec:heating}

The second way by which AGN can potentially disrupt streams even more directly
is by radiative heating. High resolution cosmological simulations with explicit
radiative transfer have shown that in dwarf galaxies ($\Mv\slsim 10^8\msun$)
during the epoch of reionization ($z>6$), photo-heating from starbursts in the 
central galaxies can heat the gas in the cosmic web filaments feeding these 
galaxies to temperatures of $T\gsim 10^{4.5}{\rm K}$ \citep{katz20}. Since the
filaments feeding these low-mass halos have virial temperatures of 
$\lsim\!10^4$K, this prevents gas from cooling to the center of 
the dark matter filament and forming the cold stream which otherwise would have
fed the central galaxy. It can even totally unbind the filament gas, thus 
effectively shutting off the cold gas supply to the central galaxy. 
This process is similar to so-called `reionization feedback' in low mass halos 
with $\Tv\slsim 10^4$, which is thought to be the primary reason that such 
low-mass halos remain dark with no appreciable star formation over cosmic 
history \citep[e.g.][]{benitez20}. 
The filaments that feed the massive 
quiescent galaxies during cosmic morning have virial temperatures of 
$\sim\! (10^{5.5} \sdash 10^6){\rm K}$ \citep{lu24,aung24}, and as such
will likely not be affected by photoheating from central star formation. 
However, AGN heating may heat the gas to much higher temperatures, in excess of
$10^5{\rm K}$ and perhaps even $10^6{\rm K}$, which would have similar
suppression effects on the cold streams feeding these massive galaxies.

\smallskip
The massive quiescent
galaxies observed at cosmic morning are typically of stellar mass 
$\sim\!10^{10.5}\msun$, suggesting central black holes of $\sim\!10^{8.5}\msun$
assuming a high BH-to-stellar mass ratio of $\sim\! 0.01$ 
\citep{harikane23_agn,maiolino24}.
The corresponding Eddington luminosity is 
$L_{\rm edd} \ssimeq 4\stimes 10^{46}\ergs$.

\begin{figure} 
\centering
\includegraphics[width=0.49\textwidth, trim={0.2cm 0.3cm 0.0cm 0.0cm},clip]
{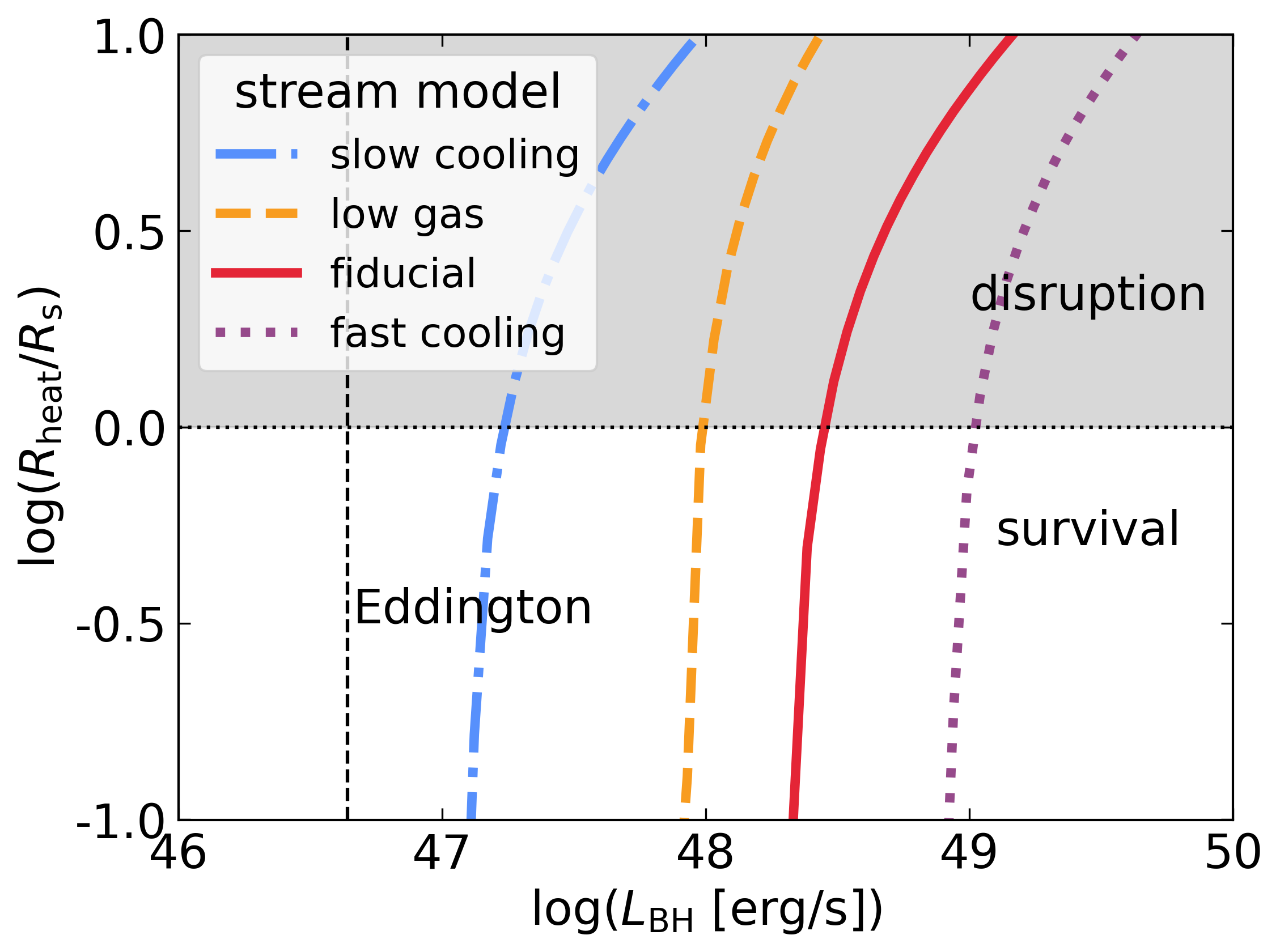}  
\vspace{-5pt}
\caption{
Direct photo-heating of a cold stream by an AGN.
Shown is the \adb{ratio $R_{\rm heat}/\Rs$}, namely,
the depth into the stream where the equilibrium temperature has 
increased to above $10^5$K with respect to the stream radius, 
as a function of AGN luminosity $\Lbh$.
\adb{This ratio, being below or above unity,
is a measure of disruption capability, as marked.}
The different cases of stream properties are listed in \tab{models} 
for $10^{12}\msun$ halos at $z \ssim 6$.
The metallicity in the stream is assumed to be $0.03 \Zsun$. 
The vertical line refers to the Eddington luminosity 
$\Lbh \seq 4 \stimes 10^{46} \ergs$ for a central black hole
of mass $M_{\rm bh} \seq 10^{8.5}\msun$. 
We learn that 
the disruption of streams by direct heating requires bursts with luminosities
significantly higher than the Eddington luminosity.
}
\vspace{-10pt}
\label{fig:heating}
\end{figure}

\subsubsection{Direct heating of the streams}

Motivated by these estimates, we consider how the radiation from a central AGN
with a given luminosity $L_{\rm bh}$ might directly heat an inflowing cold 
stream of gas number density $\ns$ and metallicity $Z$.
We consider a stream with orbital angular momentum about the central galaxy
and an associated impact
parameter, such that in practice it is irradiated from the sides.
We ignore absorption outside the stream, and assume absorption by the stream
gas as it is photoionized and photo-heated by the radiation to an equilibrium
temperature $T_{\rm eq}$.
The radiation flux at a distance $r$ from the AGN and a penetration depth $x$ 
into the stream is computed by
\be
F_{\rm bh} (r) = \frac{L_{\rm bh}}{4\pi r^2} e^{-\tau(x)} \, ,
\ee
where 
\be
\tau(x) = \int_0^{x} \ns\, \sigma\, {\dd} x \, ,
\ee
with $\sigma$ the cross section for absorption.
We assume thermal equilibrium of the stream gas
under photoionization and collisional ionization, 
and use Cloudy \citep{ferland17} to compute 
$\sigma$ (using $L_{\rm bh}$ and $Z$), and the temperature distribution
$T_{\rm eq}(r)$ (using $L_{\rm bh}$ and $\ns$). 
We define the heating penetration depth $R_{\rm heat}$ as the penetration 
distance $x$ within the stream where $T_{\rm eq}$ exceeds $10^5$K.
Even though this may be below the virial temperature of the dark matter 
filament, such that the gas does not become fully unbound, gas at temperatures 
$\sim\!10^5$K would be more vulnerable to disruption by turbulence, quasar 
winds, or hydrodynamic instabilities, making it easier for AGN to suppress cold
stream accretion.
This can be qualitatively understood by considering ${\Ts}_{,4} \sgsim 10$ 
in \equs{delta_s} to \equm{Ns}. 

\smallskip 
For an isothermal stream embedded in an isothermal halo, 
the cold stream gas density approximately follows $\ns \spropto r^{-2}$ 
\citep{mandelker18,mandelker20b}\footnote{For an NFW halo in hydrostatic 
equilibrium the density profile is even steeper \citep{aung24}.}. 
The radiation flux likewise decays as $F \spropto r^{-2}$. 
Since the equilibrium temperature is largely determined by the ionization 
parameter that is proportional to $F/n$, it is not expected to change much
with galactocentric radius. This implies that our evaluation of the 
photoheating can serve as a crude approximation at any $r$ within $\Rv$.

\smallskip
As before, we adopt the stream parameters listed in \tab{models} based 
on the analytic model of \citet{mandelker20b}. 
\Fig{heating} shows the heating penetration depth $\Rheat$ with respect to the
stream radius $\Rs$ as a function of AGN luminosity $\Lbh$
for the different models of stream properties.
The Eddington luminosity is marked at 
$L_{\rm edd} \ssimeq 4\stimes 10^{46}\ergs$ 
assuming a black-hole of mass $\sim\! 10^{8.5}\msun$.
We learn that, even for the {\it slow cooling} model, 
a luminosity of a few Eddington
luminosities is required for direct heating and disruption of the streams. 
Given the observational indications that about one third of the massive black 
holes at cosmic morning may exceed the Eddington luminosity 
\citep{trakhtenbrot11, farina22}, we conclude that
intermittent super-Eddington AGN outbursts could possibly
heat the inflowing cold streams if they are in the slow cooling regime.
The fiducial streams 
would require a luminosity higher than $L_{\rm bh} \ssim 10^{48} \ergs$, 
which is less likely.

\smallskip 
The above analysis is very preliminary and should be taken with a grain of
salt.
In particular,
absorption in the stream should be more carefully accounted for, along with
scattering and re-emission as well as the effects of ionization by the UV 
background versus self-shielding \citep{mandelker18,aung24}.
These effects might make the overall heating even weaker, because more 
radiation might be absorbed in the outer shell, thus enhancing the shielding
of the inner core.
Second, 
the claimed invariance to the distance from the AGN along the stream,
given the higher density and higher flux at smaller radii,
should be considered more carefully. 
\adb{
Here one should account for realistic stream and halo profiles which are 
not necessarily isothermal and can be steeper than $\ns \sprop r^{-2}$
\citep{aung24}.
}
Properly accounting for these effects will require detailed modeling using
semi-analytic radiative transfer methods or full simulations.

\subsubsection{Indirect disruption by slow cooling in the mixing layer}

While we found that streams in the high-density range are not likely to be 
disrupted by direct AGN heating, the AGN radiation may make the streams more 
prone to disruption by KHI. 
This is by increasing the cooling time of gas within the turbulent mixing 
layers between the cold streams and the hot CGM such that it exceeds the 
shear timescale.
Under such conditions, the entrainment of hot gas into the cold stream
would become ineffective, allowing KHI to disrupt the streams, similar to the 
case with no cooling analyzed in \citet{mandelker19a}. 
In practice, this manifests as decreasing the parameter 
$\Lambda_{\rm mix,-22.5}$ in \equ{R_s_crit} due to AGN photo-heating in the 
mixing layer.
\adb{
The mixing-layer temperature and the stream temperature are increased,
while the mixing-layer density and the stream density are decreased,
and the ionization state of the mixing layer is changed.
}

\begin{figure} 
\centering
\includegraphics[width=0.49\textwidth, trim={0.2cm 0.3cm 0.0cm 0.0cm},clip]
{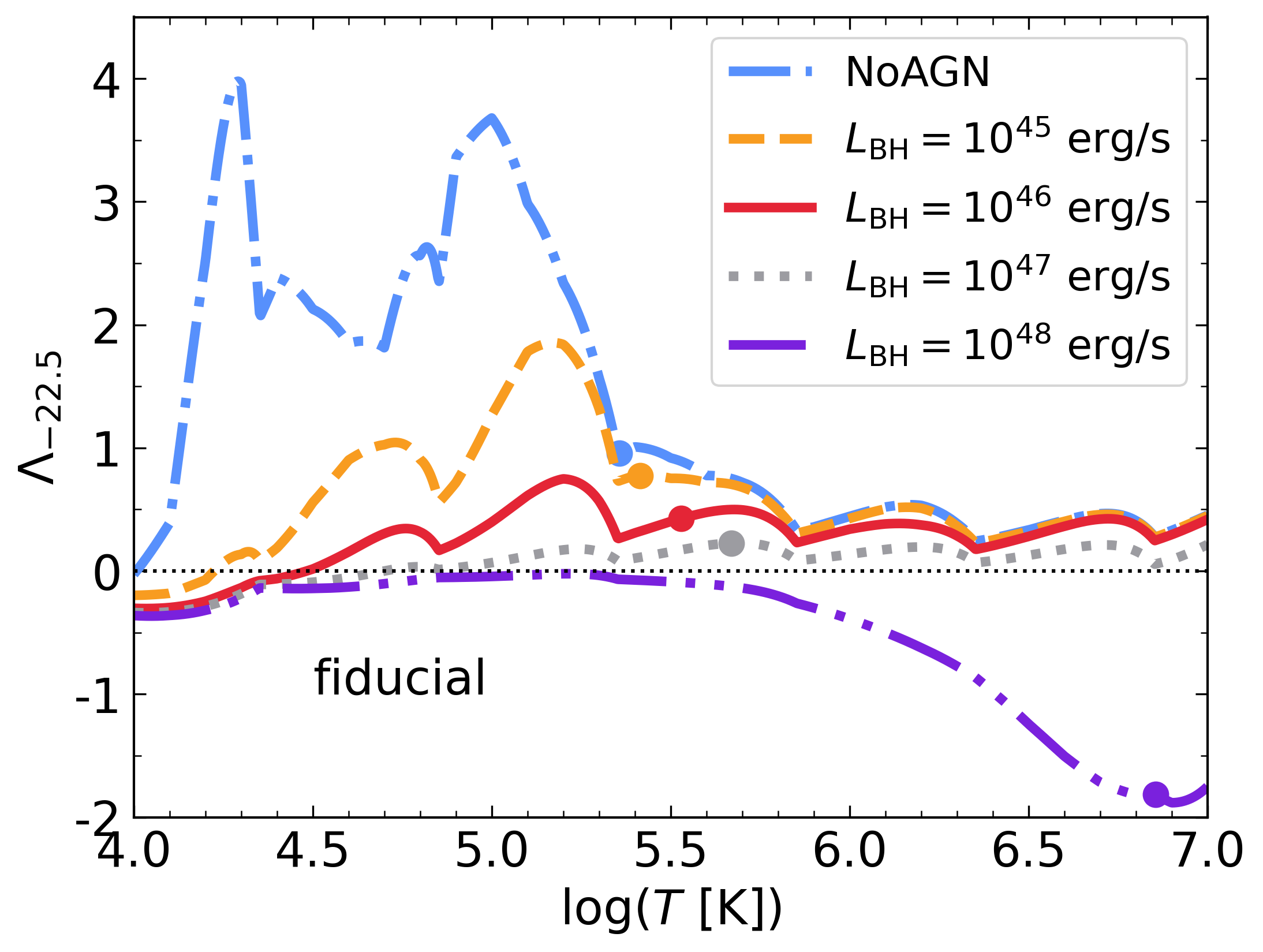}   
\vspace{-5pt}
\caption{
Indirect disruption of a cold stream by an AGN.
Shown \adb{in the stream-CGM mixing layer}
is the net cooling minus heating rate, 
$\Lambda_{\rm -22.5} \seq \Lambda(T)/10^{-22.5}{\rm \ergs \cm^{3}}$,  
as a function of temperature, for different values of AGN luminosity $\Lbh$
including no AGN (\adb{uppermost} blue lines). 
It is shown as an example for the {\it fiducial} model for stream properties
from \tab{models}.  
The cooling curves are computed assuming constant pressure, $nT \seq \ns \Ts$.
The horizontal dotted line marks thermal equilibrium, 
where the net cooling is zero. Above (below) this line, the system is
undergoing net cooling (heating). The circles on each curve show the
mixing temperature, $T_{\rm mix} \seq (T_{\rm eq}T_{\rm CGM})^{1/2}$,
with $T_{\rm eq}$ the lowest temperature where $\Lambda_{-22.5} \seq 0$.
As the AGN luminosity increases, $T_{\rm eq}$ increases while 
$\Lambda_{\rm mix} \seq \Lambda(T_{\rm mix})$ decreases, by 
up to an order of magnitude for $\Lbh \ssim L_{\rm Edd}$. This implies that the
stream and the mixing layer contain much warmer, more diffuse gas than without
the AGN, and suggests that the streams are much more prone to disruption.
    }
\vspace{-10pt}
\label{fig:heating_Lambda}
\end{figure}

\smallskip 
\Fig{heating_Lambda} shows the $\Lambda$ curves of net heating minus cooling 
as a function of temperature for different AGN luminosities, $\Lbh$, 
including the case of no AGN. 
Only the {\it fiducial} stream model from \tab{models} is shown as an example.
The cooling curves are derived at constant pressure, meaning assuming that the 
gas density scales as $n\spropto T^{-1}$, with the 
normalization set by the stream value in each model. 
The cooling function $\Lambda$ is computed for the given gas density and
temperature and the given AGN radiation intensity in
the mixing layer between the cold stream and the CGM at the virial radius,
over the temperature range from the stream edge ($\sim \! 10^4$K) to the 
CGM temperature. 
The symbols in \fig{heating_Lambda} mark the point on each curve where 
$T \seq (T_{\rm eq}\, T_{\rm max})^{1/2}$, 
where $T_{\rm eq}$ is the lowest temperature where $\Lambda \seq 0$
and $T_{\rm max}$ is either $T_{\rm CGM}$ or $2\times 10^7$K if there
is net cooling or net heating at $T_{\rm CGM}$, respectively. 
As $\Lbh$ increases, so does the temperature where the outer 
edge of the stream is in thermal equilibrium. With no AGN, this is set by 
thermal equilibrium with the UV background at 
$T_{\rm eq}\ssim (1\sdash 3)\times 10^4{\rm K}$ depending on the stream 
density. AGN increase $T_{\rm eq}$ by providing extra heating. 
At $L_{\rm Edd}$, $T_{\rm eq}$ is $\sim\! 5$ times higher, while
at $10\,L_{\rm Edd}$ it can be as high as $\sim\! 10^7{\rm K}$. 

\begin{figure} 
\centering
\includegraphics[width=0.49\textwidth, trim={0.3cm 0.3cm 0.0cm 0.0cm},clip]
{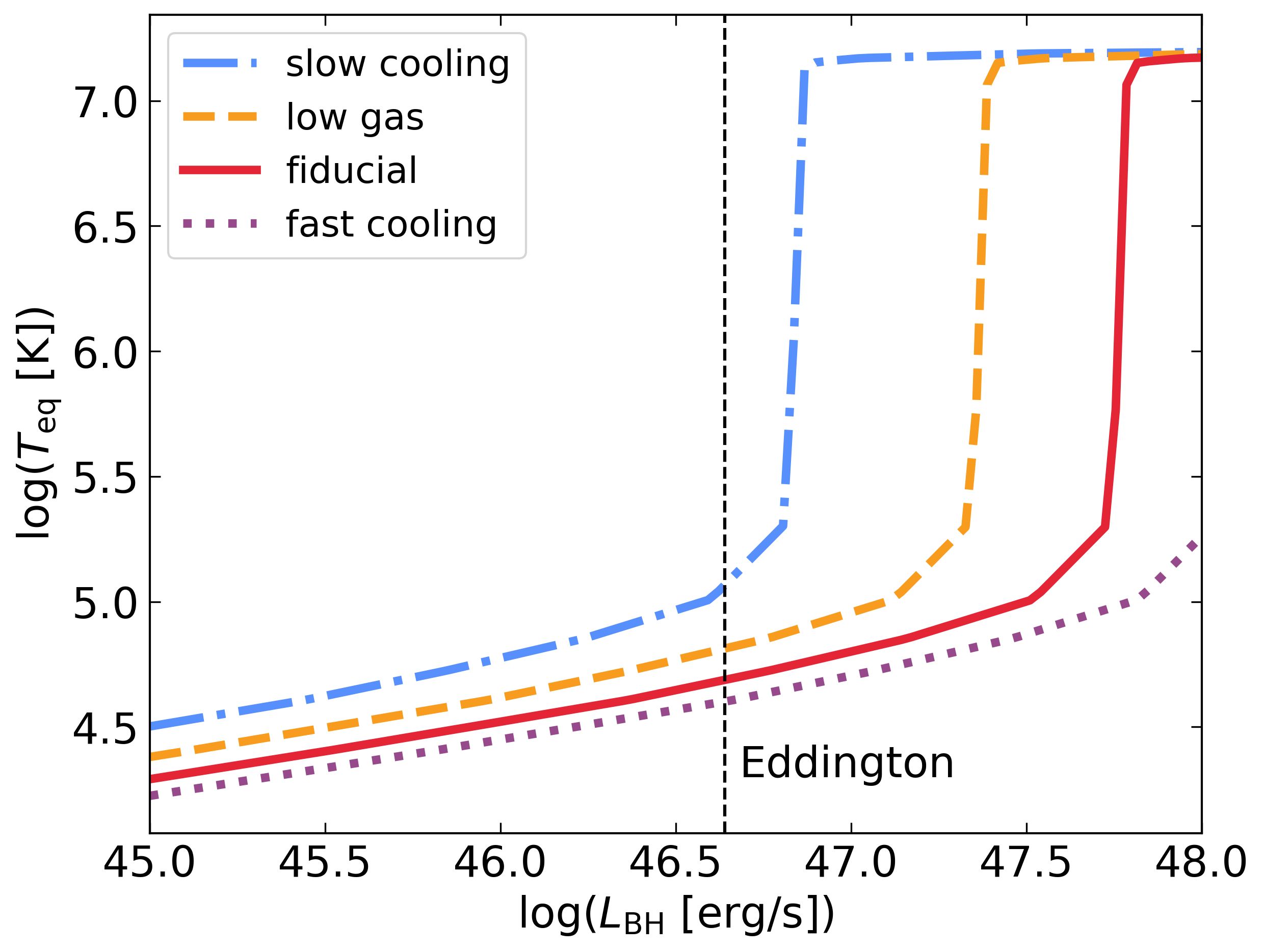}  
\vspace{-5pt}
\caption{
Indirect disruption of a cold stream by an AGN. Shown is the equilibrium  
temperature, $T_{\rm eq}$, as a function of AGN luminosity, $\Lbh$, based on
the cooling curves shown in \fig{heating_Lambda}. This is the lowest 
temperature where the net cooling minus heating vanishes and the gas is 
in thermal equilibrium. As such, this represents the temperature to which the 
outer layer of the stream will heat in the presence of an AGN from its initial
equilibrium temperature of $T_{\rm eq}\ssim (1 \sdash 3)\times 10^4{\rm K}$
when exposed only to the UV background. 
When the gas evolves at constant pressure, then $T_{\rm eq}$ 
gradually rises to $\sim 10^5{\rm K}$ at $\sim\! (1 \sdash 10)L_{\rm Edd}$ 
followed by a very rapid rise to $\sim\! 10^7{\rm K}$. (If the gas evolves at
constant density this rapid increase is absent, and the stream rarely reaches
$T_{\rm eq}>10^5{\rm K}$.)
    }
\vspace{-10pt}
\label{fig:heating_Teq}
\end{figure}

\smallskip
\Fig{heating_Teq} shows $T_{\rm eq}$ as a function of $\Lbh$ 
for our four stream models from \tab{models}.
The curves show $T_{\rm eq}$ assuming isobaric (constant pressure) heating 
and cooling as discussed above. The equilibrium temperature gradually increases 
with $\Lbh$ from $\sim\! (2\sdash 3)\times 10^4{\rm K}$ at 
$\Lbh \seq 0.01\, L_{\rm Edd}$ to $\sim\! 10^ 5{\rm K}$ at 
$\Lbh\ssim (1\sdash 10)L_{\rm Edd}$ before rapidly increasing to 
$\sim\! 10^7{\rm K}$. 

\smallskip
This increase in $T_{\rm eq}$ with increasing $\Lbh$ leads to a corresponding 
increase in the mixing temperature, which we now define as 
$T_{\rm mix}\seq (T_{\rm eq}\,T_{\rm CGM})^{1/2}$, generalizing our earlier 
definition of $(\Ts\, T_{\rm CGM})^{1/2}$. For most cases, we leave the density 
and temperature in the CGM unchanged from \tab{models}. However, if the cooling
curve shows net heating at $T_{\rm CGM}$ for given stream model and 
$L_{\rm AGN}$, we set $T_{\rm CGM} \seq 10^7{\rm K}$ and determine the density 
by enforcing constant pressure. 
For each curve in \fig{heating_Lambda}, $T_{\rm mix}$ is marked with a circle. 
For all models, 
$ \Lambda_{\rm mix} \seq \Lambda(T_{\rm mix}) \ssim 10^{-22.5} \ergs\cm^3 $ 
without AGN, and is $\gsim\! 10$ times smaller at $L_{\rm Edd}$. 
At $10\,L_{\rm Edd}$, the entire mixing layer and even the CGM itself 
may undergo net heating. 

\begin{figure} 
\centering
\includegraphics[width=0.49\textwidth, trim={0.3cm 0.3cm 0.0cm 0.0cm},clip]
{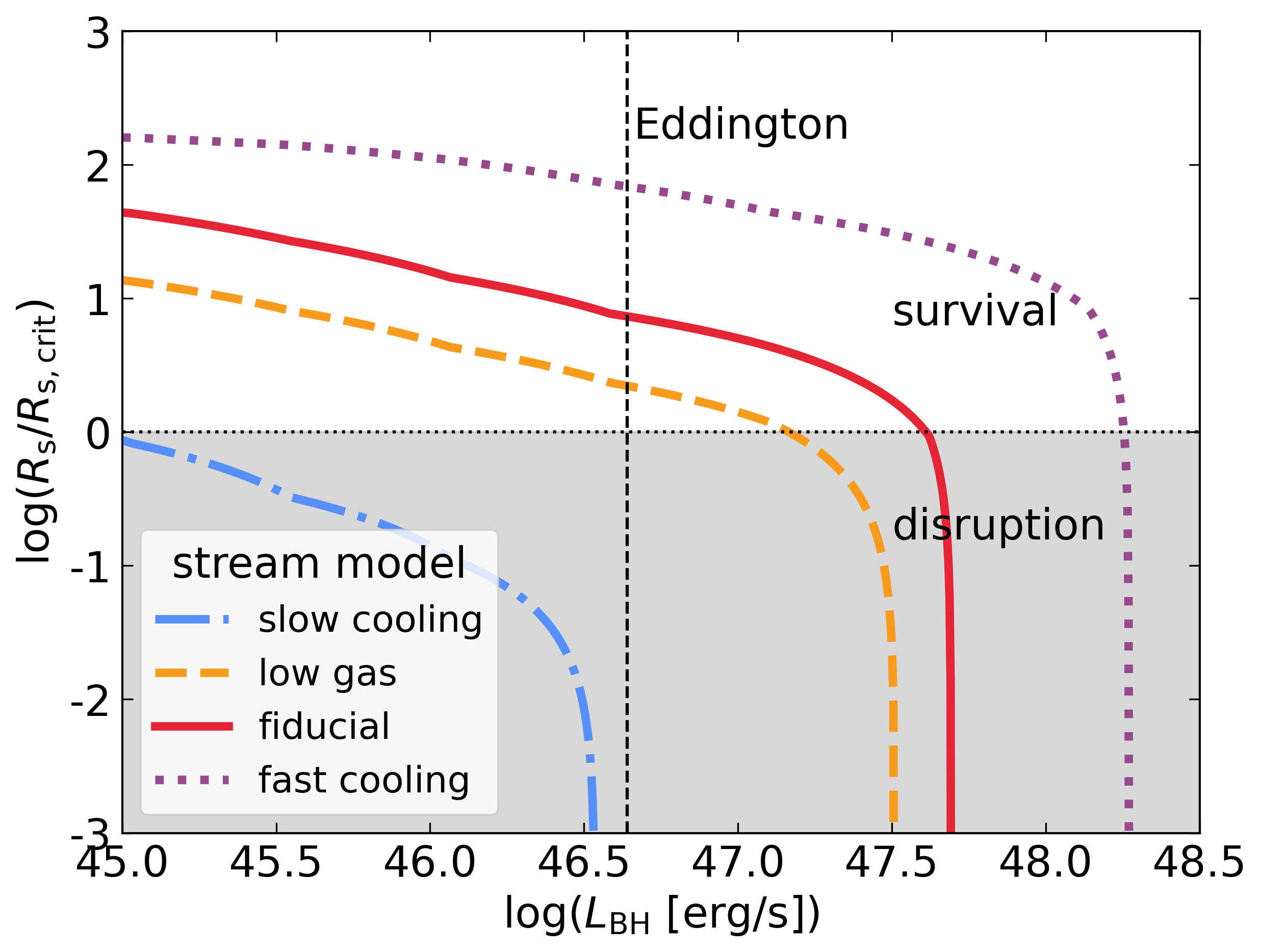}  
\vspace{-5pt}
\caption{
Indirect disruption of a cold stream by an AGN. Shown is the ratio of the
stream radius $\Rs$ 
to the critical value for disruption by KHI (\equnp{R_s_crit})
as a function of $\Lbh$ for our different stream models from \tab{models}. 
As $\Lbh$ increases, $\Lambda_{\rm mix,-22.5}$ from \equ{R_s_crit} decreases
(\fig{heating_Lambda}), while the rest of the parameters remain fixed. 
Model {\it slow cooling} reaches $\Rs/\Rscrit \slt 1$ at 
$\Lbh \ssim 0.01\,L_{\rm Edd}$, implying rapid disruption. 
While the {\it low gas} and {\it fiducial} models only enter this
regime at $\sim (10 \sdash 20)L_{\rm Edd}$, they fall below the empirically
motivated threshold of $\Rs/\Rscrit \slt 20$ at sub-Eddington luminosities, 
suggesting they will be more prone to disruption. The {\it fast cooling} model
stays above this higher threshold until $\Lbh \sgsim 30\,L_{\rm Edd}$, 
suggesting it is likely to survive the presence of strong AGN. 
    }
\vspace{-10pt}
\label{fig:heating_rcrit}
\end{figure}

\smallskip
\Fig{heating_rcrit} shows $\Rs/\Rscrit$ as a function of $\Lbh$
following \equ{R_s_crit} with $\Lambda_{\rm mix}$ computed as described above.
The decline in $\Lambda_{\rm mix}$ causes a corresponding decline in 
$\Rs/\Rscrit$.
Our {\it slow cooling} stream model enters the disruption regime, 
with $\Rs/\Rscrit \slt 1$, at very low AGN luminosities of 
$\gsim\! 0.01\,L_{\rm Edd}$. The {\it low gas} and {\it fiducial} models enter 
this regime at $\sim\! 3\, L_{\rm Edd}$ and $9\, L_{\rm Edd}$ respectively. 
However, they fall below the empirically motivated threshold of 
$\Rs/\Rscrit \slt 20$ at sub-Eddington luminosities, suggesting that they will 
be more prone to disruption. 
The inclusion of turbulence in this slow-cooling regime would make the stream 
disruption even more effective. On the other hand, our {\it fast cooling}
stream model remains above even this more-lenient threshold at least until 
$\sim\! 40\,L_{\rm Edd}$ and is likely to survive.

\subsection{Quenching by disk turbulence}
\label{sec:disk_turbulence}

If the streams do manage to penetrate through the CGM into the central disk,
they provide gas supply for star formation, but they also drive turbulence
in the disk that may suppress the disk instability and thus
slow down star formation.
\citet{gabor14}, using analytic modeling and isolated-disk simulations, 
showed that the streams are indeed expected to be effective in suppressing star
formation this way at $z \sgt 2\sdash 3$.
This is consistent with the finding of \citet{ginzburg22} that at $z \sgt 3$
the driving of disk turbulence is dominated by the incoming streams over
stellar feedback and instability-driven radial transport in the disk.
However, the preliminary estimates provide suppression of SFR by a factor
of only a few, which may not be enough for the complete quenching required.
Furthermore, the efficiency of this suppression in post-FFB galaxies
versus non-FFB galaxies is yet to be verified. 
One should also recall that the turbulence may in turn enhance the SFR by 
generating giant clumps due to excessive compressive modes of turbulence
\citep{inoue16,mandelker25,ginzburg25}.
More elaborate and accurate simulations are required in order to properly
investigate this potential SFR suppression mechanism.

\section{Post-FFB quenching}
\label{sec:q4_ffb}

\adb{Here we summarize the main quenching mechanisms that preferentially
operate in post-FFB galaxies (\se{q44_ffb}), 
and comment on the evolution of non-FFB galaxies to the peak of galaxy 
formation at cosmic noon (\se{noon}).}

\subsection{FFB contribution to quenching}
\label{sec:q44_ffb}

Considering the quenching mechanisms discussed above, one wishes to identify
the ways by which the FFB phase promotes the strong post-FFB quenching, 
in comparison to the supposedly weaker quenching in galaxies that did not 
undergo an FFB phase.
In general, given that the FFB evolutionary tracks consist of halos that
are more massive than the halos on the non-FFB tracks, 
they cross the quenching thresholds earlier, and at any given
time they find themselves higher above the thresholds. 
\adb{This} makes them obey the quenching conditions in a stronger way.

\smallskip 
A key argument for post-FFB quenching is that compaction events are more 
likely in high-sigma peaks.
Simulations \citep{dubois12,codis12} show that the rare massive halos at high 
redshifts are largely fed by low-AM streams into compact bulges, with specific 
AM an order of magnitude smaller than that of the baryons in the halo.
These streams penetrate deep inside the halo and connect to the bulge from
multiple rapidly changing directions. This causes enhanced cancellation of 
angular momentum in rarer halos, fed by more isotropically distributed streams.
A complementary effect that enhances compaction events in  
the FFB halos emerges from the fact the high-sigma peaks tend to be clustered
\citep{bbks86}.
This induces a higher than average rate of galaxy collisions, mergers,
and the associated compaction events, which cause quenching in multiple
ways as described above.
However, this effect is expected to be rather weak.

\smallskip 
A highly distinctive feature of post-FFB galaxies is the boost in BH growth
by rapid core collapse in the young, rotating star clusters and the subsequent 
inspiral in galactic disks to initial SMBH growth by mergers \citep{dekel25}. 
The massive BH formed by mergers 
can eventually lead to enhanced accretion and AGN activity that may provide 
the crucial quenching mechanism, including the driving of CGM turbulence which 
may stop the gas supply by streams. 

\smallskip 
Another source for the abrupt onset of quenching is the gas depletion from the 
inter-stellar medium (ISM) during the FFB phase. 
Then, in the absence of such local gas reservoir, quick
quenching can be achieved solely by stopping the external gas supply.

\subsection{Peak of galaxy formation at cosmic noon}
\label{sec:noon}

Along the non-FFB evolutionary track,
the highly abundant $2\sigma$-peak halos reach the upper region of the
golden mass, $\sim\!10^{12}\msun$, at cosmic noon.
In such halos the CGM is hot, fed both by feedback-driven outflows from the 
galaxy and virial-shock-heated inflow inside the halo virial radius. 
This allows the cold 
streams to be intensified by entrainment of the hot gas, thus utilizing 
recycled gas for increased SFR beyond the accretion rate of baryons onto the 
halo \citep{aung24}.  

\smallskip 
As lower-sigma peaks than halos on the FFB track, the non-FFB halos are fed by
streams with higher AM, which translates to fewer compaction events and 
less of the associated quenching that they induce.
These halos are also less clustered and tend to suffer less galaxy 
interactions and mergers, thus producing less compaction events.

\begin{figure*} 
\centering 
\includegraphics[width=0.48\textwidth]
{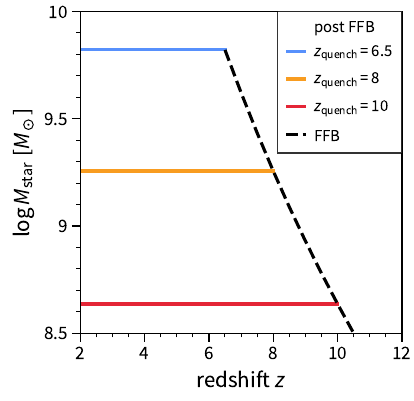}  
\includegraphics[width=0.48\textwidth]
{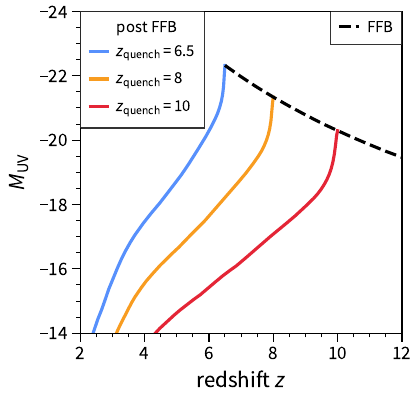} 
\vspace{-5pt}
\caption{
The predicted 
stellar mass and UV magnitude of galaxies on the fiducial FFB evolutionary
track as a function of redshift of observation.  
The onset of quenching is assumed to be at $z \seq 6.5, 8, 10$
(\adb{top, middle and bottom solid curves}), 
with the FFB and quiescent phases marked by dashed 
and solid lines respectively.   
The growth in time at high $z$ reflects the star formation in the FFB phase.
The decline of brightness in time after the onset of quenching is due to
passive evolution at the post-FFB quiescent phase.
}
\vspace{-10pt}
\label{fig:Muv}
\end{figure*}

\begin{figure} 
\centering 
\includegraphics[width=0.48\textwidth,trim={0.0cm 0.2cm 0.0cm 0.0cm},clip]
{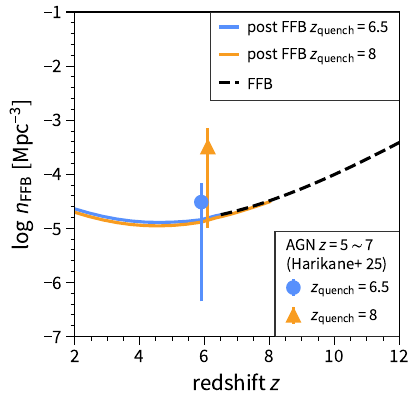}  
\vspace{-5pt}
\caption{
The predicted comoving number density of galaxies on the FFB and post-FFB 
evolutionary track, including the galaxies brighter than the fiducial 
FFB galaxies, as a function of redshift at cosmic dawn and cosmic morning.
The onset of quenching is assumed to be at $z_{\rm quench} \seq 8$ 
(orange, \adb{lower curve}) 
and $z_{\rm quench} \seq 6.5$ (blue, \adb{top curve}), 
with the FFB and quiescent phases 
marked by dashed and solid lines respectively.
We learn that the predicted number density of post-FFB quiescent galaxies at 
$z \ssim 6$ is $n \sgsim 10^{-5} \Mpc^{-3}$.
Shown for comparison \adb{(symbols with error bars)}
are number densities of AGN observed by JWST 
\citep{harikane23_agn} at $z \seq 5 \sdash 7$ in quiescent galaxies brighter 
than the corresponding UV magnitude as predicted in \fig{Muv}.
The abundance of BHs is in the ball park of that predicted for the quiescent 
galaxies when the onset of quenching is at $z \ssim 6.5$, and is higher for
higher $z_{\rm quench}$.
(See also \fig{hmf}.)
}
\vspace{-10pt}
\label{fig:n_post}
\end{figure}

\section{Observable predictions}
\label{sec:obs}

\adb{
Observable predictions concerning the FFB galaxies were presented in
\citep{li24} following \citet{dekel23}.
Here, we briefly address a morphology bimodality at cosmic dawn (\se{dawn}), 
and focus on predicted observable features for the
massive quiescent galaxies at cosmic morning,
as a guide for observations with JWST (\se{morning}).
}

\subsection{Bimodality in morphology at cosmic dawn}
\label{sec:dawn}

There are observational indications for a wide range of galaxy sizes among
the high-redshift excessively bright galaxies, especially at $z \sgeq 10$.  
\citet{harikane25} refer to this as a bimodality in morphology,
between compact galaxies on one hand and extended, clumpy galaxies on the other
hand. The actual division to two distinct types is yet to be confirmed as it is
based on a small number of galaxies, but there is clearly large variety
in size and clumpiness. \citet{ono25}, who investigating 169 galaxies at 
$z \ssim 10\sdash 16$, find a log-normal distribution of effective radii
with a mean of $\Re \ssim 133\pc$ and standard deviation 
$\sigma_{{\rm ln}{\Re}} \ssim 0.015$, the latter being comparable to disks at 
low redshifts. They identify only two extremely compact galaxies with 
$\Re \ssim 30\sdash 60\pc$.

\smallskip 
Our ongoing zoom-in cosmological simulations 
\citep[][Yao et al., in preparation, Chen et al., in preparation]{andalman25} 
indicate that a given galaxy may evolve from a compact configuration at one time
to an extended configuration at another time, and vice versa 
\citep[see also simulations by][]{ono25}.
The extended configuration tends to be associated with mergers, which in these
simulations are mostly minor mergers between galaxies that are dominated by gas.
The extended configuration could be due to capturing the incoming merging 
galaxies within the primary halo prior to the first pericenter passage 
or later near turnaround at the following apocenter prior to the final 
coalescence.
Alternatively it could be due to post-merger expansion as a result of
energy gain during the merger, including outflows and star formation in the
outskirts. This is being investigated elsewhere (Yao et al., in preparation).

\smallskip 
One may alternatively envision that the range of morphologies of the star 
forming galaxies at cosmic dawn is associated with the
bimodality discussed in the current paper between galaxies on the FFB and 
the non-FFB evolutionary tracks (\se{tracks}). 
The FFB galaxies that are associated with high-sigma peaks tend to be fed by 
streams of lower angular momentum \citep{dubois12} and thus lead to more 
compact galaxies, while the lower-sigma, non-FFB galaxies tend to be fed by 
streams of higher angular momentum that naturally result in more extended 
galaxies.

\smallskip
\adb{Furthermore,} 
the high-sigma FFB galaxies, which are expected to suffer more 
frequent and stronger compaction events, are likely to go through 
blue-nugget phases where they are of high SFR and compact.
On the other hand, the post-compaction galaxies tend to  
develop extended disks/rings from freshly accreted streams, in
which the inward radial transport is suppressed by the massive bulge
\citep{martig09,dekel20_ring}. Such galaxies may show a quiescent core
and an extended envelope forming stars at a mild rate. 
The origin of this morphological variability or bimodality and its possible
association with the bimodality between FFB and non-FFB galaxies is yet to be 
properly investigated.

\begin{figure} 
\centering 
\includegraphics[width=0.49\textwidth,trim={0.1cm 0.2cm 0.0cm 0.0cm},clip]
{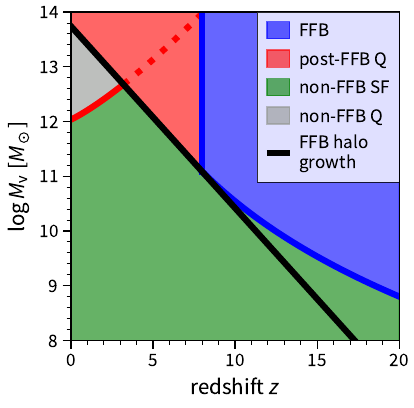}  
\vspace{-5pt}
\caption{
The division of galaxies into different zones
in the plane of halo mass and redshift, 
as the basis for the UVLFs shown in \fig{UVLF}.
The blue \adb{rightmost} curves mark the threshold for FFB.
Quenching is assumed in post-FFB galaxies at $z \slt z_{\rm quench} \seq 8$.
The black \adb{diagonal} line represents the average mass growth of the 
fiducial FFB halo, namely
the lowest-mass halo that evolves through an FFB phase (\fig{post}).  
At $z \sgt \seq 8$, all the galaxies are star forming, 
either FFB galaxies above the FFB threshold (blue shade, \adb{top-right})
or non-FFB galaxies below the FFB threshold (green shade, \adb{bottom-left}).
At $z \slt \seq 8$, the post-FFB galaxies, above the critical
mass growth curve for FFB (red shade, \adb{top left}), are quenched,
while the non-FFB galaxies are star forming (green shade, \adb{bottom-left}).
Quenching is assumed by the UM model in high halo masses   
(above the red curve, \adb{top-left}) independent of FFB, 
but the addition to the
post-FFB quenching (red shaded area, \adb{top-left}) 
is irrelevant in the redshift range of interest at cosmic morning.
}
\vspace{-10pt}
\label{fig:zones}
\end{figure}

\begin{figure*} 
\centering 
\includegraphics[width=0.99\textwidth,trim={0.2cm 0.2cm -0.1cm 0.0cm},clip]
{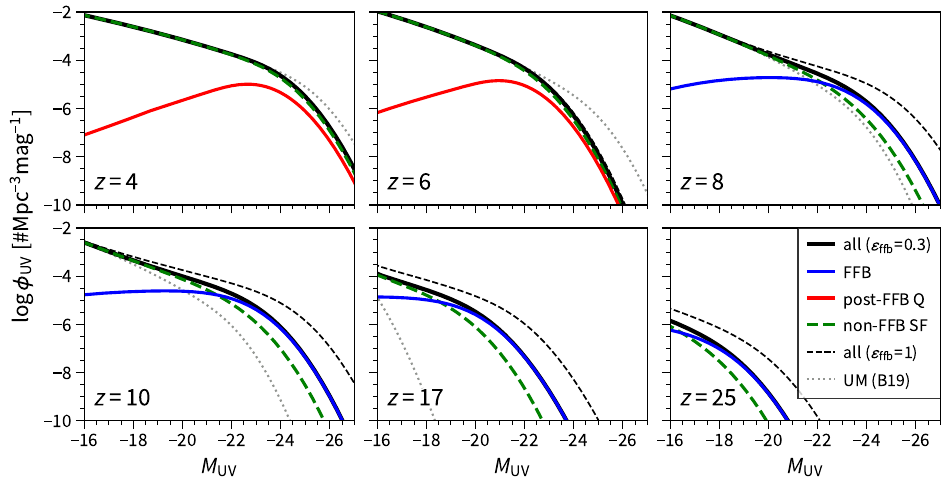}  
\vspace{-5pt}
\caption{
The predicted luminosity functions at different redshifts, at cosmic morning
($z \seq 4,6$) and at cosmic dawn ($z \seq 8,10,17,25$).
The total LF with $\epsilon_{\rm ffb}\seq 0.3$ (black \adb{thick line}) 
is the sum of the LF
for non-FFB galaxies (\adb{dashed} green) and the LF for FFB or post-FFB 
galaxies (\adb{solid} blue and solid red respectively).
Here $\zquench \seq 8$.
Shown for comparison are the UM predictions (\adb{dotted} grey), which
underestimate the LF at cosmic dawn.
At $z \sgeq \zquench$, all the galaxies are star forming and the bright end
is dominated by the FFB galaxies, above a threshold that declines from  
$\Muv \ssimeq -22$ at $z \seq 8$ to $\Muv \ssimeq -18.5$ at $z \seq 17$.
The signature of the bimodality at cosmic dawn is not explicit; it is
the bright-end excess of the LF due to the FFB galaxies.  
The total LF for $\epsilon_{\rm ffb} \seq 1$ is also shown at cosmic dawn
(thin dashed black \adb{uppermost curve}); 
it may allow a better fit to the observational estimates at 
$z \seq 17$ and $25$.
Central to this figure is the prediction that
at $z \slt \zquench$, the galaxies are divided to non-FFB star-forming galaxies
(\adb{dashed} green) and post-FFB quenched galaxies (solid red),  
with the star forming galaxies dominating the LF at all magnitudes.
The LFs of the post-FFB quiescent galaxies at cosmic morning 
peak near $\Muv \simeq -22.7$ and $\Muv \simeq -21$  
for $z \seq 4$ and $6$ respectively, where they contribute $\sim\!10\%$ of the
total LF (with $\epsilon_{\rm ffb} \seq 0.3$).  
}
\vspace{-10pt}
\label{fig:UVLF}
\end{figure*}


\subsection{Massive quiescent galaxies at cosmic morning}
\label{sec:morning}

\subsubsection{Galaxy properties and number densities}

The dark matter fraction within the stellar effective radius can be an
indicator of compaction events, which can be characteristic of galaxies on the
FFB track. Pre-compaction, one expect that the dark matter dominates the 
central regions,
but the compaction is expected to drive baryons into the center and generate
baryon-dominated centers that soon become dominated by stars \citep{lapiner23}. 
This has been seen both in JWST observations and in 
the TNG50 simulations, with the transition typically at cosmic morning, 
$z \ssim 6$ \citep{degraaff24_compaction}.

\smallskip 
The sizes of the post-FFB massive quiescent galaxies at cosmic morning
are expected to be compact in most cases. 
Forming in high-sigma-peak halos by compaction of instreaming gas of low 
angular momentum, the star forming FFB galaxies are expected to be compact, 
of sizes $\lsim\! 300\pc$ \citep[][Fig.~10]{li24}.
The stellar systems formed may be expected to remain compact in the 
subsequent quiescent phase when not many new stars are born.
We note however that those galaxies that eventually develop extended disks 
may show somewhat larger effective radii.
The current observations using JWST indicate that  
the quiescent galaxies at cosmic morning, $z \sgt 3$, 
indeed tend to be very compact \citep{ji24, degraaff24, carnall24, weibel25}.

\smallskip 
\Fig{Muv} shows the predicted stellar mass and UV magnitude of galaxies on the 
fiducial FFB evolutionary track as a function of redshift of observation.  
The onset of quenching is assumed to be at $z_{\rm quench} \seq 6.5, 8$ or $10$.
The growth in time at high $z$ prior to the onset of quenching reflects the 
star formation in the FFB phase.
The stellar mass is assumed to remain constant after $z_{\rm quench}$,
and the associated decline of brightness in time is due to
passive evolution at the post-FFB quiescent phase.

\smallskip 
\adb{The quiescent galaxies are expected to show post-starburst signatures,
depending on the end of the FFB phase, $\zquench$, and the redshift of 
observation.
For example, if the FFB ends at $z \seq 8$ and $\zquench \seq 8$, 
the stellar ages from the end of the burst are expected to be $\sim\! 300\Myr$ 
at $z \seq 6$, but as large as $\sim\!900\Myr$ by $z \seq 4$.
If $\zquench \seq 6.5$ and the sturbursts continue till that time, the stellar 
ages are expected to be $\sim\! 100\Myr$ at $z \seq 6$ and $\sim\!700 \Myr$ 
at $z \seq 4$.}

\smallskip 
\Fig{n_post} shows the predicted number density of galaxies on the FFB 
(and post-FFB) evolutionary track, those that are brighter than the fiducial 
FFB galaxies, as a function of redshift.
The onset of quenching is assumed to be either at $z_{\rm quench} \seq 8$ 
or at $z_{\rm quench} \seq 6.5$.
We learn that the predicted number density of post-FFB quiescent galaxies at
$z \ssim 6$ is $n \sgsim 10^{-5} \Mpc^{-3}$.

\smallskip 
Also shown for comparison in \fig{n_post} are number densities of AGN 
as observed by JWST \citep{harikane23_agn} at  $z \seq 5 \sdash 7$.
They are observed in quiescent galaxies brighter than the corresponding 
predicted UV magnitude shown in \fig{Muv} for galaxies on the fiducial 
FFB/non-FFB evolutionary track.
We learn that the
abundance of BHs is in the ball park of that predicted for the quiescent
galaxies when the onset of quenching is at $z \ssim 6.5$, and is higher for
higher $z_{\rm quench}$. 

\subsubsection{UVLF for quiescent and star-forming galaxies}

In order to compute the stellar mass function (SMF) and the UV luminosity 
function (UVLF) of the different galaxy
types, \fig{zones} illustrates the division of the plane of halo mass versus
redshift into three main zones: the star-forming FFB (ffb), the star-forming
non-FFB (non), and the quiescent post-FFB (post).\footnote{A small fourth 
zone of non-FFB quenched galaxies in the top-left corner of \fig{zones} is 
not relevant in the redshift range considered here.}
Quenching is assumed in post-FFB galaxies at $z \slt z_{\rm quench} \seq 8$.
At $z \sgt \zquench$, both the FFB and non-FFB galaxies are assumed to be 
star forming in their own manners, above and below the FFB threshold, 
respectively.
At  $z \slt \zquench$, the post-FFB galaxies, above the fiducial 
mass growth curve for FFB, are assumed to be quenched,
while the non-FFB galaxies are still star forming.
We allow smooth transition regions between the zones, as explained below.

\smallskip 
In order to compute the stellar mass and UV magnitude of a galaxy at a point 
$(\Mv,z)$, we derive the SFH along the halo growth trajectory leading to 
this point. For this we need the SFR at all points $(\Mv,z)$ along this 
trajectory, according to the zone they are in.
The SFR in the main body of the FFB zone, SFR$_{\rm ffb}$, is characterized 
by a global instantaneous star-formation efficiency (SFE) with respect to the 
baryonic accretion rate, $\epsilon_{\rm ffb}$, namely 
\be
{\rm SFR}_{\rm ffb} = \epsilon_{\rm ffb}\,\fb\, \Mvdot \, .
\ee
On average, as long as the mean accretion rate and SFR do not vary too strongly,
this instantaneous SFE is comparable to the global integrated SFE, namely
$\epsilon_{\rm int} \seq \Ms/(\fb \Mv)$, 
which is commonly deduced from observations.
The fiducial value is $\epsilon_{\rm ffb} \seq 0.3$, based on best fit to
JWST observations at $z \ssim 10$ \citep{li24}, but in principle it can range 
from $0.2$ to unity and vary with redshift.
The SFR in the main body of the post-FFB zone, SFR$_{\rm post}$, is assumed 
to vanish.
The SFR in the main body of the non-FFB zone, SFR$_{\rm non}$,
is adopted from the 
UniverseMachine (UM) model \citep{behroozi19} as outlined in \citet{li24}.

\smallskip 
In order to allow smooth transitions between the zones, we consider
a mixture of the components there, where the fraction of each, $f_{\rm type}$ 
is varied from unity in the main body of the associated zone to zero in the 
main body of the neighboring zone.
In the transition from FFB to non-FFB we use $f_{\rm ffb}$ from
\citet{li24}. 
In the transition from FFB to post-FFB we similarly take $f_{\rm post}$
to be a sigmoid function centered at $\zquench$ with a transition width 
$\Delta z \seq 0.2$ (full width $4\,\Delta z$).  
The SFR of the mixture at $(\Mv,z)$ is then defined by
\be
{\rm SFR} \seq f_{\rm ffb} [(1-f_{\rm post}) {\rm SFR}_{\rm ffb} 
                         + f_{\rm post} {\rm SFR}_{\rm post} ]
           +(1-f_{\rm ffb}) {\rm SFR}_{\rm non} \, .
\ee 

\smallskip 
For any given point $(\Mv,z)$ that contributes to the SMF and UVLF at $z$, 
the SFH is derived along the average halo growth
evolution track leading to this point, using \equ{Mvz}, 
making use of the SFR in the different zones along the track. 
The median stellar mass $\Ms$ is obtained by integrating the SFH over time.
The median rest-frame UV magnitude $\Muv$ at 1500 $\AA$ is calculated from 
the SFH using the Flexible Stellar Population Synthesis (FSPS) code 
\citep{conroy09},  which naturally incorporates the fact that the stars in 
younger galaxies were brighter.
We assume a Chabrier IMF \citep{chabrier03} and a metallicity of $0.1\Zsun$.  

\smallskip
\adb{It} is worth noting that
the magnitudes used here are an improvement over the magnitudes derived in 
\citet{li24} from the stellar masses using a constant $\Muv \sdash \Ms$ 
relation.  The SFH-based magnitudes are similar to the $\Ms$-based magnitudes
at $z \slt 10$, but they are higher at $z \sgeq 10$, by about one magnitude
at $z \ssim 15$ and two magnitudes at $z \ssim 25$.
This helps bringing the FFB-predicted LFs to better agreement with observed LFs
at $z\ssim 17$ and $z \ssim 25$ \citep{perez25}, though this is not the issue
in the current paper which focuses on the post-FFB quiescent galaxies at
cosmic morning.

\smallskip 
Following \citet{li24}, we introduce Gaussian-distribution scatter  
in the $\Ms \sdash \Mv$ and $\Muv \sdash \Mv$ relations about the
medians derived from the SFH, and obtain the conditional probabilities 
$p(\log\Ms|\Mv)$ and $p(\Muv|\Mv)$. 

\smallskip 
The SMF and UVLF (marked $F$) at $z$ are derived by a convolution of 
$p(F|\Mv)$
with the halo mass function $n(\Mv)$,
\be
\phi (F) = \int p(F | \Mv)\, n(\Mv)\, {\dd} \Mv \, .
\ee
This can be computed either for all the galaxies at $z$, or separately for 
the two galaxy types (FFB and non-FFB at $z>\zquench$ or post-FFB and non-FFB
at $z < \zquench$).
In the transition zones, the integrand includes in addition the corresponding 
fraction of the number of the galaxy type of interest, 
which can be smaller than unity.

\smallskip 
\Fig{UVLF} shows the predicted UV luminosity function (LF) at different
redshifts during cosmic dawn and cosmic morning. 
The total LF is the sum of the LFs for galaxies on the two evolutionary tracks,
namely, the non-FFB galaxies and the FFB/post-FFB galaxies.
The onset of post-FFB quenching is assumed here to be at $\zquench \seq 8$
and the SFE is at the fiducial value $\epsilon_{\rm ffb} \seq 0.3$.

\smallskip 
At $z \sgeq \zquench \seq 8$, all the galaxies of the two types are star 
forming and the bright end of the LF is dominated by the bursting FFB galaxies, 
above a threshold that declines from $\Muv \ssimeq -22$ at $z \seq 8$ 
to $\Muv \ssimeq -18.5$ at $z \seq 17$.
One can see that the signature of the bimodality at cosmic dawn is not 
pronounced; it is reflected in the bright-end excess of the LF due to the FFB 
galaxies.
Shown for comparison are the extrapolated UM predictions, 
which underestimate the LF at cosmic dawn.
The non-FFB LF is higher than the UM predictions due to inclusion of FFB
galaxies in the smooth transition region between the FFB and non-FFB zones.
The total LF for a higher SFE, $\epsilon_{\rm ffb} \seq 1$, is also shown at 
cosmic dawn.
Being higher than the LF for $\epsilon_{\rm ffb} \seq 0.3$, 
it may allow a better fit to the observational estimates at
$z \seq 17$ and $25$ \citep{perez25}.

\smallskip 
At $z \slt \zquench$, the galaxies are divided to non-FFB star-forming 
galaxies and post-FFB quenched galaxies. 
In this redshift range, the star-forming galaxies dominate the 
LF at all magnitudes.
The LFs of quiescent galaxies peak at $\Muv \simeq -22.7$ and 
$\Muv \simeq -21$ for $z \seq 4$ and $6$, respectively, 
and they are $\sim\! 10\%$ of the total LF at the peak in this redshift range.
The predicted LF of the quiescent galaxies from \fig{UVLF} is to be compared 
to the LF of the quiescent, red, compact galaxies as observed at cosmic 
morning (\se{quiescent}, \fig{sfh}).  

\smallskip 
Compared to the pre-JWST UM predictions, our model with 
$\epsilon_{\rm ffb} \seq 0.3$ produces a higher number density of bright 
galaxies at $z \sgt \zquench$ due to the enhanced SFE in the FFB zone and the
transition region into the non-FFB zone. 
At lower redshifts, $z \slt \zquench$, the introduction of post-FFB quenching 
suppresses the bright end of the UVLF relative to the continuous star formation in the standard UM model.

\smallskip 
The SMFs, which are not shown here, 
are qualitatively similar to the UVLFs.
The main difference is that at $z \slsim \zquench$ the post-FFB quiescent 
galaxies become the dominant component in the SMF, e.g., for 
$\Ms \sgt 10^{10.3}\msun$ at $z\seq 6$, because they retain the high
stellar masses while their luminosities begin to fade by quenching.

\section{Conclusion}
\label{sec:conc}

We argue (largely based on \fig{post})
that the early evolution of galaxies at high redshifts is bimodal
as a function of halo mass, or equivalently the associated  
height of the peaks in the density fluctuation field.
The evolutionary track of massive halos goes through an FFB phase.
It connects the observed excess of bright galaxies at 
cosmic dawn ($z \sgt 8$) to the observed excess of massive quiescent galaxies 
and the presence of supermassive black holes with high BH-to-stellar mass ratio
at cosmic morning ($z \ssim 4 \sdash 7$). 
On the other evolutionary track, consisting of lower-mass, more abundant halos,
the galaxies avoid the FFB phase.
Their stellar growth is expected to be gradual,
eventually leading to the peak of star-forming galaxies at cosmic noon 
($z \ssim 1\sdash 3$).

\smallskip
As seen in \fig{post}, the massive track consists of
dark-matter halos that are more massive than $ \sim \! 10^{10.5}\msun$ 
at $z\seq 10$ and represent $\geq\!4\sigma$ peaks in the density fluctuation 
field. These halos undergo a phase of feedback-free starbursts (FFB) with high
star-formation efficiency in thousands of dense star clusters within each
compact galaxy \citep{dekel23,li24}. 
The less massive, $2\sdash3\sigma$ peak halos 
avoid the FFB phase and form stars gradually through
cosmic dawn and cosmic morning subject to stellar feedback.

\smallskip
The key question addressed is the origin of efficient quenching of star 
formation preferably in post FFB galaxies.  
We propose that the post-FFB galaxies quench 
soon after the FFB phase and remain quiescent due to the combined effect of 
several mechanisms:

\rff
\no\bul
Gas depletion from the ISM by the FFB starbursts and the associated
outflows \citep{li24}
suppresses the available local gas supply immediately available
for further star formation.

\rff
\no\bul
The high-sigma-peak FFB halos suffer excessive compaction events 
driven by angular-momentum loss in colliding streams and enhanced frequency
of mergers \citep{dubois12}.
The compaction events drive quenching by several different ways,
including (a) further gas consumption, 
(b) formation of extended, non-contracting disk and ring, 
(c) compaction-driven black-hole growth that boosts AGN feedback, 
and (d) driving stream-disrupting turbulence in the CGM.

\rff
\no\bul
The disruption of the inflowing cold streams is a key for the required complete
quenching at cosmic morning, and it poses a non-trivial open challenge.
We assess that possible mechanisms for stream disruption are supersonic 
turbulence in the CGM (possibly generated by AGN feedback) 
and photo-heating of the stream-CGM mixing layer by AGN radiation
\adb{(as summarized in \fig{Rturb} and \fig{heating_rcrit})}. 

\rff
\no\bul
Simulations show that AGN feedback is necessary for complete quenching
\adb{(as illustrated in \fig{quenching})}. 
This could be, e.g., by generating CGM turbulence or indirect photo-heating.
The FFB phase is indeed expected to lead to rapid growth of BH seeds and 
subsequent growth to supermassive BHs that allow enhanced AGN feedback 
\citep{dekel25}.

\smallskip
The bimodality to FFB and non-FFB evolutionary tracks,
and the post-FFB quenching, are yet to be investigated in proper cosmological
simulations that resolve the FFB processes at cosmic dawn with $\sim\! 1\pc$
resolution and continue to cosmic morning while incorporating black hole 
growth and AGN feedback.

\smallskip
Based on the above discussion of post-FFB evolution, one can deduce
observable predictions for the quiescent galaxies at cosmic morning.
They are expected to be massive, compact, dominated by stars, and possibly 
showing signatures of outflows and AGN.
Their comoving number density is expected to be 
$\gsim\!  10^{-5}\Mpc^{-3}$, comparable to the density of super-bright FFB
galaxies at cosmic dawn, and to the density of black holes at cosmic morning.
%
The UV luminosity function of post-FFB quiescent galaxies at cosmic morning 
is expected to peak near $\Muv \ssim -22.7$ and $-21$
at $z \ssim 4$ and $6$, respectively,
and to contribute about $10\%$ to the total LF at these peak 
magnitudes, respectively 
\adb{(shown in \fig{UVLF})}.

\smallskip
\adb{
One should realize that the near-complete quenching of galaxies at cosmic 
morning is a non-trivial open issue. We have addressed what might be the 
relevant mechanisms, but admittedly our understanding of these mechanisms
is still rather limited. The estimates presented here should be taken as 
preliminary and serve as motivation for more detailed studies, including 
simulations. 
}

\section*{Acknowledgments}
We are grateful for stimulating discussions with Anna de Graaff, Yohan Dubois,
David Elbaz, Yuchi Harikane, Joel Primack, Volker Springel, Sandro Tacchella 
and Andrea Weibel.  Yuchi Harikane provided the number densities for BHs
shown in \fig{hmf} and \fig{n_post}, given the post-FFB UV magnitudes 
predicted.
This work was supported by the Israel Science Foundation Grant ISF 861/20,
and by the US National Science Foundation - US-Israel Bi-national Science 
Foundation grants 2023730 (with Romain Teyssier) and 2023723 (with Frank van
den Bosch).
NM acknowledges support from the Israel Science Foundation
grant ISF 3061/21.
ZL acknowledges funding from the European Union's Horizon 2020 research and 
innovation programme under the Marie Sklodowska-Curie grant agreement No 
101109759 (``CuspCore'').

\section*{DATA AVAILABILITY}

Data and results underlying
this article will be shared on reasonable request to the corresponding author.

\bibliographystyle{mnras} 
\bibliography{post_accepted}

\begin{thebibliography}{}
\makeatletter
\relax
\def\mn@urlcharsother{\let\do\@makeother \do\$\do\&\do\#\do\^\do\_\do\%\do\~}
\def\mn@doi{\begingroup\mn@urlcharsother \@ifnextchar [ {\mn@doi@}
  {\mn@doi@[]}}
\def\mn@doi@[#1]#2{\def\@tempa{#1}\ifx\@tempa\@empty \href
  {http://dx.doi.org/#2} {doi:#2}\else \href {http://dx.doi.org/#2} {#1}\fi
  \endgroup}
\def\mn@eprint#1#2{\mn@eprint@#1:#2::\@nil}
\def\mn@eprint@arXiv#1{\href {http://arxiv.org/abs/#1} {{\tt arXiv:#1}}}
\def\mn@eprint@dblp#1{\href {http://dblp.uni-trier.de/rec/bibtex/#1.xml}
  {dblp:#1}}
\def\mn@eprint@#1:#2:#3:#4\@nil{\def\@tempa {#1}\def\@tempb {#2}\def\@tempc
  {#3}\ifx \@tempc \@empty \let \@tempc \@tempb \let \@tempb \@tempa \fi \ifx
  \@tempb \@empty \def\@tempb {arXiv}\fi \@ifundefined
  {mn@eprint@\@tempb}{\@tempb:\@tempc}{\expandafter \expandafter \csname
  mn@eprint@\@tempb\endcsname \expandafter{\@tempc}}}

\bibitem[\protect\citeauthoryear{{Adamo} et~al.,}{{Adamo}
  et~al.}{2024}]{adamo24}
{Adamo} A.,  et~al., 2024, \mn@doi [\nat] {10.1038/s41586-024-07703-7}, \href
  {https://ui.adsabs.harvard.edu/abs/2024Natur.632..513A} {632, 513}

\bibitem[\protect\citeauthoryear{{Adams} et~al.,}{{Adams}
  et~al.}{2023}]{adams23}
{Adams} N.~J.,  et~al., 2023, \mn@doi [\mnras] {10.1093/mnras/stac3347}, \href
  {https://ui.adsabs.harvard.edu/abs/2023MNRAS.518.4755A} {518, 4755}

\bibitem[\protect\citeauthoryear{{Andalman}, {Teyssier}  \& {Dekel}}{{Andalman}
  et~al.}{2025}]{andalman25}
{Andalman} Z.~L.,  {Teyssier} R.,   {Dekel} A.,  2025, \mn@doi [\mnras]
  {10.1093/mnras/staf930}, \href
  {https://ui.adsabs.harvard.edu/abs/2025MNRAS.540.3350A} {540, 3350}

\bibitem[\protect\citeauthoryear{{Antwi-Danso} et~al.,}{{Antwi-Danso}
  et~al.}{2025}]{antwidanso25}
{Antwi-Danso} J.,  et~al., 2025, \mn@doi [\apj] {10.3847/1538-4357/ad8b30},
  \href {https://ui.adsabs.harvard.edu/abs/2025ApJ...978...90A} {978, 90}

\bibitem[\protect\citeauthoryear{{Armillotta}, {Fraternali}  \&
  {Marinacci}}{{Armillotta} et~al.}{2016}]{armillotta16}
{Armillotta} L.,  {Fraternali} F.,   {Marinacci} F.,  2016, \mn@doi [\mnras]
  {10.1093/mnras/stw1930}, \href
  {https://ui.adsabs.harvard.edu/abs/2016MNRAS.462.4157A} {462, 4157}

\bibitem[\protect\citeauthoryear{{Armillotta}, {Fraternali}, {Werk},
  {Prochaska}  \& {Marinacci}}{{Armillotta} et~al.}{2017}]{armillotta17}
{Armillotta} L.,  {Fraternali} F.,  {Werk} J.~K.,  {Prochaska} J.~X.,
  {Marinacci} F.,  2017, \mn@doi [\mnras] {10.1093/mnras/stx1239}, \href
  {https://ui.adsabs.harvard.edu/abs/2017MNRAS.470..114A} {470, 114}

\bibitem[\protect\citeauthoryear{{Arrabal Haro}, {Dickinson}, {Finkelstein},
  {Fujimoto}, {Fern{\'a}ndez}, {Kartaltepe}, {Jung}  \& {et al.}}{{Arrabal
  Haro} et~al.}{2023}]{arrabal-haro23}
{Arrabal Haro} P.,  {Dickinson} M.,  {Finkelstein} S.~L.,  {Fujimoto} S.,
  {Fern{\'a}ndez} V.,  {Kartaltepe} J.~S.,  {Jung} I.,   {et al.} 2023, \mn@doi
  [\apjl] {10.3847/2041-8213/acdd54}, \href
  {https://ui.adsabs.harvard.edu/abs/2023ApJ...951L..22A} {951, L22}

\bibitem[\protect\citeauthoryear{{Aung}, {Mandelker}, {Nagai}, {Dekel}  \&
  {Birnboim}}{{Aung} et~al.}{2019}]{aung19}
{Aung} H.,  {Mandelker} N.,  {Nagai} D.,  {Dekel} A.,   {Birnboim} Y.,  2019,
  \mn@doi [\mnras] {10.1093/mnras/stz1964}, \href
  {https://ui.adsabs.harvard.edu/abs/2019MNRAS.490..181A} {490, 181}

\bibitem[\protect\citeauthoryear{{Aung}, {Mandelker}, {Dekel}, {Nagai},
  {Semenov}  \& {van den Bosch}}{{Aung} et~al.}{2024}]{aung24}
{Aung} H.,  {Mandelker} N.,  {Dekel} A.,  {Nagai} D.,  {Semenov} V.,   {van den
  Bosch} F.~C.,  2024, \mn@doi [\mnras] {10.1093/mnras/stae1673}, \href
  {https://ui.adsabs.harvard.edu/abs/2024MNRAS.532.2965A} {532, 2965}

\bibitem[\protect\citeauthoryear{{Banik} \& {van den Bosch}}{{Banik} \& {van
  den Bosch}}{2021}]{banik21}
{Banik} U.,  {van den Bosch} F.~C.,  2021, \mn@doi [\apj]
  {10.3847/1538-4357/abeb6d}, \href
  {https://ui.adsabs.harvard.edu/abs/2021ApJ...912...43B} {912, 43}

\bibitem[\protect\citeauthoryear{{Bardeen}, {Bond}, {Kaiser}  \&
  {Szalay}}{{Bardeen} et~al.}{1986}]{bbks86}
{Bardeen} J.~M.,  {Bond} J.~R.,  {Kaiser} N.,   {Szalay} A.~S.,  1986, \apj,
  304, 15

\bibitem[\protect\citeauthoryear{{Barro} et~al.,}{{Barro}
  et~al.}{2013}]{barro13}
{Barro} G.,  et~al., 2013, \apj, 765, 104

\bibitem[\protect\citeauthoryear{{Barro} et~al.,}{{Barro}
  et~al.}{2017}]{barro17}
{Barro} G.,  et~al., 2017, \mn@doi [\apj] {10.3847/1538-4357/aa6b05}, \href
  {https://ui.adsabs.harvard.edu/abs/2017ApJ...840...47B} {840, 47}

\bibitem[\protect\citeauthoryear{{Begelman}, {Blandford}  \& {Rees}}{{Begelman}
  et~al.}{1980}]{begelman80}
{Begelman} M.~C.,  {Blandford} R.~D.,   {Rees} M.~J.,  1980, \mn@doi [\nat]
  {10.1038/287307a0}, \href
  {https://ui.adsabs.harvard.edu/abs/1980Natur.287..307B} {287, 307}

\bibitem[\protect\citeauthoryear{{Behroozi}, {Wechsler}  \&
  {Conroy}}{{Behroozi} et~al.}{2013}]{behroozi13}
{Behroozi} P.~S.,  {Wechsler} R.~H.,   {Conroy} C.,  2013, \apjl, 762, L31

\bibitem[\protect\citeauthoryear{{Behroozi}, {Wechsler}, {Hearin}  \&
  {Conroy}}{{Behroozi} et~al.}{2019}]{behroozi19}
{Behroozi} P.,  {Wechsler} R.~H.,  {Hearin} A.~P.,   {Conroy} C.,  2019,
  \mn@doi [\mnras] {10.1093/mnras/stz1182}, \href
  {https://ui.adsabs.harvard.edu/abs/2019MNRAS.488.3143B} {488, 3143}

\bibitem[\protect\citeauthoryear{{Behroozi} et~al.,}{{Behroozi}
  et~al.}{2020}]{behroozi20}
{Behroozi} P.,  et~al., 2020, \mn@doi [\mnras] {10.1093/mnras/staa3164}, \href
  {https://ui.adsabs.harvard.edu/abs/2020MNRAS.499.5702B} {499, 5702}

\bibitem[\protect\citeauthoryear{{Belli} et~al.,}{{Belli}
  et~al.}{2024}]{belli24}
{Belli} S.,  et~al., 2024, \mn@doi [\nat] {10.1038/s41586-024-07412-1}, \href
  {https://ui.adsabs.harvard.edu/abs/2024Natur.630...54B} {630, 54}

\bibitem[\protect\citeauthoryear{{Benitez-Llambay} \&
  {Frenk}}{{Benitez-Llambay} \& {Frenk}}{2020}]{benitez20}
{Benitez-Llambay} A.,  {Frenk} C.,  2020, \mn@doi [\mnras]
  {10.1093/mnras/staa2698}, \href
  {https://ui.adsabs.harvard.edu/abs/2020MNRAS.498.4887B} {498, 4887}

\bibitem[\protect\citeauthoryear{{Bennett} \& {Sijacki}}{{Bennett} \&
  {Sijacki}}{2020}]{bennett20}
{Bennett} J.~S.,  {Sijacki} D.,  2020, \mn@doi [\mnras]
  {10.1093/mnras/staa2835}, \href
  {https://ui.adsabs.harvard.edu/abs/2020MNRAS.499..597B} {499, 597}

\bibitem[\protect\citeauthoryear{{Berlok} \& {Pfrommer}}{{Berlok} \&
  {Pfrommer}}{2019}]{berlok19}
{Berlok} T.,  {Pfrommer} C.,  2019, \mn@doi [\mnras] {10.1093/mnras/stz2347},
  \href {https://ui.adsabs.harvard.edu/abs/2019MNRAS.489.3368B} {489, 3368}

\bibitem[\protect\citeauthoryear{{Birnboim} \& {Dekel}}{{Birnboim} \&
  {Dekel}}{2003}]{bd03}
{Birnboim} Y.,  {Dekel} A.,  2003, \mnras, 345, 349

\bibitem[\protect\citeauthoryear{{Blank}, {Macci{\`o}}  \& {Dutton}}{{Blank}
  et~al.}{2019}]{blank19}
{Blank} M.,  {Macci{\`o}} A.~V.,   {Dutton} Aaron A.and~{Obreja} A.,  2019,
  \mn@doi [\mnras] {10.1093/mnras/stz1688}, \href
  {https://ui.adsabs.harvard.edu/abs/2019MNRAS.487.5476B} {487, 5476}

\bibitem[\protect\citeauthoryear{{Blecha} \& {Loeb}}{{Blecha} \&
  {Loeb}}{2008}]{blecha08}
{Blecha} L.,  {Loeb} A.,  2008, \mn@doi [\mnras]
  {10.1111/j.1365-2966.2008.13790.x}, \href
  {https://ui.adsabs.harvard.edu/abs/2008MNRAS.390.1311B} {390, 1311}

\bibitem[\protect\citeauthoryear{{Bond}, {Cole}, {Efstathiou}  \&
  {Kaiser}}{{Bond} et~al.}{1991}]{bond91}
{Bond} J.~R.,  {Cole} S.,  {Efstathiou} G.,   {Kaiser} N.,  1991, \mn@doi
  [\apj] {10.1086/170520}, \href
  {https://ui.adsabs.harvard.edu/abs/1991ApJ...379..440B} {379, 440}

\bibitem[\protect\citeauthoryear{{Bouwens}, {Illingworth}, {Oesch}, {
  Stefanon}, {Naidu}, {van Leeuwen}  \& {Magee}}{{Bouwens}
  et~al.}{2023}]{bouwens23}
{Bouwens} R.,  {Illingworth} G.,  {Oesch} P.,  { Stefanon} M.,  {Naidu} R.,
  {van Leeuwen} I.,   {Magee} D.,  2023, \mn@doi [\mnras]
  {10.1093/mnras/stad1014}, \href
  {https://ui.adsabs.harvard.edu/abs/2023MNRAS.523.1009B} {523, 1009}

\bibitem[\protect\citeauthoryear{{Boylan-Kolchin}}{{Boylan-Kolchin}}{2023}]{boylan23}
{Boylan-Kolchin} M.,  2023, \mn@doi [Nature Astronomy]
  {10.1038/s41550-023-01937-7}, \href
  {https://ui.adsabs.harvard.edu/abs/2023NatAs...7..731B} {7, 731}

\bibitem[\protect\citeauthoryear{{Carnall} et~al.,}{{Carnall}
  et~al.}{2023a}]{carnall23_z35}
{Carnall} A.~C.,  et~al., 2023a, \mn@doi [\mnras] {10.1093/mnras/stad369},
  \href {https://ui.adsabs.harvard.edu/abs/2023MNRAS.520.3974C} {520, 3974}

\bibitem[\protect\citeauthoryear{{Carnall} et~al.,}{{Carnall}
  et~al.}{2023b}]{carnall23}
{Carnall} A.~C.,  et~al., 2023b, \mn@doi [\nat] {10.1038/s41586-023-06158-6},
  \href {https://ui.adsabs.harvard.edu/abs/2023Natur.619..716C} {619, 716}

\bibitem[\protect\citeauthoryear{{Carnall} et~al.,}{{Carnall}
  et~al.}{2024}]{carnall24}
{Carnall} A.~C.,  et~al., 2024, \mn@doi [\mnras] {10.1093/mnras/stae2092},
  \href {https://ui.adsabs.harvard.edu/abs/2024MNRAS.534..325C} {534, 325}

\bibitem[\protect\citeauthoryear{{Ceverino} \& {Klypin}}{{Ceverino} \&
  {Klypin}}{2009}]{ceverino09}
{Ceverino} D.,  {Klypin} A.,  2009, \apj, 695, 292

\bibitem[\protect\citeauthoryear{{Ceverino}, {Klypin}, {Klimek},
  {Trujillo-Gomez}, {Churchill}, {Primack}  \& {Dekel}}{{Ceverino}
  et~al.}{2014}]{ceverino14}
{Ceverino} D.,  {Klypin} A.,  {Klimek} E.~S.,  {Trujillo-Gomez} S.,
  {Churchill} C.~W.,  {Primack} J.,   {Dekel} A.,  2014, \mnras, 442, 1545

\bibitem[\protect\citeauthoryear{{Chabrier}}{{Chabrier}}{2003}]{chabrier03}
{Chabrier} G.,  2003, \mn@doi [\pasp] {10.1086/376392}, \href
  {http://adsabs.harvard.edu/abs/2003PASP..115..763C} {115, 763}

\bibitem[\protect\citeauthoryear{{Claeyssens}, {Adamo}, {Richard}, {Mahler},
  {Messa}  \& {Dessauges-Zavadsky}}{{Claeyssens} et~al.}{2023}]{claeyssens23}
{Claeyssens} A.,  {Adamo} A.,  {Richard} J.,  {Mahler} G.,  {Messa} M.,
  {Dessauges-Zavadsky} M.,  2023, \mn@doi [\mnras] {10.1093/mnras/stac3791},
  \href {https://ui.adsabs.harvard.edu/abs/2023MNRAS.520.2180C} {520, 2180}

\bibitem[\protect\citeauthoryear{{Codis}, {Pichon}, {Devriendt}, {Slyz},
  {Pogosyan}, {Dubois}  \& {Sousbie}}{{Codis} et~al.}{2012}]{codis12}
{Codis} S.,  {Pichon} C.,  {Devriendt} J.,  {Slyz} A.,  {Pogosyan} D.,
  {Dubois} Y.,   {Sousbie} T.,  2012, \mnras, 427, 3320

\bibitem[\protect\citeauthoryear{{Codis}, {Pogosyan}  \& {Pichon}}{{Codis}
  et~al.}{2018}]{codis18}
{Codis} S.,  {Pogosyan} D.,   {Pichon} C.,  2018, \mn@doi [\mnras]
  {10.1093/mnras/sty1643}, \href
  {https://ui.adsabs.harvard.edu/abs/2018MNRAS.479..973C} {479, 973}

\bibitem[\protect\citeauthoryear{{Conroy}, {Gunn}  \& {White}}{{Conroy}
  et~al.}{2009}]{conroy09}
{Conroy} C.,  {Gunn} J.~E.,   {White} M.,  2009, \mn@doi [\apj]
  {10.1088/0004-637X/699/1/486}, \href
  {https://ui.adsabs.harvard.edu/abs/2009ApJ...699..486C} {699, 486}

\bibitem[\protect\citeauthoryear{{Costa}, {Sijacki}  \& {Haehnelt}}{{Costa}
  et~al.}{2014}]{costa14}
{Costa} T.,  {Sijacki} D.,   {Haehnelt} M.~G.,  2014, \mn@doi [\mnras]
  {10.1093/mnras/stu1632}, \href
  {https://ui.adsabs.harvard.edu/abs/2014MNRAS.444.2355C} {444, 2355}

\bibitem[\protect\citeauthoryear{{Costa}, {Arrigoni Battaia}, {Farina},
  {Keating}, {Rosdahl}  \& {Kimm}}{{Costa} et~al.}{2022}]{costa22}
{Costa} T.,  {Arrigoni Battaia} F.,  {Farina} E.~P.,  {Keating} L.~C.,
  {Rosdahl} J.,   {Kimm} T.,  2022, \mn@doi [\mnras] {10.1093/mnras/stac2432},
  \href {https://ui.adsabs.harvard.edu/abs/2022MNRAS.517.1767C} {517, 1767}

\bibitem[\protect\citeauthoryear{{D'Eugenio} et~al.,}{{D'Eugenio}
  et~al.}{2024}]{deugenio24}
{D'Eugenio} F.,  et~al., 2024, \mn@doi [Nature Astronomy]
  {10.1038/s41550-024-02345-1}, \href
  {https://ui.adsabs.harvard.edu/abs/2024NatAs...8.1443D} {8, 1443}

\bibitem[\protect\citeauthoryear{{Daddi} et~al.,}{{Daddi}
  et~al.}{2022a}]{daddi22a}
{Daddi} E.,  et~al., 2022a, \mn@doi [\aap] {10.1051/0004-6361/202243574}, \href
  {https://ui.adsabs.harvard.edu/abs/2022A&A...661L...7D} {661, L7}

\bibitem[\protect\citeauthoryear{{Daddi} et~al.,}{{Daddi}
  et~al.}{2022b}]{daddi22b}
{Daddi} E.,  et~al., 2022b, \mn@doi [\apjl] {10.3847/2041-8213/ac531f}, \href
  {https://ui.adsabs.harvard.edu/abs/2022ApJ...926L..21D} {926, L21}

\bibitem[\protect\citeauthoryear{{Dekel} \& {Birnboim}}{{Dekel} \&
  {Birnboim}}{2006}]{db06}
{Dekel} A.,  {Birnboim} Y.,  2006, \mnras, 368, 2

\bibitem[\protect\citeauthoryear{{Dekel} \& {Silk}}{{Dekel} \&
  {Silk}}{1986}]{ds86}
{Dekel} A.,  {Silk} J.,  1986, \apj, 303, 39

\bibitem[\protect\citeauthoryear{{Dekel} et~al.,}{{Dekel}
  et~al.}{2009}]{dekel09}
{Dekel} A.,  et~al., 2009, \nat, 457, 451

\bibitem[\protect\citeauthoryear{{Dekel}, {Zolotov}, {Tweed}, {Cacciato},
  {Ceverino}  \& {Primack}}{{Dekel} et~al.}{2013}]{dekel13}
{Dekel} A.,  {Zolotov} A.,  {Tweed} D.,  {Cacciato} M.,  {Ceverino} D.,
  {Primack} J.~R.,  2013, \mnras, 435, 999

\bibitem[\protect\citeauthoryear{{Dekel}, {Lapiner}  \& {Dubois}}{{Dekel}
  et~al.}{2019}]{dekel19_gold}
{Dekel} A.,  {Lapiner} S.,   {Dubois} Y.,  2019, \mn@doi [arXiv e-prints]
  {10.48550/arXiv.1904.08431}, \href
  {https://ui.adsabs.harvard.edu/abs/2019arXiv190408431D} {p. arXiv:1904.08431}

\bibitem[\protect\citeauthoryear{{Dekel}, {Ginzburg}, {Jiang}, {Freundlich},
  {Lapiner}, {Ceverino}  \& {Primack}}{{Dekel} et~al.}{2020a}]{dekel20_flip}
{Dekel} A.,  {Ginzburg} O.,  {Jiang} F.,  {Freundlich} J.,  {Lapiner} S.,
  {Ceverino} D.,   {Primack} J.,  2020a, \mn@doi [\mnras]
  {10.1093/mnras/staa470}, \href
  {https://ui.adsabs.harvard.edu/abs/2020MNRAS.493.4126D} {493, 4126}

\bibitem[\protect\citeauthoryear{{Dekel} et~al.,}{{Dekel}
  et~al.}{2020b}]{dekel20_ring}
{Dekel} A.,  et~al., 2020b, \mn@doi [\mnras] {10.1093/mnras/staa1713}, \href
  {https://ui.adsabs.harvard.edu/abs/2020MNRAS.496.5372D} {496, 5372}

\bibitem[\protect\citeauthoryear{{Dekel}, {Sarkar}, {Birnboim}, {Mandelker}  \&
  {Li}}{{Dekel} et~al.}{2023}]{dekel23}
{Dekel} A.,  {Sarkar} K.~C.,  {Birnboim} Y.,  {Mandelker} N.,   {Li} Z.,  2023,
  \mn@doi [\mnras] {10.1093/mnras/stad1557}, \href
  {https://ui.adsabs.harvard.edu/abs/2023MNRAS.523.3201D} {523, 3201}

\bibitem[\protect\citeauthoryear{{Dekel}, {Stone}, {Dutta Chowdhury},
  {Gilbaum}, {Li}, {Mandelker}  \& {van den Bosch}}{{Dekel}
  et~al.}{2025}]{dekel25}
{Dekel} A.,  {Stone} N.~C.,  {Dutta Chowdhury} D.,  {Gilbaum} S.,  {Li} Z.,
  {Mandelker} N.,   {van den Bosch} F.~C.,  2025, \aap, \href
  {https://ui.adsabs.harvard.edu/abs/2025A&A...695A..97D} {695, A97}

\bibitem[\protect\citeauthoryear{{Donnan} et~al.,}{{Donnan}
  et~al.}{2023a}]{donnan23b}
{Donnan} C.~T.,  et~al., 2023a, \mn@doi [\mnras] {10.1093/mnras/stac3472},
  \href {https://ui.adsabs.harvard.edu/abs/2023MNRAS.518.6011D} {518, 6011}

\bibitem[\protect\citeauthoryear{{Donnan}, {McLeod}, {McLure}, {Dunlop},
  {Carnall}, {Cullen}  \& {Magee}}{{Donnan} et~al.}{2023b}]{donnan23a}
{Donnan} C.~T.,  {McLeod} D.~J.,  {McLure} R.~J.,  {Dunlop} J.~S.,  {Carnall}
  A.~C.,  {Cullen} F.,   {Magee} D.,  2023b, \mn@doi [\mnras]
  {10.1093/mnras/stad471}, \href
  {https://ui.adsabs.harvard.edu/abs/2023MNRAS.520.4554D} {520, 4554}

\bibitem[\protect\citeauthoryear{{Dubois}, {Pichon}, {Haehnelt}, {Kimm},
  {Slyz}, {Devriendt}  \& {Pogosyan}}{{Dubois} et~al.}{2012}]{dubois12}
{Dubois} Y.,  {Pichon} C.,  {Haehnelt} M.,  {Kimm} T.,  {Slyz} A.,  {Devriendt}
  J.,   {Pogosyan} D.,  2012, \mn@doi [\mnras]
  {10.1111/j.1365-2966.2012.21160.x10.1002/asna.19141990903}, \href
  {https://ui.adsabs.harvard.edu/abs/2012MNRAS.423.3616D} {423, 3616}

\bibitem[\protect\citeauthoryear{{Dubois}, {Pichon}, {Devriendt}, {Silk},
  {Haehnelt}, {Kimm}  \& {Slyz}}{{Dubois} et~al.}{2013}]{dubois13}
{Dubois} Y.,  {Pichon} C.,  {Devriendt} J.,  {Silk} J.,  {Haehnelt} M.,  {Kimm}
  T.,   {Slyz} A.,  2013, \mn@doi [\mnras] {10.1093/mnras/sts224}, \href
  {https://ui.adsabs.harvard.edu/abs/2013MNRAS.428.2885D} {428, 2885}

\bibitem[\protect\citeauthoryear{{Dubois}, {Peirani}, {Pichon}, {Devriendt},
  {Gavazzi}  \& {Welker}}{{Dubois} et~al.}{2016}]{dubois16}
{Dubois} Y.,  {Peirani} S.,  {Pichon} h.,  {Devriendt} J.,  {Gavazzi} R.,
  {Welker} Chard~{Volonteri} M.,  2016, \mn@doi [\mnras]
  {10.1093/mnras/stw2265}, \href
  {https://ui.adsabs.harvard.edu/abs/2016MNRAS.463.3948D} {463, 3948}

\bibitem[\protect\citeauthoryear{{Dubois} et~al.,}{{Dubois}
  et~al.}{2021}]{dubois21}
{Dubois} Y.,  et~al., 2021, \mn@doi [\aap] {10.1051/0004-6361/202039429}, \href
  {https://ui.adsabs.harvard.edu/abs/2021A&A...651A.109D} {651, A109}

\bibitem[\protect\citeauthoryear{{Dutta Chowdhury}, {Dekel}, {Mandelker},
  {Ginzburg}  \& {Genzel}}{{Dutta Chowdhury} et~al.}{2024}]{dutta25}
{Dutta Chowdhury} D.,  {Dekel} A.,  {Mandelker} N.,  {Ginzburg} O.,   {Genzel}
  R.,  2024, \mn@doi [arXiv e-prints] {10.48550/arXiv.2409.01589}, \href
  {https://ui.adsabs.harvard.edu/abs/2024arXiv240901589D} {p. arXiv:2409.01589}

\bibitem[\protect\citeauthoryear{{Fall} \& {Efstathiou}}{{Fall} \&
  {Efstathiou}}{1980}]{fall80}
{Fall} S.~M.,  {Efstathiou} G.,  1980, \mn@doi [\mnras]
  {10.1093/mnras/193.2.189}, \href
  {https://ui.adsabs.harvard.edu/abs/1980MNRAS.193..189F} {193, 189}

\bibitem[\protect\citeauthoryear{{Farina} et~al.,}{{Farina}
  et~al.}{2022}]{farina22}
{Farina} E.~P.,  et~al., 2022, \mn@doi [\apj] {10.3847/1538-4357/ac9626}, \href
  {https://ui.adsabs.harvard.edu/abs/2022ApJ...941..106F} {941, 106}

\bibitem[\protect\citeauthoryear{{Ferland} et~al.,}{{Ferland}
  et~al.}{2017}]{ferland17}
{Ferland} G.~J.,  et~al., 2017, \mn@doi [\rmxaa] {10.48550/arXiv.1705.10877},
  \href {https://ui.adsabs.harvard.edu/abs/2017RMxAA..53..385F} {53, 385}

\bibitem[\protect\citeauthoryear{{Ferrara}, {Pallottini}  \& {Dayal}}{{Ferrara}
  et~al.}{2023}]{ferrara23}
{Ferrara} A.,  {Pallottini} A.,   {Dayal} P.,  2023, \mn@doi [\mnras]
  {10.1093/mnras/stad1095}, \href
  {https://ui.adsabs.harvard.edu/abs/2023MNRAS.522.3986F} {522, 3986}

\bibitem[\protect\citeauthoryear{{Fielding}, {Ostriker}, {Bryan}  \&
  {Jermyn}}{{Fielding} et~al.}{2020}]{fielding20}
{Fielding} D.~B.,  {Ostriker} E.~C.,  {Bryan} G.~L.,   {Jermyn} A.~S.,  2020,
  \mn@doi [\apjl] {10.3847/2041-8213/ab8d2c}, \href
  {https://ui.adsabs.harvard.edu/abs/2020ApJ...894L..24F} {894, L24}

\bibitem[\protect\citeauthoryear{{Finkelstein}, {Bagley}, {Arrabal Haro}  \&
  {CEERS}}{{Finkelstein} et~al.}{2022}]{finkelstein22b}
{Finkelstein} S.~L.,  {Bagley} M.~B.,  {Arrabal Haro} P.,   {CEERS} 2022,
  \mn@doi [\apjl] {10.3847/2041-8213/ac966e}, \href
  {https://ui.adsabs.harvard.edu/abs/2022ApJ...940L..55F} {940, L55}

\bibitem[\protect\citeauthoryear{{Finkelstein}, {Bagley}, {Ferguson}  \&
  {CEERS}}{{Finkelstein} et~al.}{2023}]{finkelstein23}
{Finkelstein} S.~L.,  {Bagley} M.~B.,  {Ferguson} H.~C.,   {CEERS} 2023,
  \mn@doi [\apjl] {10.3847/2041-8213/acade4}, \href
  {https://ui.adsabs.harvard.edu/abs/2023ApJ...946L..13F} {946, L13}

\bibitem[\protect\citeauthoryear{{Fiore}, {Ferrara}, {Bischetti}, {Feruglio}
  \& {Travascio}}{{Fiore} et~al.}{2023}]{fiore23}
{Fiore} F.,  {Ferrara} A.,  {Bischetti} M.,  {Feruglio} C.,   {Travascio} A.,
  2023, \mn@doi [\apjl] {10.3847/2041-8213/acb5f2}, \href
  {https://ui.adsabs.harvard.edu/abs/2023ApJ...943L..27F} {943, L27}

\bibitem[\protect\citeauthoryear{{F{\"o}rster Schreiber}, {Renzini}, {Mancini},
  {Genzel}, {Bouch{\'e}}, {Cresci}, {Hicks}  \& {et al.}}{{F{\"o}rster
  Schreiber} et~al.}{2018}]{forster18}
{F{\"o}rster Schreiber} N.~M.,  {Renzini} A.,  {Mancini} C.,  {Genzel} R.,
  {Bouch{\'e}} N.,  {Cresci} G.,  {Hicks} E.~K.~S.,   {et al.} 2018, \mn@doi
  [\apjs] {10.3847/1538-4365/aadd49}, \href
  {http://adsabs.harvard.edu/abs/2018ApJS..238...21F} {238, 21}

\bibitem[\protect\citeauthoryear{{F{\"o}rster Schreiber} et~al.,}{{F{\"o}rster
  Schreiber} et~al.}{2019}]{forster19}
{F{\"o}rster Schreiber} N.~M.,  et~al., 2019, \mn@doi [\apj]
  {10.3847/1538-4357/ab0ca2}, \href
  {https://ui.adsabs.harvard.edu/abs/2019ApJ...875...21F} {875, 21}

\bibitem[\protect\citeauthoryear{{Fujimoto} et~al.,}{{Fujimoto}
  et~al.}{2025}]{fujimoto25}
{Fujimoto} S.,  et~al., 2025, \mn@doi [Nature Astronomy]
  {10.1038/s41550-025-02592-w}, \href
  {https://ui.adsabs.harvard.edu/abs/2025NatAs.tmp..157F} {}

\bibitem[\protect\citeauthoryear{{Gabor} \& {Bournaud}}{{Gabor} \&
  {Bournaud}}{2014}]{gabor14}
{Gabor} J.~M.,  {Bournaud} F.,  2014, \mn@doi [\mnras] {10.1093/mnrasl/slt139},
  \href {https://ui.adsabs.harvard.edu/abs/2014MNRAS.437L..56G} {437, L56}

\bibitem[\protect\citeauthoryear{{Ginzburg}, {Dekel}, {Mandelker}  \&
  {Krumholz}}{{Ginzburg} et~al.}{2022}]{ginzburg22}
{Ginzburg} O.,  {Dekel} A.,  {Mandelker} N.,   {Krumholz} M.~R.,  2022, \mn@doi
  [\mnras] {10.1093/mnras/stac1324}, \href
  {https://ui.adsabs.harvard.edu/abs/2022MNRAS.513.6177G} {513, 6177}

\bibitem[\protect\citeauthoryear{{Ginzburg}, {Dekel}, {Mandelker}, {Dutta
  Chowdhury}, {Bournaud}, {Ceverino}  \& {Primack}}{{Ginzburg}
  et~al.}{2025}]{ginzburg25}
{Ginzburg} O.,  {Dekel} A.,  {Mandelker} N.,  {Dutta Chowdhury} D.,  {Bournaud}
  F.,  {Ceverino} D.,   {Primack} J.,  2025, \mn@doi [\aap]
  {10.1051/0004-6361/202553729}, \href
  {https://ui.adsabs.harvard.edu/abs/2025A&A...698A.110G} {698, A110}

\bibitem[\protect\citeauthoryear{{Goerdt}, {Moore}, {Read}, {Stadel}  \&
  {Zemp}}{{Goerdt} et~al.}{2006}]{goerdt06}
{Goerdt} T.,  {Moore} B.,  {Read} J.~I.,  {Stadel} J.,   {Zemp} M.,  2006,
  \mn@doi [\mnras] {10.1111/j.1365-2966.2006.10182.x}, \href
  {http://adsabs.harvard.edu/abs/2006MNRAS.368.1073G} {368, 1073}

\bibitem[\protect\citeauthoryear{{Gronke} \& {Oh}}{{Gronke} \&
  {Oh}}{2018}]{gronke18}
{Gronke} M.,  {Oh} S.~P.,  2018, \mn@doi [\mnras] {10.1093/mnrasl/sly131},
  \href {https://ui.adsabs.harvard.edu/abs/2018MNRAS.480L.111G} {480, L111}

\bibitem[\protect\citeauthoryear{{Gronke} \& {Oh}}{{Gronke} \&
  {Oh}}{2020a}]{gronke20b}
{Gronke} M.,  {Oh} S.~P.,  2020a, \mn@doi [\mnras] {10.1093/mnras/stz3332},
  \href {https://ui.adsabs.harvard.edu/abs/2020MNRAS.492.1970G} {492, 1970}

\bibitem[\protect\citeauthoryear{{Gronke} \& {Oh}}{{Gronke} \&
  {Oh}}{2020b}]{gronke20}
{Gronke} M.,  {Oh} S.~P.,  2020b, \mn@doi [\mnras] {10.1093/mnrasl/slaa033},
  \href {https://ui.adsabs.harvard.edu/abs/2020MNRAS.494L..27G} {494, L27}

\bibitem[\protect\citeauthoryear{{Gronke}, {Oh}, {Ji}  \& {Norman}}{{Gronke}
  et~al.}{2022}]{gronke22}
{Gronke} M.,  {Oh} S.~P.,  {Ji} S.,   {Norman} C.,  2022, \mn@doi [\mnras]
  {10.1093/mnras/stab3351}, \href
  {https://ui.adsabs.harvard.edu/abs/2022MNRAS.511..859G} {511, 859}

\bibitem[\protect\citeauthoryear{{Grudi{\'c}} \& {Hopkins}}{{Grudi{\'c}} \&
  {Hopkins}}{2023}]{grudic23}
{Grudi{\'c}} M.~Y.,  {Hopkins} P.~F.,  2023, \mn@doi [arXiv e-prints]
  {10.48550/arXiv.2308.16268}, \href
  {https://ui.adsabs.harvard.edu/abs/2023arXiv230816268G} {p. arXiv:2308.16268}

\bibitem[\protect\citeauthoryear{{Hachisu}}{{Hachisu}}{1979}]{hachisu79}
{Hachisu} I.,  1979, \pasj, \href
  {https://ui.adsabs.harvard.edu/abs/1979PASJ...31..523H} {31, 523}

\bibitem[\protect\citeauthoryear{{Harikane} et~al.,}{{Harikane}
  et~al.}{2023a}]{harikane23}
{Harikane} Y.,  et~al., 2023a, \mn@doi [\apjs] {10.3847/1538-4365/acaaa9},
  \href {https://ui.adsabs.harvard.edu/abs/2023ApJS..265....5H} {265, 5}

\bibitem[\protect\citeauthoryear{{Harikane} et~al.,}{{Harikane}
  et~al.}{2023b}]{harikane23_agn}
{Harikane} Y.,  et~al., 2023b, \mn@doi [\apj] {10.3847/1538-4357/ad029e}, \href
  {https://ui.adsabs.harvard.edu/abs/2023ApJ...959...39H} {959, 39}

\bibitem[\protect\citeauthoryear{{Harikane} et~al.,}{{Harikane}
  et~al.}{2025}]{harikane25}
{Harikane} Y.,  et~al., 2025, \mn@doi [\apj] {10.3847/1538-4357/ad9b2c}, \href
  {https://ui.adsabs.harvard.edu/abs/2025ApJ...980..138H} {980, 138}

\bibitem[\protect\citeauthoryear{{Haslbauer}, {Kroupa}, {Zonoozi}  \&
  {Haghi}}{{Haslbauer} et~al.}{2022}]{haslbauer22}
{Haslbauer} M.,  {Kroupa} P.,  {Zonoozi} A.~H.,   {Haghi} H.,  2022, \mn@doi
  [\apjl] {10.3847/2041-8213/ac9a50}, \href
  {https://ui.adsabs.harvard.edu/abs/2022ApJ...939L..31H} {939, L31}

\bibitem[\protect\citeauthoryear{{Heckman} \& {Best}}{{Heckman} \&
  {Best}}{2014}]{heckman14}
{Heckman} T.~M.,  {Best} P.~N.,  2014, \mn@doi [\araa]
  {10.1146/annurev-astro-081913-035722}, \href
  {https://ui.adsabs.harvard.edu/abs/2014ARA&A..52..589H} {52, 589}

\bibitem[\protect\citeauthoryear{{Huertas-Company} et~al.,}{{Huertas-Company}
  et~al.}{2018}]{huertas18}
{Huertas-Company} M.,  et~al., 2018, \mn@doi [\apj] {10.3847/1538-4357/aabfed},
  \href {https://ui.adsabs.harvard.edu/abs/2018ApJ...858..114H} {858, 114}

\bibitem[\protect\citeauthoryear{{Inayoshi}, {Harikane}, {Inoue}, {Li}  \&
  {Ho}}{{Inayoshi} et~al.}{2022}]{inayoshi22}
{Inayoshi} K.,  {Harikane} Y.,  {Inoue} A.~K.,  {Li} W.,   {Ho} L.~C.,  2022,
  \mn@doi [\apjl] {10.3847/2041-8213/ac9310}, \href
  {https://ui.adsabs.harvard.edu/abs/2022ApJ...938L..10I} {938, L10}

\bibitem[\protect\citeauthoryear{{Inoue}, {Dekel}, {Mandelker}, {Ceverino},
  {Bournaud}  \& {Primack}}{{Inoue} et~al.}{2016}]{inoue16}
{Inoue} S.,  {Dekel} A.,  {Mandelker} N.,  {Ceverino} D.,  {Bournaud} F.,
  {Primack} J.,  2016, \mn@doi [\mnras] {10.1093/mnras/stv2793}, \href
  {http://adsabs.harvard.edu/abs/2016MNRAS.456.2052I} {456, 2052}

\bibitem[\protect\citeauthoryear{{Jennings}, {Beckmann}, {Sijacki}  \&
  {Dubois}}{{Jennings} et~al.}{2023}]{jennings23}
{Jennings} F.,  {Beckmann} R.~S.,  {Sijacki} D.,   {Dubois} Y.,  2023, \mn@doi
  [\mnras] {10.1093/mnras/stac3426}, \href
  {https://ui.adsabs.harvard.edu/abs/2023MNRAS.518.5215J} {518, 5215}

\bibitem[\protect\citeauthoryear{{Ji}, {Oh}  \& {Masterson}}{{Ji}
  et~al.}{2019}]{ji19}
{Ji} S.,  {Oh} S.~P.,   {Masterson} P.,  2019, \mn@doi [\mnras]
  {10.1093/mnras/stz1248}, \href
  {https://ui.adsabs.harvard.edu/abs/2019MNRAS.487..737J} {487, 737}

\bibitem[\protect\citeauthoryear{{Ji} et~al.,}{{Ji} et~al.}{2024}]{ji24}
{Ji} Z.,  et~al., 2024, \mn@doi [arXiv e-prints] {10.48550/arXiv.2401.00934},
  \href {https://ui.adsabs.harvard.edu/abs/2024arXiv240100934J} {p.
  arXiv:2401.00934}

\bibitem[\protect\citeauthoryear{{Jiang} et~al.,}{{Jiang}
  et~al.}{2019}]{jiang19}
{Jiang} F.,  et~al., 2019, \mn@doi [\mnras] {10.1093/mnras/stz1952}, \href
  {https://ui.adsabs.harvard.edu/abs/2019MNRAS.488.4801J} {488, 4801}

\bibitem[\protect\citeauthoryear{{Kanjilal}, {Dutta}  \& {Sharma}}{{Kanjilal}
  et~al.}{2021}]{kanjilal21}
{Kanjilal} V.,  {Dutta} A.,   {Sharma} P.,  2021, \mn@doi [\mnras]
  {10.1093/mnras/staa3610}, \href
  {https://ui.adsabs.harvard.edu/abs/2021MNRAS.501.1143K} {501, 1143}

\bibitem[\protect\citeauthoryear{{Katz} et~al.,}{{Katz} et~al.}{2020}]{katz20}
{Katz} H.,  et~al., 2020, \mn@doi [\mnras] {10.1093/mnras/staa639}, \href
  {https://ui.adsabs.harvard.edu/abs/2020MNRAS.494.2200K} {494, 2200}

\bibitem[\protect\citeauthoryear{{Kaul}, {Tan}, {Oh}  \& {Mandelker}}{{Kaul}
  et~al.}{2025}]{kaul25}
{Kaul} I.,  {Tan} B.,  {Oh} S.~P.,   {Mandelker} N.,  2025, \mn@doi [\mnras]
  {10.1093/mnras/staf706}, \href
  {https://ui.adsabs.harvard.edu/abs/2025MNRAS.tmp..664K} {}

\bibitem[\protect\citeauthoryear{{Kaur} \& {Sridhar}}{{Kaur} \&
  {Sridhar}}{2018}]{kaur18}
{Kaur} K.,  {Sridhar} S.,  2018, \mn@doi [\apj] {10.3847/1538-4357/aaeacf},
  \href {https://ui.adsabs.harvard.edu/abs/2018ApJ...868..134K} {868, 134}

\bibitem[\protect\citeauthoryear{{Kere{\v s}}, {Katz}, {Weinberg}  \&
  {Dav{\'e}}}{{Kere{\v s}} et~al.}{2005}]{keres05}
{Kere{\v s}} D.,  {Katz} N.,  {Weinberg} D.~H.,   {Dav{\'e}} R.,  2005, \mnras,
  363, 2

\bibitem[\protect\citeauthoryear{{Kere{\v s}}, {Katz}, {Fardal}, {Dav{\'e}}  \&
  {Weinberg}}{{Kere{\v s}} et~al.}{2009}]{keres09}
{Kere{\v s}} D.,  {Katz} N.,  {Fardal} M.,  {Dav{\'e}} R.,   {Weinberg} D.~H.,
  2009, \mnras, 395, 160

\bibitem[\protect\citeauthoryear{{Klypin} et~al.,}{{Klypin}
  et~al.}{2021}]{klypin21}
{Klypin} A.,  et~al., 2021, \mn@doi [\mnras] {10.1093/mnras/stab769}, \href
  {https://ui.adsabs.harvard.edu/abs/2021MNRAS.504..769K} {504, 769}

\bibitem[\protect\citeauthoryear{{Kravtsov}, {Klypin}  \&
  {Khokhlov}}{{Kravtsov} et~al.}{1997}]{krav97}
{Kravtsov} A.~V.,  {Klypin} A.~A.,   {Khokhlov} A.~M.,  1997, \apjs, 111, 73

\bibitem[\protect\citeauthoryear{{Labb{\'e}} et~al.,}{{Labb{\'e}}
  et~al.}{2023}]{labbe23}
{Labb{\'e}} I.,  et~al., 2023, \mn@doi [\nat] {10.1038/s41586-023-05786-2},
  \href {https://ui.adsabs.harvard.edu/abs/2023Natur.616..266L} {616, 266}

\bibitem[\protect\citeauthoryear{{Lapiner}, {Dekel}  \& {Dubois}}{{Lapiner}
  et~al.}{2021}]{lapiner21}
{Lapiner} S.,  {Dekel} A.,   {Dubois} Y.,  2021, \mn@doi [\mnras]
  {10.1093/mnras/stab1205}, \href
  {https://ui.adsabs.harvard.edu/abs/2021MNRAS.505..172L} {505, 172}

\bibitem[\protect\citeauthoryear{{Lapiner} et~al.,}{{Lapiner}
  et~al.}{2023}]{lapiner23}
{Lapiner} S.,  et~al., 2023, \mn@doi [\mnras] {10.1093/mnras/stad1263}, \href
  {https://ui.adsabs.harvard.edu/abs/2023MNRAS.522.4515L} {522, 4515}

\bibitem[\protect\citeauthoryear{{Li}, {Dekel}, {Sarkar}, {Aung}, {Giavalisco},
  {Mandelker}  \& {Tacchella}}{{Li} et~al.}{2024}]{li24}
{Li} Z.,  {Dekel} A.,  {Sarkar} K.~C.,  {Aung} H.,  {Giavalisco} M.,
  {Mandelker} N.,   {Tacchella} S.,  2024, \mn@doi [\aap]
  {10.1051/0004-6361/202348727}, \href
  {https://ui.adsabs.harvard.edu/abs/2024A&A...690A.108L} {690, A108}

\bibitem[\protect\citeauthoryear{{Lochhaas}, {Tumlinson}, {O'Shea}, {Peeples},
  {Smith}, {Werk}, {Augustin}  \& {Simons}}{{Lochhaas}
  et~al.}{2021}]{lochhaas21}
{Lochhaas} C.,  {Tumlinson} J.,  {O'Shea} B.~W.,  {Peeples} M.~S.,  {Smith}
  B.~D.,  {Werk} J.~K.,  {Augustin} R.,   {Simons} R.~C.,  2021, \mn@doi [\apj]
  {10.3847/1538-4357/ac2496}, \href
  {https://ui.adsabs.harvard.edu/abs/2021ApJ...922..121L} {922, 121}

\bibitem[\protect\citeauthoryear{{Lovell}, {Harrison}, {Harikane}, {Tacchella}
  \& {Wilkins}}{{Lovell} et~al.}{2023}]{lovell23}
{Lovell} C.~C.,  {Harrison} I.,  {Harikane} Y.,  {Tacchella} S.,   {Wilkins}
  S.~M.,  2023, \mn@doi [\mnras] {10.1093/mnras/stac3224}, \href
  {https://ui.adsabs.harvard.edu/abs/2023MNRAS.518.2511L} {518, 2511}

\bibitem[\protect\citeauthoryear{{Lu}, {Mandelker}, {Oh}, {Dekel}, {van den
  Bosch}, {Springel}, {Nagai}  \& {van de Voort}}{{Lu} et~al.}{2024}]{lu24}
{Lu} Y.~S.,  {Mandelker} N.,  {Oh} S.~P.,  {Dekel} A.,  {van den Bosch} F.~C.,
  {Springel} V.,  {Nagai} D.,   {van de Voort} F.,  2024, \mn@doi [\mnras]
  {10.1093/mnras/stad3779}, \href
  {https://ui.adsabs.harvard.edu/abs/2024MNRAS.52711256L} {527, 11256}

\bibitem[\protect\citeauthoryear{{Madau} \& {Dickinson}}{{Madau} \&
  {Dickinson}}{2014}]{madau14}
{Madau} P.,  {Dickinson} M.,  2014, \mn@doi [\araa]
  {10.1146/annurev-astro-081811-125615}, \href
  {https://ui.adsabs.harvard.edu/abs/2014ARA&A..52..415M} {52, 415}

\bibitem[\protect\citeauthoryear{{Madau} \& {Quataert}}{{Madau} \&
  {Quataert}}{2004}]{madau04}
{Madau} P.,  {Quataert} E.,  2004, \mn@doi [\apjl] {10.1086/421017}, \href
  {https://ui.adsabs.harvard.edu/abs/2004ApJ...606L..17M} {606, L17}

\bibitem[\protect\citeauthoryear{{Maiolino} et~al.,}{{Maiolino}
  et~al.}{2024}]{maiolino24}
{Maiolino} R.,  et~al., 2024, \mn@doi [\aap] {10.1051/0004-6361/202347640},
  \href {https://ui.adsabs.harvard.edu/abs/2024A&A...691A.145M} {691, A145}

\bibitem[\protect\citeauthoryear{{Mandelker}, {Padnos}, {Dekel}, {Birnboim},
  {Burkert}, {Krumholz}  \& {Steinberg}}{{Mandelker}
  et~al.}{2016}]{mandelker16}
{Mandelker} N.,  {Padnos} D.,  {Dekel} A.,  {Birnboim} Y.,  {Burkert} A.,
  {Krumholz} M.~R.,   {Steinberg} E.,  2016, \mn@doi [\mnras]
  {10.1093/mnras/stw2267}, \href
  {http://adsabs.harvard.edu/abs/2016MNRAS.463.3921M} {463, 3921}

\bibitem[\protect\citeauthoryear{{Mandelker}, {van Dokkum}, {Brodie}, {van den
  Bosch}  \& {Ceverino}}{{Mandelker} et~al.}{2018}]{mandelker18}
{Mandelker} N.,  {van Dokkum} P.~G.,  {Brodie} J.~P.,  {van den Bosch} F.~C.,
  {Ceverino} D.,  2018, \mn@doi [\apj] {10.3847/1538-4357/aaca98}, \href
  {https://ui.adsabs.harvard.edu/abs/2018ApJ...861..148M} {861, 148}

\bibitem[\protect\citeauthoryear{{Mandelker}, {Nagai}, {Aung}, {Dekel},
  {Padnos}  \& {Birnboim}}{{Mandelker} et~al.}{2019}]{mandelker19a}
{Mandelker} N.,  {Nagai} D.,  {Aung} H.,  {Dekel} A.,  {Padnos} D.,
  {Birnboim} Y.,  2019, \mn@doi [\mnras] {10.1093/mnras/stz012}, \href
  {https://ui.adsabs.harvard.edu/abs/2019MNRAS.484.1100M} {484, 1100}

\bibitem[\protect\citeauthoryear{{Mandelker}, {Nagai}, {Aung}, {Dekel},
  {Birnboim}  \& {van den Bosch}}{{Mandelker} et~al.}{2020a}]{mandelker20a}
{Mandelker} N.,  {Nagai} D.,  {Aung} H.,  {Dekel} A.,  {Birnboim} Y.,   {van
  den Bosch} F.~C.,  2020a, \mn@doi [\mnras] {10.1093/mnras/staa812}, \href
  {https://ui.adsabs.harvard.edu/abs/2020MNRAS.494.2641M} {494, 2641}

\bibitem[\protect\citeauthoryear{{Mandelker}, {van den Bosch}, {Nagai},
  {Dekel}, {Birnboim}  \& {Aung}}{{Mandelker} et~al.}{2020b}]{mandelker20b}
{Mandelker} N.,  {van den Bosch} F.~C.,  {Nagai} D.,  {Dekel} A.,  {Birnboim}
  Y.,   {Aung} H.,  2020b, \mn@doi [\mnras] {10.1093/mnras/staa2421}, \href
  {https://ui.adsabs.harvard.edu/abs/2020MNRAS.498.2415M} {498, 2415}

\bibitem[\protect\citeauthoryear{{Mandelker}, {van den Bosch}, {Springel}, {van
  de Voort}, {Burchett}, {Butsky}, {Nagai}  \& {Oh}}{{Mandelker}
  et~al.}{2021}]{mandelker21}
{Mandelker} N.,  {van den Bosch} F.~C.,  {Springel} V.,  {van de Voort} F.,
  {Burchett} J.~N.,  {Butsky} I.~S.,  {Nagai} D.,   {Oh} S.~P.,  2021, \mn@doi
  [\apj] {10.3847/1538-4357/ac2d29}, \href
  {https://ui.adsabs.harvard.edu/abs/2021ApJ...923..115M} {923, 115}

\bibitem[\protect\citeauthoryear{{Mandelker}, {Ginzburg}, {Dekel}, {Bournaud},
  {Krumholz}, {Ceverino}  \& {Primack}}{{Mandelker} et~al.}{2025}]{mandelker25}
{Mandelker} N.,  {Ginzburg} O.,  {Dekel} A.,  {Bournaud} F.,  {Krumholz} M.~R.,
   {Ceverino} D.,   {Primack} J.,  2025, \mn@doi [\mnras]
  {10.1093/mnrasl/slae122}, \href
  {https://ui.adsabs.harvard.edu/abs/2025MNRAS.538L...9M} {538, L9}

\bibitem[\protect\citeauthoryear{{Martig}, {Bournaud}, {Teyssier}  \&
  {Dekel}}{{Martig} et~al.}{2009}]{martig09}
{Martig} M.,  {Bournaud} F.,  {Teyssier} R.,   {Dekel} A.,  2009, \apj, 707,
  250

\bibitem[\protect\citeauthoryear{{Mason}, {Trenti}  \& {Treu}}{{Mason}
  et~al.}{2023}]{mason23}
{Mason} C.~A.,  {Trenti} M.,   {Treu} T.,  2023, \mn@doi [\mnras]
  {10.1093/mnras/stad035}, \href
  {https://ui.adsabs.harvard.edu/abs/2023MNRAS.521..497M} {521, 497}

\bibitem[\protect\citeauthoryear{{Mayer}, {van Donkelaar}, {Messa}, {Capelo}
  \& {Adamo}}{{Mayer} et~al.}{2025}]{mayer25}
{Mayer} L.,  {van Donkelaar} F.,  {Messa} M.,  {Capelo} P.~R.,   {Adamo} A.,
  2025, \mn@doi [\apjl] {10.3847/2041-8213/adadfe}, \href
  {https://ui.adsabs.harvard.edu/abs/2025ApJ...981L..28M} {981, L28}

\bibitem[\protect\citeauthoryear{{McCourt}, {Oh}, {O'Leary}  \&
  {Madigan}}{{McCourt} et~al.}{2018}]{mccourt18}
{McCourt} M.,  {Oh} S.~P.,  {O'Leary} R.,   {Madigan} A.-M.,  2018, \mn@doi
  [\mnras] {10.1093/mnras/stx2687}, \href
  {https://ui.adsabs.harvard.edu/abs/2018MNRAS.473.5407M} {473, 5407}

\bibitem[\protect\citeauthoryear{{Menon}, {Federrath}  \& {Krumholz}}{{Menon}
  et~al.}{2023}]{menon23}
{Menon} S.~H.,  {Federrath} C.,   {Krumholz} M.~R.,  2023, \mn@doi [\mnras]
  {10.1093/mnras/stad856}, \href
  {https://ui.adsabs.harvard.edu/abs/2023MNRAS.521.5160M} {521, 5160}

\bibitem[\protect\citeauthoryear{{Menon}, {Lancaster}, {Burkhart},
  {Somerville}, {Dekel}  \& {Krumholz}}{{Menon} et~al.}{2024}]{menon24}
{Menon} S.~H.,  {Lancaster} L.,  {Burkhart} B.,  {Somerville} R.~S.,  {Dekel}
  A.,   {Krumholz} M.~R.,  2024, \mn@doi [\apjl] {10.3847/2041-8213/ad462d},
  \href {https://ui.adsabs.harvard.edu/abs/2024ApJ...967L..28M} {967, L28}

\bibitem[\protect\citeauthoryear{{Messa}, {Dessauges-Zavadsky}, {Adamo},
  {Richard}  \& {Claeyssens}}{{Messa} et~al.}{2024}]{messa24}
{Messa} M.,  {Dessauges-Zavadsky} M.,  {Adamo} A.,  {Richard} J.,
  {Claeyssens} A.,  2024, \mn@doi [\mnras] {10.1093/mnras/stae565}, \href
  {https://ui.adsabs.harvard.edu/abs/2024MNRAS.529.2162M} {529, 2162}

\bibitem[\protect\citeauthoryear{{Moster}, {Naab}  \& {White}}{{Moster}
  et~al.}{2018}]{moster18}
{Moster} B.~P.,  {Naab} T.,   {White} S. D.~M.,  2018, \mn@doi [\mnras]
  {10.1093/mnras/sty655}, \href
  {https://ui.adsabs.harvard.edu/abs/2018MNRAS.477.1822M} {477, 1822}

\bibitem[\protect\citeauthoryear{{Moster}, {Naab}  \& {White}}{{Moster}
  et~al.}{2020}]{moster20}
{Moster} B.~P.,  {Naab} T.,   {White} S. D.~M.,  2020, \mn@doi [\mnras]
  {10.1093/mnras/staa3019}, \href
  {https://ui.adsabs.harvard.edu/abs/2020MNRAS.499.4748M} {499, 4748}

\bibitem[\protect\citeauthoryear{{Mowla}}{{Mowla}}{2024}]{mowla24_gems}
{Mowla} L.,  2024, \mn@doi [\nat] {10.1038/d41586-024-02438-x}, \href
  {https://ui.adsabs.harvard.edu/abs/2024Natur.632..505M} {632, 505}

\bibitem[\protect\citeauthoryear{{Mowla} et~al.,}{{Mowla}
  et~al.}{2024a}]{mowla24}
{Mowla} L.,  et~al., 2024a, \mn@doi [arXiv e-prints]
  {10.48550/arXiv.2402.08696}, \href
  {https://ui.adsabs.harvard.edu/abs/2024arXiv240208696M} {p. arXiv:2402.08696}

\bibitem[\protect\citeauthoryear{{Mowla} et~al.,}{{Mowla}
  et~al.}{2024b}]{mowla24_nat}
{Mowla} L.,  et~al., 2024b, \mn@doi [\nat] {10.1038/s41586-024-08293-0}, \href
  {https://ui.adsabs.harvard.edu/abs/2024Natur.636..332M} {636, 332}

\bibitem[\protect\citeauthoryear{{Naidu} et~al.,}{{Naidu}
  et~al.}{2022}]{naidu22}
{Naidu} R.~P.,  et~al., 2022, \mn@doi [\apjl] {10.3847/2041-8213/ac9b22}, \href
  {https://ui.adsabs.harvard.edu/abs/2022ApJ...940L..14N} {940, L14}

\bibitem[\protect\citeauthoryear{{Ono}, {Ouchi}, {Harikane}, {Yajima},
  {Nakajima}, {Fujimoto}, {Nakane}  \& {Xu}}{{Ono} et~al.}{2025}]{ono25}
{Ono} Y.,  {Ouchi} M.,  {Harikane} Y.,  {Yajima} H.,  {Nakajima} K.,
  {Fujimoto} S.,  {Nakane} M.,   {Xu} Y.,  2025, \mn@doi [arXiv e-prints]
  {10.48550/arXiv.2502.08885}, \href
  {https://ui.adsabs.harvard.edu/abs/2025arXiv250208885O} {p. arXiv:2502.08885}

\bibitem[\protect\citeauthoryear{{Onoue} et~al.,}{{Onoue}
  et~al.}{2024}]{onoue24}
{Onoue} M.,  et~al., 2024, \mn@doi [arXiv e-prints]
  {10.48550/arXiv.2409.07113}, \href
  {https://ui.adsabs.harvard.edu/abs/2024arXiv240907113O} {p. arXiv:2409.07113}

\bibitem[\protect\citeauthoryear{{Ostriker}}{{Ostriker}}{1964}]{ostriker64}
{Ostriker} J.,  1964, \mn@doi [\apj] {10.1086/148005}, \href
  {https://ui.adsabs.harvard.edu/abs/1964ApJ...140.1056O} {140, 1056}

\bibitem[\protect\citeauthoryear{{Padnos}, {Mandelker}, {Birnboim}, {Dekel},
  {Krumholz}  \& {Steinberg}}{{Padnos} et~al.}{2018}]{padnos18}
{Padnos} D.,  {Mandelker} N.,  {Birnboim} Y.,  {Dekel} A.,  {Krumholz} M.~R.,
  {Steinberg} E.,  2018, \mn@doi [\mnras] {10.1093/mnras/sty789}, \href
  {https://ui.adsabs.harvard.edu/abs/2018MNRAS.477.3293P} {477, 3293}

\bibitem[\protect\citeauthoryear{{P{\'e}rez-Gonz{\'a}lez}, {Costantin},
  {Langero odi}, {Rinaldi}, {Annunziatella}, {Ilbert}, {Colina}  \& {et
  al.}}{{P{\'e}rez-Gonz{\'a}lez} et~al.}{2023}]{perez23}
{P{\'e}rez-Gonz{\'a}lez} P.~G.,  {Costantin} L.,  {Langero odi} D.,  {Rinaldi}
  P.,  {Annunziatella} M.,  {Ilbert} O.,  {Colina} L.,   {et al.} 2023, \mn@doi
  [\apjl] {10.3847/2041-8213/acd9d0}, \href
  {https://ui.adsabs.harvard.edu/abs/2023ApJ...951L...1P} {951, L1}

\bibitem[\protect\citeauthoryear{{P{\'e}rez-Gonz{\'a}lez}
  et~al.,}{{P{\'e}rez-Gonz{\'a}lez} et~al.}{2025}]{perez25}
{P{\'e}rez-Gonz{\'a}lez} P.~G.,  et~al., 2025, \mn@doi [arXiv e-prints]
  {10.48550/arXiv.2503.15594}, \href
  {https://ui.adsabs.harvard.edu/abs/2025arXiv250315594P} {p. arXiv:2503.15594}

\bibitem[\protect\citeauthoryear{{Press} \& {Schechter}}{{Press} \&
  {Schechter}}{1974}]{press74}
{Press} W.~H.,  {Schechter} P.,  1974, \apj, 187, 425

\bibitem[\protect\citeauthoryear{{Read}, {Goerdt}, {Moore}, {Pontzen}, {Stadel}
   \& {Lake}}{{Read} et~al.}{2006}]{read06}
{Read} J.~I.,  {Goerdt} T.,  {Moore} B.,  {Pontzen} A.~P.,  {Stadel} J.,
  {Lake} G.,  2006, \mn@doi [\mnras] {10.1111/j.1365-2966.2006.11022.x}, \href
  {https://ui.adsabs.harvard.edu/abs/2006MNRAS.373.1451R} {373, 1451}

\bibitem[\protect\citeauthoryear{{Reines} \& {Volonteri}}{{Reines} \&
  {Volonteri}}{2015}]{reines15}
{Reines} A.~E.,  {Volonteri} M.,  2015, \mn@doi [\apj]
  {10.1088/0004-637X/813/2/82}, \href
  {https://ui.adsabs.harvard.edu/abs/2015ApJ...813...82R} {813, 82}

\bibitem[\protect\citeauthoryear{{Setton} et~al.,}{{Setton}
  et~al.}{2024}]{setton24}
{Setton} D.~J.,  et~al., 2024, \mn@doi [\apj] {10.3847/1538-4357/ad6a18}, \href
  {https://ui.adsabs.harvard.edu/abs/2024ApJ...974..145S} {974, 145}

\bibitem[\protect\citeauthoryear{{Shuntov} et~al.,}{{Shuntov}
  et~al.}{2022}]{shuntov22}
{Shuntov} M.,  et~al., 2022, \mn@doi [\aap] {10.1051/0004-6361/202243136},
  \href {https://ui.adsabs.harvard.edu/abs/2022A&A...664A..61S} {664, A61}

\bibitem[\protect\citeauthoryear{{Sparre}, {Pfrommer}  \& {Ehlert}}{{Sparre}
  et~al.}{2020}]{sparre20}
{Sparre} M.,  {Pfrommer} C.,   {Ehlert} K.,  2020, \mn@doi [\mnras]
  {10.1093/mnras/staa3177}, \href
  {https://ui.adsabs.harvard.edu/abs/2020MNRAS.499.4261S} {499, 4261}

\bibitem[\protect\citeauthoryear{{Stefanon}, {Bouwens}, {Labb{\'e}},
  {Illingworth}, {Gonzalez}  \& {Oesch}}{{Stefanon} et~al.}{2021}]{stefanon21}
{Stefanon} M.,  {Bouwens} R.~J.,  {Labb{\'e}} I.,  {Illingworth} G.~D.,
  {Gonzalez} V.,   {Oesch} P.~A.,  2021, \mn@doi [\apj]
  {10.3847/1538-4357/ac1bb6}, \href
  {https://ui.adsabs.harvard.edu/abs/2021ApJ...922...29S} {922, 29}

\bibitem[\protect\citeauthoryear{{Steinhardt}, {Kokorev}, {Rusakov}, {Garcia}
  \& {Sneppen}}{{Steinhardt} et~al.}{2023}]{steinhardt23}
{Steinhardt} C.~L.,  {Kokorev} V.,  {Rusakov} V.,  {Garcia} E.,   {Sneppen} A.,
   2023, \mn@doi [\apjl] {10.3847/2041-8213/acdef6}, \href
  {https://ui.adsabs.harvard.edu/abs/2023ApJ...951L..40S} {951, L40}

\bibitem[\protect\citeauthoryear{{Sun}, {Faucher-Gigu{\`e}re}, {Hayward}  \&
  {Shen}}{{Sun} et~al.}{2023a}]{sun23a}
{Sun} G.,  {Faucher-Gigu{\`e}re} C.-A.,  {Hayward} C.~C.,   {Shen} X.,  2023a,
  \mn@doi [\mnras] {10.1093/mnras/stad2902}, \href
  {https://ui.adsabs.harvard.edu/abs/2023MNRAS.526.2665S} {526, 2665}

\bibitem[\protect\citeauthoryear{{Sun}, {Faucher-Gigu{\`e}re}, {Hayward},
  {Shen}, {Wetzel}  \& {Cochrane }}{{Sun} et~al.}{2023b}]{sun23b}
{Sun} G.,  {Faucher-Gigu{\`e}re} C.-A.,  {Hayward} C.~C.,  {Shen} X.,  {Wetzel}
  A.,   {Cochrane } R.~K.,  2023b, \mn@doi [\apjl] {10.3847/2041-8213/acf85a},
  \href {https://ui.adsabs.harvard.edu/abs/2023ApJ...955L..35S} {955, L35}

\bibitem[\protect\citeauthoryear{{Tacchella}, {Dekel}, {Carollo}, {Ceverino},
  {DeGraf}, {Lapiner}, {Mandelker}  \& {Primack Joel}}{{Tacchella}
  et~al.}{2016a}]{tacchella16_ms}
{Tacchella} S.,  {Dekel} A.,  {Carollo} C.~M.,  {Ceverino} D.,  {DeGraf} C.,
  {Lapiner} S.,  {Mandelker} N.,   {Primack Joel} R.,  2016a, \mn@doi [\mnras]
  {10.1093/mnras/stw131}, \href
  {http://adsabs.harvard.edu/abs/2016MNRAS.457.2790T} {457, 2790}

\bibitem[\protect\citeauthoryear{{Tacchella}, {Dekel}, {Carollo}, {Ceverino},
  {DeGraf}, {Lapiner}, {Mandelker}  \& {Primack}}{{Tacchella}
  et~al.}{2016b}]{tacchella16_prof}
{Tacchella} S.,  {Dekel} A.,  {Carollo} C.~M.,  {Ceverino} D.,  {DeGraf} C.,
  {Lapiner} S.,  {Mandelker} N.,   {Primack} J.~R.,  2016b, \mn@doi [\mnras]
  {10.1093/mnras/stw303}, \href
  {http://adsabs.harvard.edu/abs/2016MNRAS.458..242T} {458, 242}

\bibitem[\protect\citeauthoryear{{Tacchella} et~al.,}{{Tacchella}
  et~al.}{2023}]{tacchella23b}
{Tacchella} S.,  et~al., 2023, \mn@doi [\apj] {10.3847/1538-4357/acdbc6}, \href
  {https://ui.adsabs.harvard.edu/abs/2023ApJ...952...74T} {952, 74}

\bibitem[\protect\citeauthoryear{{Tan}, {Oh}  \& {Gronke}}{{Tan}
  et~al.}{2021}]{tan21}
{Tan} B.,  {Oh} S.~P.,   {Gronke} M.,  2021, \mn@doi [\mnras]
  {10.1093/mnras/stab053}, \href
  {https://ui.adsabs.harvard.edu/abs/2021MNRAS.502.3179T} {502, 3179}

\bibitem[\protect\citeauthoryear{{Teyssier}}{{Teyssier}}{2002}]{teyssier02}
{Teyssier} R.,  2002, \aap, 385, 337

\bibitem[\protect\citeauthoryear{{Tomassetti} et~al.,}{{Tomassetti}
  et~al.}{2016}]{tomassetti16}
{Tomassetti} M.,  et~al., 2016, \mn@doi [\mnras] {10.1093/mnras/stw606}, \href
  {http://adsabs.harvard.edu/abs/2016MNRAS.458.4477T} {458, 4477}

\bibitem[\protect\citeauthoryear{{Trakhtenbrot}, {Netzer}, {Lira}  \&
  {Shemmer}}{{Trakhtenbrot} et~al.}{2011}]{trakhtenbrot11}
{Trakhtenbrot} B.,  {Netzer} H.,  {Lira} P.,   {Shemmer} O.,  2011, \mn@doi
  [\apj] {10.1088/0004-637X/730/1/7}, \href
  {https://ui.adsabs.harvard.edu/abs/2011ApJ...730....7T} {730, 7}

\bibitem[\protect\citeauthoryear{{Turner} et~al.,}{{Turner}
  et~al.}{2025}]{turner25}
{Turner} C.,  et~al., 2025, \mn@doi [\mnras] {10.1093/mnras/staf128}, \href
  {https://ui.adsabs.harvard.edu/abs/2025MNRAS.537.1826T} {537, 1826}

\bibitem[\protect\citeauthoryear{{{\"U}bler} et~al.,}{{{\"U}bler}
  et~al.}{2023}]{ubler23}
{{\"U}bler} H.,  et~al., 2023, \mn@doi [\aap] {10.1051/0004-6361/202346137},
  \href {https://ui.adsabs.harvard.edu/abs/2023A&A...677A.145U} {677, A145}

\bibitem[\protect\citeauthoryear{{Valentino} et~al.,}{{Valentino}
  et~al.}{2025}]{valentino25}
{Valentino} F.,  et~al., 2025, \mn@doi [\aap] {10.1051/0004-6361/202553908},
  \href {https://ui.adsabs.harvard.edu/abs/2025A&A...699A.358V} {699, A358}

\bibitem[\protect\citeauthoryear{{Vanzella} et~al.,}{{Vanzella}
  et~al.}{2023}]{vanzella23}
{Vanzella} E.,  et~al., 2023, \mn@doi [\apj] {10.3847/1538-4357/acb59a}, \href
  {https://ui.adsabs.harvard.edu/abs/2023ApJ...945...53V} {945, 53}

\bibitem[\protect\citeauthoryear{{Wang} et~al.,}{{Wang} et~al.}{2024}]{wang24}
{Wang} B.,  et~al., 2024, \mn@doi [\apjl] {10.3847/2041-8213/ad55f7}, \href
  {https://ui.adsabs.harvard.edu/abs/2024ApJ...969L..13W} {969, L13}

\bibitem[\protect\citeauthoryear{{Watson}, {Iliev}, {D'Aloisio}, {Knebe},
  {Shapiro}  \& {Yepes}}{{Watson} et~al.}{2013}]{watson13}
{Watson} W.~A.,  {Iliev} I.~T.,  {D'Aloisio} A.,  {Knebe} A.,  {Shapiro} P.~R.,
    {Yepes} G.,  2013, \mn@doi [\mnras] {10.1093/mnras/stt791}, \href
  {https://ui.adsabs.harvard.edu/abs/2013MNRAS.433.1230W} {433, 1230}

\bibitem[\protect\citeauthoryear{{Wechsler}, {Bullock}, {Primack}, {Kravtsov}
  \& {Dekel}}{{Wechsler} et~al.}{2002}]{wechsler02}
{Wechsler} R.~H.,  {Bullock} J.~S.,  {Primack} J.~R.,  {Kravtsov} A.~V.,
  {Dekel} A.,  2002, \apj, 568, 52

\bibitem[\protect\citeauthoryear{{Weibel} et~al.,}{{Weibel}
  et~al.}{2024}]{weibel24_smf}
{Weibel} A.,  et~al., 2024, \mn@doi [\mnras] {10.1093/mnras/stae1891}, \href
  {https://ui.adsabs.harvard.edu/abs/2024MNRAS.533.1808W} {533, 1808}

\bibitem[\protect\citeauthoryear{{Weibel} et~al.,}{{Weibel}
  et~al.}{2025}]{weibel25}
{Weibel} A.,  et~al., 2025, \mn@doi [\apj] {10.3847/1538-4357/adab7a}, \href
  {https://ui.adsabs.harvard.edu/abs/2025ApJ...983...11W} {983, 11}

\bibitem[\protect\citeauthoryear{{Wilkins} et~al.,}{{Wilkins}
  et~al.}{2023}]{wilkins23}
{Wilkins} S.~M.,  et~al., 2023, \mn@doi [\mnras] {10.1093/mnras/stac3280},
  \href {https://ui.adsabs.harvard.edu/abs/2023MNRAS.519.3118W} {519, 3118}

\bibitem[\protect\citeauthoryear{{Xie} et~al.,}{{Xie} et~al.}{2024}]{xie24}
{Xie} L.,  et~al., 2024, \mn@doi [\apjl] {10.3847/2041-8213/ad380a}, \href
  {https://ui.adsabs.harvard.edu/abs/2024ApJ...966L...2X} {966, L2}

\bibitem[\protect\citeauthoryear{{Yang} \& {Ji}}{{Yang} \& {Ji}}{2023}]{yang23}
{Yang} Y.,  {Ji} S.,  2023, \mn@doi [\mnras] {10.1093/mnras/stad264}, \href
  {https://ui.adsabs.harvard.edu/abs/2023MNRAS.520.2148Y} {520, 2148}

\bibitem[\protect\citeauthoryear{{Yao}, {Mandelker}, {Oh}, {Aung}  \&
  {Dekel}}{{Yao} et~al.}{2025}]{yao25}
{Yao} Z.,  {Mandelker} N.,  {Oh} S.~P.,  {Aung} H.,   {Dekel} A.,  2025,
  \mn@doi [\mnras] {10.1093/mnras/stae2771}, \href
  {https://ui.adsabs.harvard.edu/abs/2025MNRAS.536.3053Y} {536, 3053}

\bibitem[\protect\citeauthoryear{{Yung}, {Somerville}, {Finkelstein}, {Wilkins}
   \& {Gardner}}{{Yung} et~al.}{2024}]{yung24}
{Yung} L.~Y.~A.,  {Somerville} R.~S.,  {Finkelstein} S.~L.,  {Wilkins} S.~M.,
  {Gardner} J.~P.,  2024, \mn@doi [\mnras] {10.1093/mnras/stad3484}, \href
  {https://ui.adsabs.harvard.edu/abs/2024MNRAS.527.5929Y} {527, 5929}

\bibitem[\protect\citeauthoryear{{Zackrisson}, {Rydberg}, {Schaerer},
  {{\"O}stlin}  \& {Tuli}}{{Zackrisson} et~al.}{2011}]{zackrisson11}
{Zackrisson} E.,  {Rydberg} C.-E.,  {Schaerer} D.,  {{\"O}stlin} G.,   {Tuli}
  M.,  2011, \mn@doi [\apj] {10.1088/0004-637X/740/1/13}, \href
  {https://ui.adsabs.harvard.edu/abs/2011ApJ...740...13Z} {740, 13}

\bibitem[\protect\citeauthoryear{{Zhao}, {Jing}, {Mo}  \& {B{\"o}rner}}{{Zhao}
  et~al.}{2009}]{zhao09}
{Zhao} D.~H.,  {Jing} Y.~P.,  {Mo} H.~J.,   {B{\"o}rner} G.,  2009, \mn@doi
  [\apj] {10.1088/0004-637X/707/1/354}, \href
  {http://adsabs.harvard.edu/abs/2009ApJ...707..354Z} {707, 354}

\bibitem[\protect\citeauthoryear{{Ziparo}, {Ferrara}  \& {Sommovigo}}{{Ziparo}
  et~al.}{2023}]{ziparo23}
{Ziparo} F.,  {Ferrara} A.,   {Sommovigo} Laura an d~{Kohandel} M.,  2023,
  \mn@doi [\mnras] {10.1093/mnras/stad125}, \href
  {https://ui.adsabs.harvard.edu/abs/2023MNRAS.520.2445Z} {520, 2445}

\bibitem[\protect\citeauthoryear{{Zolotov} et~al.,}{{Zolotov}
  et~al.}{2015}]{zolotov15}
{Zolotov} A.,  et~al., 2015, \mn@doi [\mnras] {10.1093/mnras/stv740}, \href
  {http://adsabs.harvard.edu/abs/2015MNRAS.450.2327Z} {450, 2327}

\bibitem[\protect\citeauthoryear{{de Graaff} et~al.,}{{de Graaff}
  et~al.}{2024a}]{degraaff24}
{de Graaff} A.,  et~al., 2024a, \mn@doi [Nature Astronomy]
  {10.1038/s41550-024-02424-3}, \href
  {https://ui.adsabs.harvard.edu/abs/2024NatAs.tmp..284D} {}

\bibitem[\protect\citeauthoryear{{de Graaff}, {Pillepich}  \& {Rix}}{{de
  Graaff} et~al.}{2024b}]{degraaff24_compaction}
{de Graaff} A.,  {Pillepich} A.,   {Rix} H.-W.,  2024b, \mn@doi [\apjl]
  {10.3847/2041-8213/ad4c65}, \href
  {https://ui.adsabs.harvard.edu/abs/2024ApJ...967L..40D} {967, L40}

\bibitem[\protect\citeauthoryear{{van Dokkum} et~al.,}{{van Dokkum}
  et~al.}{2015}]{dokkum15}
{van Dokkum} P.~G.,  et~al., 2015, \mn@doi [\apj] {10.1088/0004-637X/813/1/23},
  \href {http://adsabs.harvard.edu/abs/2015ApJ...813...23V} {813, 23}

\makeatother
\end{thebibliography}



\label{lastpage}
\end{document}